\documentclass[twocolumn,twocolappendix]{aastex701}
\usepackage{graphicx,natbib,color,bm,url,times,hyperref,orcidlink}

\usepackage[T1]{fontenc}
\usepackage{amsmath}
\usepackage{longtable}

\received{2025 December 31}
\revised{2026 March 24}
\accepted{2026 March 30}
\published{2026 May 18}

\begin{document}

\title[show]{The Longest-period Young Transiting Exoplanets---A Duo of Puffy Giants inside 
a Debris Disk\footnote{This study uses CHEOPS data observed as part of the Guest Observers (GO) programmes CH\_PR240025 (PI del Burgo) and CH\_PR250012 (PI del Burgo), and the Guaranteed Time Observation (GTO) programmes CH\_PR100017 and CH\_PR140079.}}

\author[orcid=0000-0002-8949-5200,sname='del Burgo']{Carlos del Burgo}
\affil{Instituto de Astrof\'\i sica de Canarias, V\'\i a L\'actea S/N, La Laguna, E-38200, Tenerife, Spain}
\affil{Departamento de Astrof\'\i sica, Universidad de la Laguna, La Laguna, E-38200, Tenerife, Spain}
\email[show]{cburgo@ll.edu.es}
\author[orcid=0000-0002-3814-5323, sname='Suarez Mascare\~no']{Alejandro Suarez Mascare\~no}
\affil{Instituto de Astrof\'\i sica de Canarias, V\'\i a L\'actea S/N, La Laguna, E-38200, Tenerife, Spain}
\affil{Departamento de Astrof\'\i sica, Universidad de la Laguna, La Laguna, E-38200, Tenerife, Spain}
\email[show]{alejandro.suarez.mascareno@iac.es}
\author[orcid=0000-0002-6342-9600, sname='Heras']{Ana Heras}
\affil{European Space Agency (ESA), European Space Research and Technology Centre (ESTEC), Keplerlaan 1, 2201 AZ Noordwijk, The Netherlands
}
\email[show]{Ana.Heras@esa.int}
\author[orcid=0000-0001-6208-1801, sname='Marshall']{Jonathan P. Marshall}
\affil{Academia Sinica Institute of Astronomy and Astrophysics, 11F of AS/NTU Astronomy-Mathematics Building, No.1, Sect. 4, Roosevelt Rd, Taipei 106319, Taiwan}
\email[]{jmarshall@asiaa.sinica.edu.tw}
\author[orcid=0000-0003-1452-2240, sname='Wheatley']{Peter J.\ Wheatley}
\affil{Department of Physics, University of Warwick, Gibbet Hill Road, Coventry CV4 7AL, UK}
\affil{Centre for Exoplanets and Habitability, University of Warwick, Gibbet Hill Road, Coventry CV4 7AL, UK}
\email[]{p.j.wheatley@warwick.ac.uk}
\author[orcid=0000-0001-7904-4441, sname='Bryant']{Edward M. Bryant}
\affil{Department of Physics, University of Warwick, Gibbet Hill Road, Coventry CV4 7AL, UK}
\affil{Centre for Exoplanets and Habitability, University of Warwick, Gibbet Hill Road, Coventry CV4 7AL, UK}
\email[]{edward.m.bryant@warwick.ac.uk}
\author[orcid=0000-0002-4259-0155, sname='Gill']{Samuel Gill}
\affil{Department of Physics, University of Warwick, Gibbet Hill Road, Coventry CV4 7AL, UK}
\affil{Centre for Exoplanets and Habitability, University of Warwick, Gibbet Hill Road, Coventry CV4 7AL, UK}
\email[]{Samuel.Gill@warwick.ac.uk}
\author[orcid=, sname='Fern\'andez Fern\'andez']{Jorge Fern\'andez Fern\'andez}
\affil{Department of Physics, University of Warwick, Gibbet Hill Road, Coventry CV4 7AL, UK}
\affil{Centre for Exoplanets and Habitability, University of Warwick, Gibbet Hill Road, Coventry CV4 7AL, UK}
\email[]{Jorge.Fernandez-Fernandez.2@warwick.ac.uk}
\author[orcid=0000-0001-7416-7522, sname='Anderson']{David R. Anderson}
\affil{Instituto de Astronom\'\i a, Universidad Cat\'olica del Norte, Angamos 0610, 1270709, Antofagasta, Chile}
\email[]{david.anderson@ucn.cl}
\author[orcid=0000-0002-1357-9774, sname='Battley']{Matthew P. Battley}
\affil{Astronomy Unit, Queen Mary University of London, London E1 4NS, UK}
\affil{Observatoire Astronomique de l’Universit\'e de Gen\`eve, Chemin Pegasi 51, CH-1290 Versoix, Switzerland}
\email[]{m.battley@qmul.ac.uk}
\author[orcid=0000-0003-2851-3070, sname='Gillen']{Edward Gillen}
\affil{Astronomy Unit, Queen Mary University of London, London E1 4NS, UK}
\email[]{e.gillen@qmul.ac.uk}
\author[orcid=0000-0003-2417-7006, sname='Ulmer-Moll']{Sol\`ene Ulmer-Moll}
\affil{Observatoire Astronomique de l’Universit\'e de Gen\`eve, Chemin Pegasi 51, CH-1290 Versoix, Switzerland}
\affil{Leiden Observatory, Leiden University, P.O. Box 9513, 2300 RA Leiden, The Netherlands}
\email[]{ulmer-moll@strw.leidenuniv.nl}
\author[orcid=0000-0003-1631-4170, sname='McCormac']{James McCormac}
\affil{Department of Physics, University of Warwick, Gibbet Hill Road, Coventry CV4 7AL, UK}
\affil{Centre for Exoplanets and Habitability, University of Warwick, Gibbet Hill Road, Coventry CV4 7AL, UK}
\email[]{J.J.McCormac@warwick.ac.uk}
\author[orcid=0000-0001-9699-1459, sname='Lendl']{Monika Lendl}
\affil{Observatoire Astronomique de l’Universit\'e de Gen\`eve, Chemin Pegasi 51, CH-1290 Versoix, Switzerland}
\email[]{monika.lendl@unige.ch}
\author[orcid=0009-0004-7473-4573, sname='Apergis']{Ioannis Apergis}
\affil{Department of Physics, University of Warwick, Gibbet Hill Road, Coventry CV4 7AL, UK}
\affil{Centre for Exoplanets and Habitability, University of Warwick, Gibbet Hill Road, Coventry CV4 7AL, UK}
\email[]{ioannis.apergis@warwick.ac.uk}
\author[orcid=0000-0002-8675-182X, sname='Hawthorn']{Faith Hawthorn}
\affil{Department of Physics, University of Warwick, Gibbet Hill Road, Coventry CV4 7AL, UK}
\affil{Rugby School, Lawrence Sheriff St, Rugby, Warwickshire, CV22 5EH, UK}
\email[]{faha@rugbyschool.net}
\author[orcid=0000-0003-2733-8725, sname='Jenkins']{James S. Jenkins}
\affil{Instituto de Estudios Astrof\'\i sicos, Facultad de Ingenier\'\i a y Ciencias, Universidad Diego Portales, Av. Ej\'ercito Libertador 441, Santiago, Chile}
\affil{Centro de Excelencia en Astrof\'\i sica y Tecnolog\'\i as Afines (CATA), Camino El Observatorio 1515, Las Condes, Santiago, Chile}
\email[]{james.jenkins@mail.udp.cl}
\author[orcid=0000-0002-7927-9555, sname='Moyano']{Maximiliano Moyano}
\affil{Instituto de Astronom\'\i a, Universidad Cat\'olica del Norte, Angamos 0610, 1270709, Antofagasta, Chile}
\email[]{mmoyano@ucn.cl}
\author[orcid=0000-0002-5254-2499, sname='Nielsen']{Louise D. Nielsen}
\affil{University Observatory, Faculty of Physics, Ludwig-Maximilians-Universität München, Scheinerstr. 1, 81679 Munich, Germany}
\email[]{Louise.Nielsen@lmu.de}
\author[orcid=0000-0002-2386-4341, sname='Smith']{Alexis M. S. Smith}
\affil{Institute of Space Research, German Aerospace Center (DLR), Rutherfordstr. 2, 12489, Berlin, Germany}
\email[]{alexis.smith@dlr.de}
\author[orcid=0000-0001-8018-0264, sname='Saha']{Suman Saha}
\affil{Instituto de Estudios Astrof\'\i sicos, Facultad de Ingenier\'\i a y Ciencias, Universidad Diego Portales, Av. Ej\'ercito Libertador 441, Santiago, Chile}
\affil{Centro de Excelencia en Astrof\'\i sica y Tecnolog\'\i as Afines (CATA), Camino El Observatorio 1515, Las Condes, Santiago, Chile}
\email[]{suman.saha@mail.udp.cl}
\author[orcid=0000-0001-7576-6236, sname='Udry']{Stéphane Udry}
\affil{Observatoire Astronomique de l’Universit\'e de Gen\`eve, Chemin Pegasi 51, CH-1290 Versoix, Switzerland}
\email[]{stephane.udry@unige.ch}
\author[orcid=0000-0002-1896-2377, sname='Vines']{Jose I. Vines}
\affil{Instituto de Astronom\'\i a, Universidad Cat\'olica del Norte, Angamos 0610, 1270709, Antofagasta, Chile}
\email[]{jose.vines.l@gmail.com}
\author[orcid=0000-0001-6604-5533, sname='West']{Richard G.\ West}
\affil{Department of Physics, University of Warwick, Gibbet Hill Road, Coventry CV4 7AL, UK}
\affil{Centre for Exoplanets and Habitability, University of Warwick, Gibbet Hill Road, Coventry CV4 7AL, UK}
\email[]{richard.west@warwick.ac.uk}
\author[orcid=0000-0001-6023-1335, sname='Bayliss']{Daniel Bayliss}
\affil{Department of Physics, University of Warwick, Gibbet Hill Road, Coventry CV4 7AL, UK}
\affil{Centre for Exoplanets and Habitability, University of Warwick, Gibbet Hill Road, Coventry CV4 7AL, UK}
\email[]{d.bayliss@warwick.ac.uk}
\author[orcid=0000-0002-4047-4724, sname='Osborn']{Hugh P.\ Osborn}
\affil{Center for Space and Habitability, University of Bern, Gesellschaftsstrasse 6, 3012 Bern, Switzerland}
\affil{ETH Zurich, Department of Physics, Wolfgang-Pauli-Strasse 2, CH8093 Zurich, Switzerland}
\email{hugh.osborn@unibe.ch}
\author[0000-0002-7188-8428]{Tristan Guillot}
\affil{Observatoire de la C\^ote d'Azur, Universit\'e C\^ote d'Azur, CNRS, Laboratoire Lagrange, Bd de l'Observatoire, CS 34229, 06304 Nice cedex 4, France}
\email[]{tristan.guillot@oca.eu}
\author[0000-0002-5510-8751]{Amaury H. M. J. Triaud}
\affil{School of Physics \& Astronomy, University of Birmingham, Edgbaston, Birmingham B15 2TT, UK} \email[]{a.triaud@bham.ac.uk}
\author[0000-0002-3503-3617]{Olga Suarez}
\affil{Observatoire de la C\^ote d'Azur, Universit\'e C\^ote d'Azur, CNRS, Laboratoire Lagrange, Bd de l'Observatoire, CS 34229, 06304 Nice cedex 4, France}
\email[]{olga.suarez@oca.eu}
\author[orcid=0009-0006-2606-3271, sname='Beltrame']{Matteo Beltrame}
\affil{Istituto di Scienze Polari del CNR (ISP-CNR), Università Ca' Foscari, Via Torino n. 155, 30172 Venezia Mestre (VE), Italy}
\affil{Programma Nazionale di Ricerche in Antartide (PNRA), Institut polaire français Paul-Émile Victor (IPEV), France}
\email[]{matteo.beltrame95@gmail.com}
\author[0000-0001-7948-6493]{Abdelkrim Agabi}
\affil{Observatoire de la C\^ote d'Azur, Universit\'e C\^ote d'Azur, CNRS, Laboratoire Lagrange, Bd de l'Observatoire, CS 34229, 06304 Nice cedex 4, France}
\email[]{karim.agabi@unice.fr}
\author[orcid=0000-0001-9573-4928, sname='Pagano']{Isabella Pagano}
\affil{Istituto Nazionale di Astrofisica, 00136 Roma, Italy}
\email[]{isabella.pagano@inaf.it}
\author[orcid=0000-0003-0030-332X, sname='Hooton']{Matthew J. Hooton}
\affil{Cavendish Laboratory, JJ Thomson Avenue, Cambridge CB3 0HE, UK}
\email[]{mh2143@cam.ac.uk}
\author[orcid=0000-0003-0684-7803, sname='Burleigh']{Matthew R. Burleigh}
\affil{School of Physics and Astronomy, University of Leicester, Leicester LE1 7RH, UK}
\email[]{mrb1@le.ac.uk}
\author[orcid=0000-0002-0856-4527, sname='Abe']{Lyu Abe}
\affil{Observatoire de la C\^ote d'Azur, Universit\'e C\^ote d'Azur, CNRS, Laboratoire Lagrange, Bd de l'Observatoire, CS 34229, 06304 Nice cedex 4, France}
\email[]{lyu.abe@oca.eu}
\author[orcid=0000-0002-4278-1437, sname='Bendjoya']{Philippe Bendjoya}
\affil{Observatoire de la C\^ote d'Azur, Universit\'e C\^ote d'Azur, CNRS, Laboratoire Lagrange, Bd de l'Observatoire, CS 34229, 06304 Nice cedex 4, France}
\email[]{philippe.bendjoya@oca.eu}
\author[orcid=0000-0002-3937-630X, sname='Dransfield']{{Georgina} Dransfield}
\affil{School of Physics \& Astronomy, University of Birmingham, Edgbaston, Birmingham B15 2TT, UK} 
\email[]{g.dransfield@bham.ac.uk}
\affil{Department of Astrophysics, University of Oxford, Denys Wilkinson Building, Keble Road, Oxford OX1 3RH, UK}
\affil{Magdalen College, University of Oxford, Oxford OX1 4AU, UK}
\author[orcid=0000-0001-5000-7292, sname='M\’ekarnia']{Djamel M\'ekarnia}
\affil{Observatoire de la C\^ote d'Azur, Universit\'e C\^ote d’Azur, CNRS, Laboratoire Lagrange, Bd de l'Observatoire, CS 34229, 06304 Nice cedex 4, France}
\email[]{mekarnia@oca.eu}

\begin{abstract}

We identify two large-radius planets around the F-type star HD 114082 as the longest-period young transiting exoplanets known. 
From the first transit, detected by NASA's Transiting Exoplanet Survey Satellite (TESS), and a second dip, spotted by the Next-Generation Transit Survey (NGTS), we predicted mid-transit times for HD 114082 b (planet b). We pinpoint its orbit (period $P_b$= 225.5504$\pm$0.0004 days) from a third transit captured with the ESA's CHaracterising ExOplanet Satellite and the upgraded Antarctic Search for Transiting ExoPlanets telescope (ASTEP$+$), alongside orbit-discriminating observations.
Another dimming partly covered by ASTEP$+$ completes the four-transit series. We  support with dynamical evidence the planetary nature of a deeper transit detected with TESS and NGTS, identifying planet c. Additionally, we reexamine the debris disk, fitting its excess emission with two dust components.
Fundamental stellar parameters are inferred from stellar evolution models, while a joint modeling of photometric and radial-velocity time series yields the planetary parameters, with masses further constrained using an N-body code. 
For planet b, the semimajor axis $a_b$= 0.791$\pm$0.008 au, eccentricity $e_b$$\approx$ 0, inclination $i_b$= 89$^\circ$.791$\pm$0.014, radius $R_b$= 1.046$\pm$0.014 $R_{\rm J}$, and 95\% confidence upper limit on its mass $M_{\rm 95\%,b}$= 1.6 $M_{\rm J}$. 
For planet c, $a_{\rm c}$= 0.99$^{+0.03}_{-0.04}$ au, $e_c$$\approx$ 0, $i_{\rm c}$= 89$^\circ$.701$\pm$0.011, $R_{\rm c}$= 1.36$\pm$0.03 $R_{\rm J}$, and $M_{\rm 95\%,c}$= 2.0 $M_{\rm J}$ (0.24 $M_{\rm J}$ if adding transit timing variation constrains). They seem to be moderate-to-low-mass giants in nearly resonant, coplanar, circular orbits that formed in situ, or beyond the snowline, and migrated inwards, shaping the disk. 

\end{abstract}



\section{Introduction}

The standard model for giant-planet formation \citep[e.g.,][]{armitage2024,ikoma2025} posits that giants form in gas-rich protoplanetary disks via a bottom-up process: submicron dust grains acquire ice mantles and coalesce into pebbles, planetesimals, and planetary cores that grow to their crossover mass, initiating runaway gas accretion. Core formation becomes challenging at semimajor axes $a$ $\gtrsim$ 30 au, where giants could form via gravitational instabilities in the disk. Accretion may occur in situ for planets as massive as a few Jupiter masses before being halted by a gas dispersal mechanism. The resulting system consists mainly of newly formed planets and a disk of dust and planetesimals (asteroids and comets), known as a debris disk, with a fractional dust luminosity $L_{\rm dust}/L_{\star}$$<$ 1\% and an age exceeding a few Ma \citep[][]{hughes2018,hasegawa2022,bergez-casalou2024,marino2026}.

The bodies producing dust in debris disks are dynamically excited into a collisional cascade by planetary companion(s), the most massive planetesimals in the disk, or some combination thereof \citep[e.g.,][]{mustill2009,krivovbooth2018,munoz2023,marshall2025}. 
However, the mechanisms and conditions driving planetary formation and evolution are not yet fully understood \citep[e.g.,][]{drazkowska2023,youdin2025}. 

Classical models suggest a crossover mass for runaway gas accretion of $\sim$20--30 $M_\oplus$, although $\sim$100 $M_\oplus$ has also been proposed \citep[][]{helled2023}. Theory predicts that infant giant planets contract to their final size over hundreds of million of years, while recent observations indicate planetary evolution may yield mature giants 
on timescales roughly 10 times shorter, though this remains debated \citep[e.g.][]{benatti2021,suarezmascareno2022,barragan2024}. 

Hydrodynamic simulations show that two giant planets, if they begin accreting simultaneously, end up with nearly identical masses after 0.5 Ma, irrespective of disk viscosity; whereas if one begins accreting later, their mass ratio initially peaks higher but quickly settles below 2 \citep[][]{bergezcasalou2023}. This accounts for most observed systems with multiple gas-giant planets. 

Through dynamical interactions, planets shape the radial and vertical structure of the underlying planetesimal belt \citep[e.g. ][]{terrill2023,pearce2024,marino2026}, 
regardless of whether they undergo disk or tidal migration, dictating the formation and evolution of inner rocky planets \citep[][]{garcia2023,best2024,kong2024,sandhaus2025}. Thus, planetary masses can be constrained from the spatially resolved architecture of a disk. Most evident in single-star systems, chaotic, resonant interactions between giant planets amplify destabilizations. Their scattering can disrupt planetary orbits, triggering migration, ejection, and engulfment by the star \citep[][]{carrera2019}. 

Hot Jupiters represent $\approx$15\% of known exoplanets, as their short orbital periods, large masses, and usually inflated radii enhance their detectability---despite a truly low occurrence rate of $\approx$2\%; warm Jupiters appear to be as common as hot Jupiters, while the occurrence rate abruptly rises by about a factor of 4 for cold Jupiters, at $a$ $\gtrsim$1 au, reaching ratios per 100 stars of 14.1$^{+2.0}_{-1.8}$ and 8.9$^{+3.0}_{-2.4}$ for $a$ in the ranges of 2--8 au and 8--32 au, respectively \citep[][]{fulton2021}. Thus, giant planets are rare, especially in pairs \citep[][]{wittenmyer2020}, and challenging to detect if orbiting young stars, because of strong stellar variability. For ages $\lesssim$ 100 Ma, most planets tend to be puffed-up, Neptune-mass planets, which formed in situ \citep[][]{karalis2025}. 

The discovery and study of a statistically meaningful sample of planets as a function of age is still necessary to understanding their origin and evolution. In particular, transiting planets in multiple systems offer a wealth of information, as their masses and radii can be accurately determined, constraining their composition.

HD 114082, a young F-type star, originally was reported to host a Jupiter-sized planet (hereafter, planet b) with an orbital period $P$= 109.8$\pm$0.4 d ($a$$\approx$ 0.5 au), eccentricity $e$= 0.40$\pm$0.04, and mass $M$= 8.0$\pm$1.0 $M_{\rm J}$ \citep{zakhozhay2022}. These results pointed out that this could be the longest-period and most massive among the youngest known transiting exoplanets. \citet{engler2023} revised  planet b's orbital period to be $P$= 197$^{+171}_{-109}$ d, based on the Transiting Exoplanet Survey Satellite (TESS) Sector 38 light curve (with a monotransit) and additional data. HD 114082 also hosts an unresolved dusty structure with an ill-constrained stellocentric radius \citep[0.8--18 au;][]{chenchristine2014,jangcondell2015,mittal2015} and an external ($\approx$38 au), nearly edge-on CO-free exo-Kuiper Belt, with a small vertical scale height in scattered light \citep[][]{wahhaj2016,kral2020,engler2023,matra2025}. The planes of the debris disk and planet b's orbit form an angle $>$6$^{\circ}$, suggesting the known planet is dynamically interacting with other bodies within the belt \citep[][]{engler2023}. Indeed, the NASA Exoplanet Archive{\footnote{https://exoplanetarchive.ipac.caltech.edu} \citep[][]{christiansen2025}} lists a candidate planet, TOI-6697.02 (hereafter planet c), identified from a monotransit in TESS Sector 64. 

In this Letter, we present the first results of a multi-telescope follow-up campaign, which started with the Next Generation Transit Survey (NGTS), after the TESS detection of a monotransit in  HD 114082's light curve, and continued with the CHaracterising ExOplanet Satellite (CHEOPS), the Las Cumbres Observatory (LCO) network, and the Antarctic Search for Transiting ExoPlanets telescope (ASTEP$+$), aimed at constraining the characteristics of the two planets in the system. Section \ref{thehost} is devoted to the host star. Section \ref{data} describes the observations and data preparation, including ancillary data. Section \ref{analysisandresults} shows our analysis and results, Section \ref{discussion} discusses them, and Section \ref{conclusions} lists our conclusions. Appendices \ref{appendixa}, \ref{appendixb}, \ref{appendixc}, \ref{appendixd}, and \ref{appendixe} respectively include further
details on the stellar and planetary characterization, transit timing variations (TTVs), nature of planet c, and the system architecture, 
while Appendix \ref{appendixf} provides additional tables. 

\section{The host star} \label{thehost}

HD 114082 is a bright ($G$= 8.1060$\pm$0.0028), young (15$\pm$3 Ma), fast-rotating ($v\sin i$= 39.2$\pm$0.5 km s$^{-1}$) F3 star 
with an effective temperature of 6600$\pm$70\,K, at a distance of 95.06$\pm$0.20 pc, 
being part of Lower Centaurus Crux subgroup in the Scorpius-Centaurus OB Association, which hosts low-density, 
unbound, star-forming regions \citep[][]{gaia2016,gagne2018,avallone2022,zakhozhay2022,gaia2023,squicciarini2025}. 
Key parameters of HD 114082 are presented in Appendix \ref{appendixa}, including those derived to characterize its planets. 

\section{Observations and data preparation}
\label{data}

Observations from several facilities and ancillary public data are used to yield precise photometric time series of HD 114082 (Sections \ref{tessdata}--\ref{lcodata}). 
Every light curve provides a reliable measurement of a broadband flux change over time, enabling the detection of small variations. For ground-based observations, we applied differential photometry relative to nearby non-variable comparison stars, which minimises atmospheric and instrumental effects on the observed flux, isolating astrophysical signals and improving photometric precision. 
In contrast, space observations are free from scintillation noise and are unaffected by day--night cycle or cloud cover. 
Table \ref{tab:instruments} includes the photometric precision achieved with each facility before and after light-curve binning. 
We also use public time-series spectra of our target (see \S\ref{rvdata}).

\subsection{TESS}
\label{tessdata}

NASA's TESS satellite \citep[][]{ricker2014} employs four widefield cameras with a 10 cm effective aperture size to perform a survey of 85\% of the sky degrees 
in 30 partially overlapping sectors of 24$^\circ\times$96$^\circ$ (combined field of view or FOV), each observed for two orbits in the 600-1000 nm bandpass. 
The pixel size is 21$^{\prime\prime}$ and the FWHM of the point-spread function (PSF) is nearly 30$^{\prime\prime}$.  

We use public data of Sectors 38 (from 2021 April 28 to 2021 May 26), 64 and 65 (from 2023 April 6 to 2023 June 2), and 99 (from 2026 January 5 to 2026 February 2), corresponding to HD 114082 (TESS magnitude of 7.8), with a 2 minutes cadence, and discard data from Sector 11, because the star's signal is obscured. Light curves are generated by the TESS Science Processing Operations Center (SPOC). 
The SPOC pipeline \citep[][]{Jenkins2016} first calibrates science data by orbit and then by sector, extracting high-precision time-series photometry for selected targets \citep[][]{caldwell2020}. 
TESS light-curve files provide time stamps and fluxes from the simple aperture photometry (SAP) and the presearch data conditioning corrected SAP  
\citep[PDCSAP;][]{Smith2012, Stumpe2012, Stumpe2014}. The SAP flux is calculated 
by summing all calibrated pixels in the optimal (highest signal-to-noise ratio or S/N) aperture, whereas PDCSAP flux is corrected for instrumental long-term trends. 
Our analysis is based on the PDCSAP flux. Aperture photometry on the crowded field around our target yields a transit depth dilution that was corrected to avoid underestimating the planetary radii (see Section \ref{analysisandresults:gp} and Appendix \ref{appendixb}). 

\begin{table*}\centering
\caption{For Each Facility Used, Median Cadence, Number of Points (N), rms (Average over All Epochs), and Uncertainty ($\sigma$), for Original and 5 minutes Binned Light Curves, 
    as Well as the rms after Model Subtraction (rms$_{res}$) \label{tab:instruments}}
    \begin{tabular}{lccccccccc}
    \hline\hline 
        Facility                & Cadence &      N       & RMS         & $\sigma$ & Cadence   & N$_{bin}$ & RMS$_{bin}$ & $\sigma_{bin}$ & RMS$_{res}$ \\
                                &    s    &              & ppt         &    ppt   &   s       & ppt       &    ppt      & ppt            & ppt         \\
    \hline
        CHEOPS                  &    19.8 &    14,126     & 1.4         &  0.23    &   300     & 1212      &  1.5        &  0.06          & 0.21        \\
        TESS                    &    120  &    70,941     & 0.7         &  0.6     &   300     & 28,494     &  0.7        &  0.4           & 0.25        \\ 
        NGTS                    &    13   &    175,886    & 6           &  6       &   300     & 7920      &  2.0        &  1.2           & 1.1         \\
        ASTEP$+$ (R)            &    7.4  &    38,756     & 7           &  1.7     &   300     & 1189      &  1.7        &  0.28          & 1.1         \\
        LCO                     &    4.9  &    7205     & 5           &  1.0     &   300     & 257       &  2.0        &  0.8           & 0.8         \\
    \hline
    \end{tabular}
\end{table*}

\subsection{CHEOPS}

ESA's first small-class mission, CHEOPS \citep[][]{rando2020, benz2021}, is conducting high-precision photometry 
in the 330-1100 nm bandpass using a 30 cm telescope, from a Sun-synchronous, dusk-dawn orbit at 700 km above Earth. 
The FOV of 19$^{\prime} \times$19$^{\prime}$ rotates around the line of sight once per orbit ($P$= 98.7 min), the pixel size is 1$^{\prime\prime}$, and 
90\% of the total energy of the defocused PSF is within a radius of 16 pixels. This avoids the saturation of bright stars onto the detector 
and enables the performing of high-precision photometry of one target at a time without significant contamination. 

We present Fourth Announcement of Opportunity (AO-4) cycle observations of Guest Observers (GO) program PR240025 (PI: C. del Burgo) 
and AO-5 cycle GO program PR250012 (PI: C. del Burgo), aimed at determining the key parameters of planet b. 
Critical time observations at 30 s cadence were made on 2024 June 5, targeting an orbital period alias identified from an 
NGTS transit detection on 2023 March 11 (see Section \ref{sec-ngts}).
In addition, we present public data from Guaranteed Time Observers (GTO) programmes PR100017 and PR140079, to complement the target's observations. 
Data were processed at the Science Operations Centre (SOC) from the CHEOPS Data Reduction Pipeline \citep[][]{hoyer2020}. 
Appendix \ref{appendixf} lists the CHEOPS programmes. To produce a roll angle detrended light curve, we followed the procedure outlined by~\citet{maxted2022}. 

\subsection{NGTS}
\label{sec-ngts}
NGTS \citep[][]{wheatley2018}, located at the ESO Paranal Observatory in Chile, uses an array of 12 robotic 20 cm telescopes
to perform high-precision photometry of exoplanet transits in the 520--890\,nm bandpass. 

In order to follow up the TESS detection of a single transit of planet b, one NGTS telescope monitored HD 114082 during 167 nights 
between 2023 February 17 and 2023 September 6. The second transit of planet b was captured on 2023 March 11, and the first transit of planet c was detected, concurrently with TESS, on 2023 April 29. NGTS observations resumed on 2023 December 6, with another 47 nights of observations accumulated by 2024 August 19. 

Due to the star's brightness, the telescope was defocused to produce a donut-shaped PSF with a radius of approximately 6 pixels (30$^{\prime\prime}$), preventing detector saturation. Observations with 10\,s exposure time and $\sim$13\,s cadence were made continuously for as long as the target was visible above an elevation of 30$^\circ$. 
Precise autoguiding was carried out using the science images and the DONUTS algorithm \citep{DONUTS}. 

Differential photometry was performed using a 6\,pixel radius aperture for HD 114082 as well as 100 comparison stars from across the 8 deg$^2$ field 
that were selected to match the target brightness and color as closely as possible. The resulting light curves were filtered for cloudy intervals and detrended 
against airmass to account for color differences between the target and the averaged comparison stars. 

\subsection{ASTEP$+$}

ASTEP$+$ \citep[][]{crouzet2020,dransfield2022,schmider2022} is a 40 cm aperture telescope located near Concordia Station, at Dome C (on the Antarctic Plateau, at an elevation of 3234 m). 
In 2021, the telescope received a major upgrade of its camera system \citep{schmider2022}. It now operates with two high-sensitive cameras, 
providing simultaneous photometry in a blue channel (central wavelength $\lambda_{\rm c}$ of 555 nm and bandwidth $\Delta\lambda$ of 293 nm) and in a red channel ($\lambda_{\rm c}$: 850 nm; $\Delta\lambda$: 276 nm). Real-time guiding is ensured in the blue channel. The FOV is nearly 40$^{\prime}$ wide and the image scale is 1.39$^{\prime\prime}$ pixel$^{-1}$. 
Concordia Station experiences a near-continuous night (polar night) from mid May to late July, which benefits target's monitoring. 
Time-critical observations were acquired on 2024 June 6 and 2024 September 25--26, and on 2025 May 8--10 and 2025 August 30, with exposure times of 60 s and 1 s for the blue and red channels, respectively. 
They were conducted under a polar night sky with a new moon in June (total darkness), or under nonpolar night conditions with significant lunar illumination on the three other dates (two shortly after the austral winter). Wind speeds ranged from 2 to 9 m s$^{-1}$. Here, we adopt the red (R) channel data, taken at a 7.4\,s cadence. 

Automated on-site data processing, including detrending, was conducted and promptly verified by inspecting the target's light curve 
and the target flux and systematics. These enable a near-instantaneous identification of transit events. The largest data products 
files---containing aperture photometry for up to 1000 stars in the field, observation metadata, and the stacked image generated during reduction---were later transferred to a local server. Data acquired in 2024 were finally transmitted to Europe, being reanalysed. 

\subsection{The LCO Network}
\label{lcodata}

We use data from the LCO global telescope network \citep{Brown2013}. 
A Director's Discretionary Time program (DDT2025A-006; PI: C. del Burgo) was granted 30 hr to monitor HD 114082 on 2025 May 8--10, aimed at testing a candidate orbital period for planet b. The observations were performed at Cerro Tololo Inter-American Observatory with the 0.4 m telescope, equipped with QHY600 CMOS cameras. A 2 s exposure time in the V band and the default readout mode in a $30^\prime\times30^\prime$ FOV yielded sufficient comparison stars to calibrate the 7 s cadence light curve of HD 114082. 
Bad weather hindered observations most of the time. LCO archival data obtained in 2022 and 2023 complement the dataset. 
Appendix \ref{appendixf} provides detailed information of the different observing campaigns. 
All retrieved data products were processed using the {\sc banzai} pipeline \citep[][]{mccully2018}. We derive the light curves using the {\sc eloy} package\footnote{https://github.com/lgrcia/eloy}, which implements the algorithm from \citet{broeg2005}.  

All available LCO time-series segments cover total durations shorter than the transits of planets b and c and do not show flux decreases that would indicate a transit ingress or egress. We select those sets of consecutive observations, and common filters for a given telescope, on which differential photometry could be performed, so that the resulting normalized fluxes represent relative variations within each set. Table~\ref{tab:lco_extended} provides details of these observations. 
With gaps between consecutive observations exceeding the characteristic transit durations of planets b and c, the uniform relative flux levels confirm that all LCO data were acquired out of transit.
These measurements provide additional constraints on the orbital periods of planets b (via the period alias testing) and c. In particular, the LCO light curves, together with those from NGTS, TESS, and CHEOPS, narrow the probability density function (PDF) of midtransit times for planet c and, thus, remarkably constrain its predicted orbital period (see Appendix \ref{appendixb}).

\subsection{Fibre-fed Extended Range Optical Spectrograph and High Accuracy Radial velocity planet Searcher Spectra}
\label{rvdata}
 
We use the 4 year monitoring campaign time-series spectra of HD 114082 obtained with the Fibre-fed Extended Range Optical Spectrograph \citep[FEROS;][]{kaufer1999}, conducted by \citet{zakhozhay2022}. 
We also employ the public spectra of the High Accuracy Radial velocity planet Searcher \citep[HARPS;][]{mayor2003} previously used by \citet{zakhozhay2022}. We downloaded 54 FEROS spectra from the ESO archive\footnote{https://archive.eso.org}, while 18 HARPS spectra were retrieved from the Data \& Analysis Center for Exoplanets\footnote{https://dace.unige.ch}. 

Both FEROS and HARPS are equipped with pipelines that yield extracted and wavelength-calibrated spectra. These reduced data were used to derive the radial velocity (RV) and bisector span curves with a Gaussian fit to the cross-correlation function (CCF) of each spectrum with the HARPS G2 mask \citep[][]{Fellgett1955, Baranne1996, Pepe2000}. Given the fast stellar rotation (causing significant line broadening), a RV range of $\pm$200 km s$^{-1}$ around the systemic velocity is used. Then, we sigma-clipped outliers with RVs larger than 500 m s$^{-1}$, yielding 48 FEROS RV data points (6 rejected) and 18 HARPS RV data points (none rejected). A quick look at the FEROS data revealed a weak correlation of the resulting velocities with ambient pressure. We detrended the RV measurements using a third order polynomial against pressure. The resulting RVs show an RMS of 118 m s$^{-1}$, with a median uncertainty of 160 m s$^{-1}$, for FEROS, and 52 m s$^{-1}$, with a median uncertainty of 20 m s$^{-1}$, for the HARPS data. The fitting residuals are 18 m s$^{-1}$ (HARPS) and 114 m s$^{-1}$ (FEROS). 

\begin{figure*}[]
    \resizebox{\hsize}{!}{\includegraphics{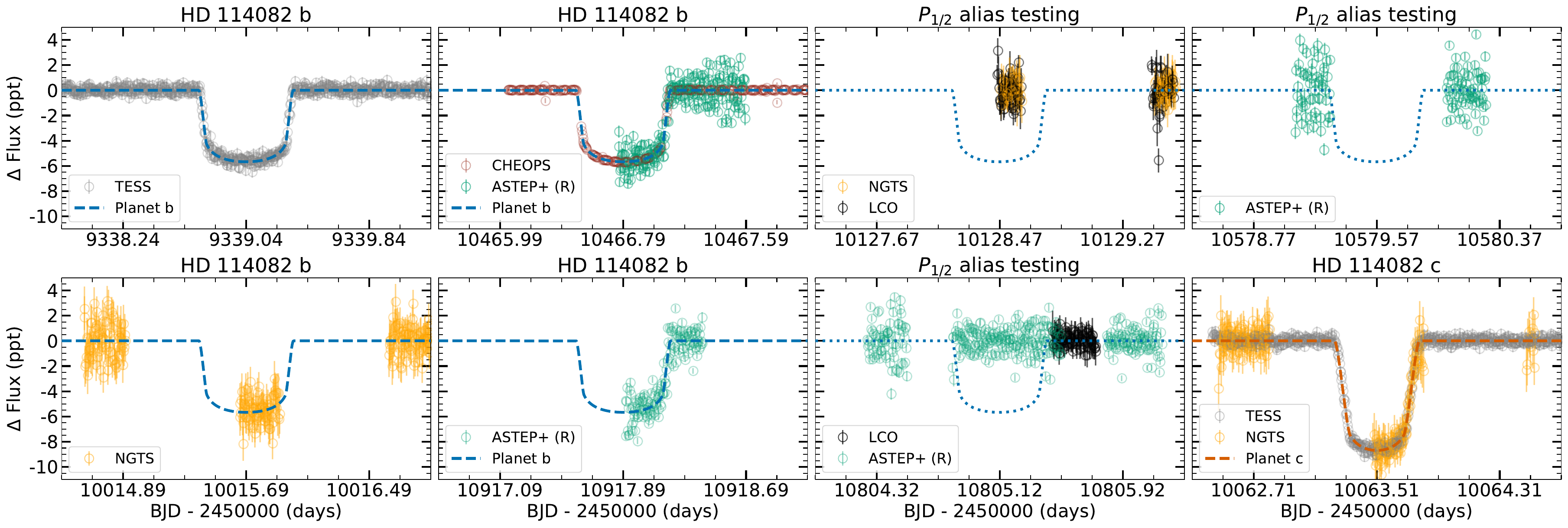}}
    \caption{Binned light curves of HD 114082 during eight decisive events; symbols and colors denote the facilities used. 
    {\it Left:} Four non-consecutive transits of planet b, fully detected by TESS (1st) and CHEOPS (3rd), and partly ($\approx$50\%) observed by NGTS (2nd) and ASTEP$+$ (3rd, 4th), with fitted model (blue dashed line). 
    {\it Centre-right and top-right:} Period alias testing con\-duct\-ed using observations from ASTEP$+$, LCO, and NGTS, 
    which ruled out the orbital period $P_{1/2}$ ($\approx$112.78 d), with the expected dips (blue dotted lines) as a reference. 
    {\it Bottom-right:} Transit of planet c observed by TESS (fully) and NGTS (over half), with fitted model (red dashed line). The data used to generate this figure are available in machine-readable format in the online journal.} 
    \label{fig:fig1}
\end{figure*}

\section{Analysis and Results}
\label{analysisandresults}

\subsection{Planets b and c: Transit Observations}
\label{planetb}

TESS detected two transits different in depth ($\delta$) and duration ($T_{\rm 14}$) in HD 114082. The first of these, owing to planet b, was registered on 2021 May 4 in Sector 38. We inferred a series of candidate orbital periods for planet b from the lapse between this midtransit time (BJD 2459339.0380$^{+0.0012}_{-0.0011}$) 
and that estimated from another dip of the same depth---detected by NGTS on 2023 March 11. We then predicted and observed a third transit---fully with CHEOPS and over half coverage with ASTEP$+$---on 2024 June 5, consistent with orbital period $P$= 225.5504$\pm$0.0004 days and its alias $P_{1/2}$= 112.77520$\pm$0.00022 days. 
While the observing conditions on 2024 September 26 prevented a confident conclusion with ASTEP$+$, no dip was detected either on 2023 July 2 by NGTS and LCO or on 2025 May 8--10 by ASTEP$+$ and LCO, ruling out $P_{1/2}$, irrespective of any potential TTVs (see Appendix \ref{appendixc}). Subsequently, we observed another half a transit of planet b with ASTEP$+$, completing a series of four transits. See Figure \ref{fig:fig1}. 

The second TESS monotransit, due to planet c, occurred on 2023 April 29 (midtransit at BJD 2460063.5113 $\pm$ 0.0012) in Sector 64. The same transit was observed (over half) by NGTS, enhancing its planetary nature (see Appendix \ref{appendixd}), further supported by dynamical constraints as well as the observed alignment between the two planetary orbits and their misalignment with the outer belt's plane, reflecting planet-planet interactions. 

\subsection{Planetary System Parameters}
\label{analysisandresults:gp}

We construct a global model of photometry and RV---incorporating a floating zero-point offset for each instrument, a linear trend for the RV curve, and linear trends for the RV against bisector span---and apply it to all available photometric and spectroscopic data. Short-term variability (both stellar and systematic) is described using Gaussian Process (GP) regression \citep[][]{Rasmussen2006} with the S$+$LEAF code \citep{Delisle2022}, while planetary transit signatures and RV signals are represented using Pytransit  \citep[][see also \citealt{MandelAgol2002} for a description of the quadratic limb-darkening law]{Parviainen2015} and sinusoidal functions, respectively. Given the large TESS pixel scale, we include a flux dilution factor. To avoid biasing the results towards any specific orbital period, we conduct a blind period search. 
The parameters are optimized using Bayesian inference with the nested sampling \citep{Skilling2004,Skilling2006} code Dynesty \citep{Speagle2020,Koposov2023}. 

\begin{table}[h!]
    \footnotesize
    \centering
    \caption{\footnotesize Orbital and Physical Parameters of Planets b and c for the Preferred Solution (Circular Model) 
    \label{tab:planetary_parameters}}
    \begin{tabular}{lcc}
    \hline\hline 
        Parameter & Planet b & Planet c \\
    \hline
        $\delta$ (\%)               &      0.569$\pm$0.003                           &   0.87$\pm$0.03               \\
        $T_{\rm 14}$ (h)              &    14.69$\pm$0.05                       &   13.18$\pm$0.09  \\
        $T_0$ (BJD)         & 2460466.7909$\pm$0.0017 & 2460063.5114$\pm$0.0009 \\
        $P$ (d)                                & 225.5504$\pm$0.0004    & 314$^{+11}_{-18}$ \\         
        $e$                           & 0  (fixed)                                                   & 0 (fixed) \\
        $b$                                     & 0.421$^{+0.022}_{-0.024}$                              & 0.75$\pm$0.10  \\
        $i$ (deg)                             & 89.791$\pm$0.014                         &  89.701$\pm$0.011  \\
        $R/R_\star$                              & 0.0714$\pm$0.0005                       &  0.0928$\pm$0.0014 \\ 
        $a/R_\star$         & 115.6$\pm$1.3 & 144$^{+5}_{-6}$ \\
        $a$ (au)                                      &  0.791$\pm$0.008                                &  0.99$^{+0.03}_{-0.04}$ \\           
        $R$ ($R_{\rm J}$)                       &  1.046$\pm$0.014              &  1.36$\pm$0.03         \\  
        $M_{\rm 95\%}$ ($M_{\rm J}$)         & $<$1.6                                              &  $<$2.0 \\
        $K_{\rm 95\%}$ (m s$^{-1}$)    & $<$44                                            & $<$49 \\
        $\rho_{\rm 95\%}$ (g cm$^{-3}$)           & $<$1.7                                          &  $<$1.0   \\ 
        $log\,g_{p,\rm 95\%}$ (cgs) & $<$3.1 & $<$2.3\\
        $\rho_\star$ (g cm$^{-3}$)       &  \multicolumn{2}{c}{0.57$\pm$0.02}             \\
    \hline
       \multicolumn{3}{p{8.25cm}}{{\bf Note.} $R_\star$ and $M_\star$ from Appendix \ref{appendixa} were used to derive $R$ and $a$, and $M_{95\%}$ and $K_{95\%}$, respectively.} \\
    \end{tabular}
\end{table}
\normalsize

The model converges on a solution with two significant planetary signals that is consistent with zero eccentricity ($e$$=$0) for both planets. 
Table \ref{tab:planetary_parameters} lists the key planetary parameters for this solution, including the midpoint time of transit ($T_0$), orbital period ($P$), impact parameter ($b$), inclination ($i$), radius ($R$), and the 95\% confidence upper limit on the mass $M_{95\%}$, linked to the RV semiamplitude ($K$). 
The transits of planet b, combined with our period alias testing, firmly establish its period $P_{\rm b}$. 
For planet c, the shape and duration of its transit, the stellar parameters, 
and the constraints from nontransit data reduce the 1$\sigma$ uncertainty of $P_{\rm c}$ ($\Delta P_{\rm c}$) to $<$6\% for the preferred circular model, although 
this uncertainty is higher for the eccentric model (Table~\ref{tab:prior_posterior} in Appendix \ref{appendixf}). 
For $e$$=$0, $P_{\rm c}/P_{\rm b}$ = 1.39$^{+0.05}_{-0.08}$, which encloses within 1$\sigma$ the second-order 7:5 mean-motion resonance (MMR) and within 3$\sigma$ the first-order 3:2 MMR. The latter is dynamically stronger and more stable, but the uncertainty in $P_{\rm c}$ prevents the establishing of an unequivocal link to any of these two MMRs,  
with $\Delta P_{\rm c}$ $\approx$ (0.5--0.8) ($P_{\frac{3}{2}}$-$P_{\frac{7}{5}}$). The planets could be nearly locked or locked, 
which shall be clarified eventually with further observations. 

For a more detailed description of the applied method, see Appendix \ref{appendixb}. Table~\ref{tab:prior_posterior} in Appendix \ref{appendixf} lists the full set of parameters and priors of the models, including that for nonzero eccentricity.

\begin{figure}[h!]
	\resizebox{\hsize}{!}{\includegraphics{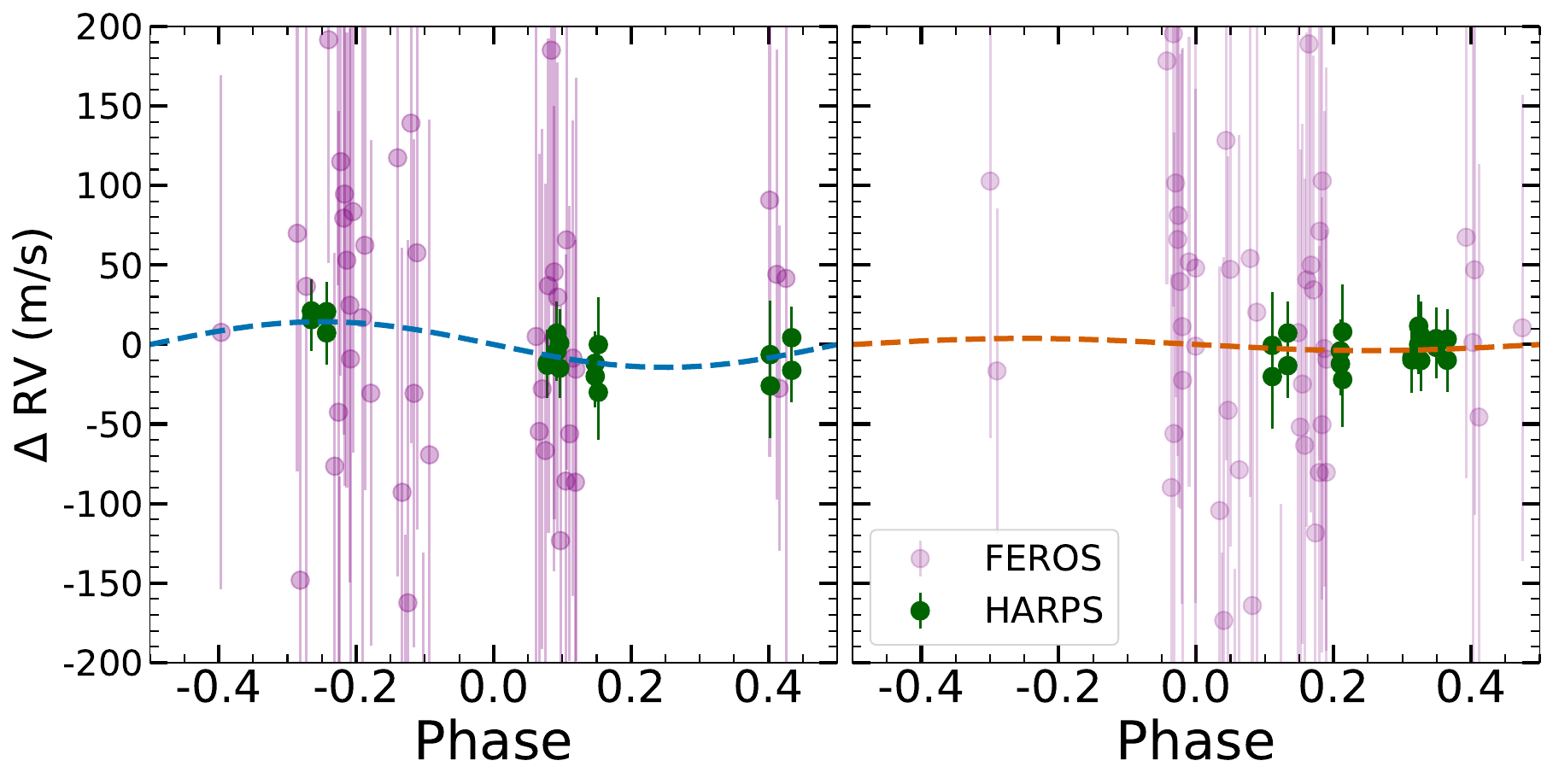}}
    \caption{
    Phased-detrended RVs of planet b ({\it left}) and planet c ({\it right}), inferred from our GP modeling---applied to the full dataset. 
    The RV points in green and purple stand for HARPS and FEROS, respectively. The data used to generate this figure are available in machine-readable format.\\
    (The data used to create this figure are available in the \href{https://iopscience.iop.org/article/10.3847/2041-8213/ae63bd}{online article})
    }
    \label{fig:fig2}
\end{figure}

Figure \ref{fig:fig2} shows phased RV data with planetary models derived from median masses of the posterior distribution. 

\subsection{System Architecture} 
\label{systemarch}

\begin{figure}[h!]
	\resizebox{\hsize}{!}{\includegraphics[trim={0 2.5cm 0 0},clip]{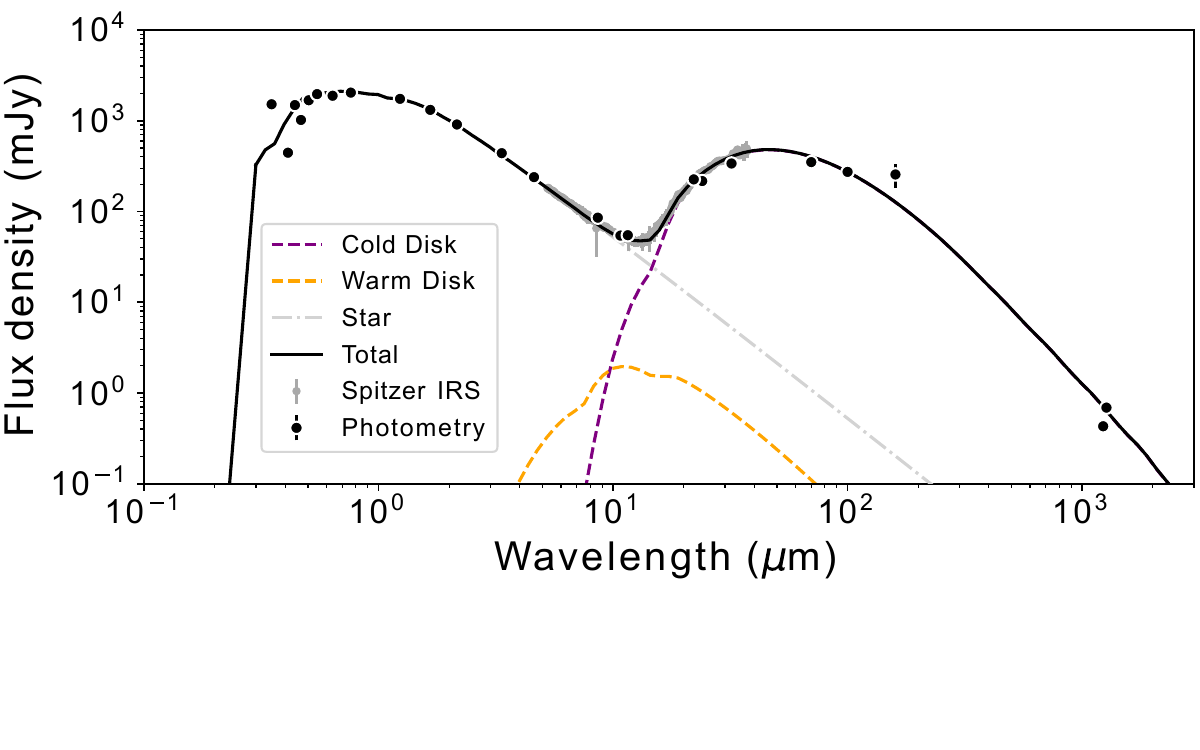}}
	\resizebox{\hsize}{!}{\includegraphics[trim={-.1cm 0 0 0},clip]{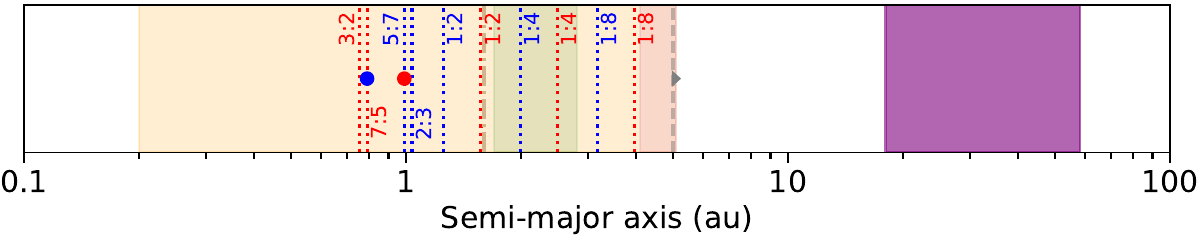}}
    \caption{
    {\it Top:} SED of HD114082, showing 
    stellar photosphere (dotted-dashed gray curve), warm dust (dashed orange curve), cold dust (dashed purple curve), and total emission (solid black line) models, derived from literature photometry (black points) and Spitzer/IRS spectrum (gray line). 
    {\it Bottom:} schematic view of the HD114082 system architecture, showing planet b (blue point) and planet c (red point), respective MMRs and secular resonances (for $e$=0; the red and blue dotted lines), the lower limit of the H$_2$O snowline (dashed gray line; estimated), the habitable zone for a 5 $M_\oplus$ planet from \citet[][shaded green region]{kopparapu2014}, and the locations of the inner warm belt (shaded orange region; estimated) 
    and the spatially constrained exo-Kuiper Belt \citep[shaded purple region:][]{matra2025} are displayed. Also, the maximum stellocentric radius for a potential transit (dotted-dashed line).
    }
    \label{fig:fig3}
\end{figure}

We seek to constrain the location of the unresolved, warm debris disk of HD 114082 \citep[][]{chenchristine2014,mittal2015,jangcondell2015} 
using the spectral energy distribution (SED) in conjunction with the spatially constrained extent of the outer belt \citep[][]{engler2023,matra2025}. 
We adopt the stellar parameters from Appendix \ref{appendixa} to select the PHOENIX stellar photosphere model \citep[][]{Hauschildt1999} and scale it to the photometry in the optical and near-infrared from Tycho BV, Gaia, 2MASS, and WISE \citep[][]{hog2000,gaia2016,gaia2023,skrutskie2006,wright2010}, after proving the excellent match. 
Figure \ref{fig:fig3} shows all the photometry used in our analysis; ultraviolet photometry \cite[][]{paunzen2015} is included for completeness. 

The excess emission is modeled as two dust components fitted to the mid-infrared to millimetre photometry taken from AKARI \citep[][]{ishihara2010}, 
the Spitzer Infrared Spectrograph (IRS) and MIPS \citep[][]{chenchristine2014}, Herschel PACS \citep[][]{marton2024}, and the Atacame Large Millimeter/submillimeter Array \citep[ALMA;][]{liemansifry2016,matra2025}. 
We employ the CASSIS low-resolution spectrum from the IRS onboard the Spitzer Space Telescope \citep[][]{lebouteiller2011}.
See Appendix \ref{appendixe} for more details.

Because of growth and destruction processes, dust grains cover a broad size range, exhibiting different wavelength-dependent absorption, emission, and scattering behaviour. We adopt a power-law distribution for the number density n(s) $\propto$ $s^{-q}$ \citep[][]{dohnanyi1969}, with $s$ being the size of a spherical grain. Our model yields a lower boundary of the size distribution $s_{\rm min} = 1.41^{+0.37}_{-0.21}~\mu$m and $q = 3.52\pm0.04$, with dust masses of $5.5^{+69.2}_{-0.4} \times 10^{-5} M_{\oplus}$ (inner belt) and (4.6$\pm$0.2) $\times$ 10$^{-2}$ $M_{\oplus}$ (outer belt). 
Although weakly constrained at $1.3^{+3.8}_{-1.1}$ au, the inner belt's location indicates that it should be markedly influenced by dynamical interactions with planets b and c \citep[potentially carving Kirkwood-like gaps via large, resonant eccentric increases;][]{horner2020,yoshikawa1989}. MMR chains naturally arise from disk-driven planet migration, forming fragile ones for distant planets, especially around young stars \citep[][]{petit2025}. We tentatively conclude that dust grains are spatially segregated according to their size in the inner belt, whereas asteroids are concentrated on its outskirts. A sketch of the system architecture is shown in Figure \ref{fig:fig3}.  

\section{Discussion} \label{discussion}

\subsection{The Pair of Planets around HD 114082}

\citet{zakhozhay2022} adopted a one-planet Keplerian model to fit a RV curve affected by sparse, irregular sampling and processing artifacts, yielding erroneous parameters for planet b. FEROS data provide broad time coverage but show a high scatter, while the more precise HARPS data are clustered. We collected more public time-series spectra (spread over 6.2 years), enhancing data processing and markedly reducing the RV scatter. Using this refined dataset, our joint model yields a 95\% confidence upper limit on the mass of planet b, which is one-fifth of planet b's value in \citet{zakhozhay2022}, while evidencing the errors in the orbital period and eccentricity therein. 

\begin{figure}[h!]
	\resizebox{\hsize}{!}{\includegraphics{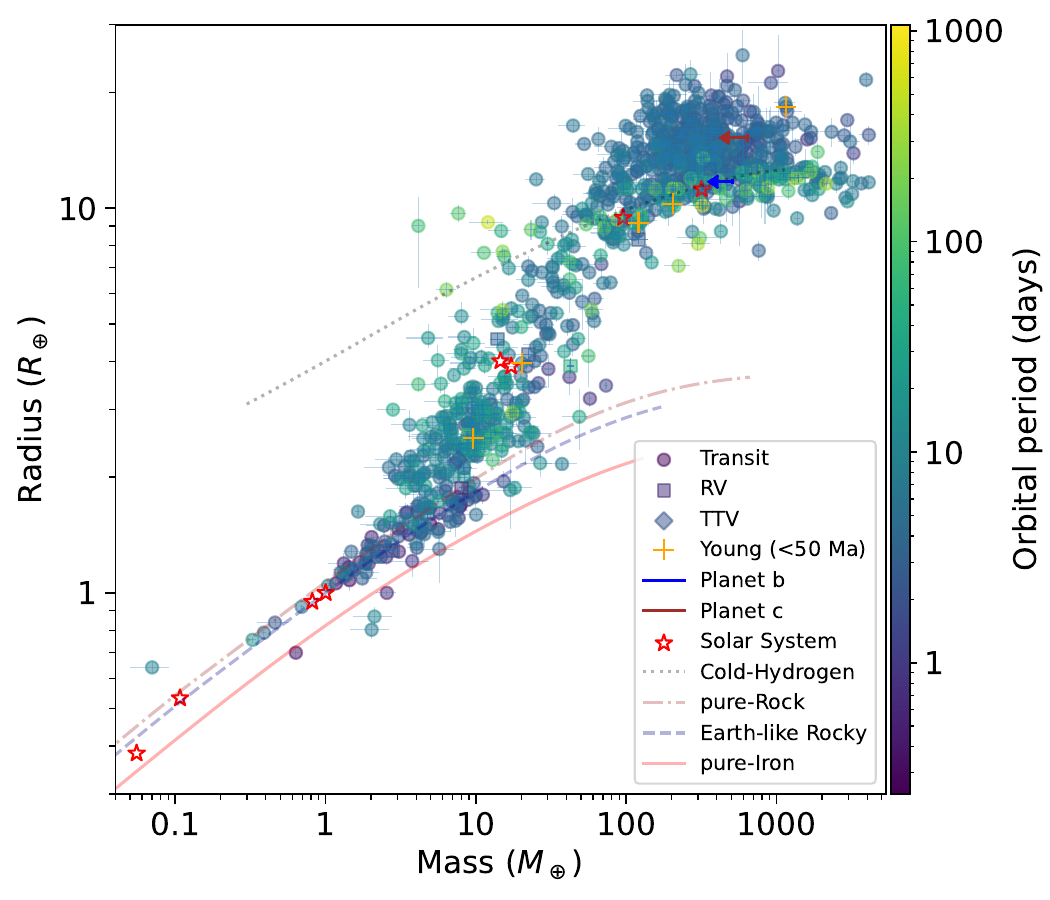}}
    \caption{The $R$--$M$ diagram with planets b and c ($R$ and $M_{95\%}$ from Table \ref{tab:planetary_parameters}) showing, as per the NASA Exoplanet Archive \citep[][]{christiansen2025}, planets discovered by the transit, radial velocity, and TTV methods, with $\ge$3$\sigma$ significant masses and radii; they are color-coded by orbital period, with crosses to mark young cases (ages $\lesssim$50 Ma). Solar System planets are denoted by red five-pointed stars. $R$--$M$ relations for different compositions \citep[][]{zeng2019} are shown as a reference.}
    \label{fig:fig4}
\end{figure}

Figure \ref{fig:fig4} shows the locations of planets b and c in the planetary radius vs mass diagram.

\subsection{Formation and Evolution of the System}
\label{pairformation}

The observed properties of the HD114082 system enable us to draw several inferences regarding its formation and evolution. The host star lies within the $\delta$ Scuti instability strip \citep[][]{zakhozhay2022,gautam2025} and above the Kraft break \citep[][]{kraft1967,beyer2024}. With  
a predominantly radiative envelope and weak magnetic braking, 
the young star retains its primordial rotation and preserves any spin--orbit misalignment from planet migration or scattering. 
The almost coplanar, nearly circular orbits of moderate-to-low mass planets b and c, seemingly in near resonance and likely aligned with the inner dusty belt but inclined by $\sim$7$^\circ$ to the outer exo-Kuiper Belt, favor the core accretion scenario for the planets' formation. The subsequent question is where it occurred.

Similarly to the super-puff Kepler-51 d, which accreted and retains a substantial H/He envelope despite its small semimajor axis \citep[][]{tang2025}, planets b and c could have formed in-situ and grown to substantial sizes but limited masses. Alternatively, they could have emerged beyond the H$_2$O snowline ($>$5 au), from cores that were assembled in the protoplanetary disk and migrated inwards into the observed orbits, while opening a gap in the debris disk. 
This migration could have been accelerated by tidal heating \citep[inflated eccentric-migration;][]{glanz2022}, which could explain 
why these planets are so large for their masses. 

The location of the high-albedo ($\sim$60\%), low polarization ($\sim$17\%) exo-Kuiper Belt could coincide with the CO$_2$ snowline \citep[][]{engler2023}. The vertical scale height of the outer belt in scattered light \citep[$\leq 0.01$;][]{engler2023} lies below the `minimum' expected level for a collisionally stirred disk \citep[0.04~$\pm$~0.02;][]{thebault2009}. This suggests that the outer belt is primordial and essentially undisturbed. 
By contrast, the inner planetesimal belt’s dynamics must be driven by interaction with the pair of planets b and c. 
They likely have driven planetesimal orbits within the inner belt to the same inclination within the secular evolution timescale \citep[e.g.][]{wyatt1999,sefilian2025}. Thus, our findings suggest that the inner belt is coplanar with the planetary orbits, having evolved together.

\section{Conclusions}
\label{conclusions}

Follow-up observations and ancillary data enable us to study the multiplanet system around F-type star HD 114082 with greater details and confidence. Orbital parameters, radii, and 95\% confidence upper limits on the masses of planets b and c are fairly constrained. Thus, we yield an accurate and precise orbital period for planet b and majorly categorise planet c, advancing it toward confirmation---pending one more transit detection. In addition, we present a two-component model that accounts for the infrared excess from the inner Asteroid belt and the exo-Kuiper Belt of HD 114082. 

We highlight that planets b and c are the two longest-period known young transiting exoplanets. They are low-density giants with an orbital period ratio consistent with two commensurabilities, the mean-motion resonances (MMRs) 7:5 and 3:2. 
These planets have almost coplanar, circular orbits, misaligned with the outer, likely primordial exo-Kuiper Belt. They could have formed by core accretion and reach the observed orbits after migrating from the formation site, undergoing tidal heating while growing further from the inner belt, clearing it all the way through, and shelving it dynamically. Alternatively, these giants could be Neptune-mass planets that formed in situ and experienced a combination of boil-off and photoevaporative mass loss, as occurs for the super-puffs. 

We intend to precisely determine the period of planet c and perform a full analysis of TTVs from transit observations in the next years, aiming at further characterizing the system, including the determination of the planetary masses and the discernment of any resonance. 

\begin{acknowledgments}
{We thank the referee for providing useful comments.} 
We thank CHEOPS staff, in particular Bruno Mer\'\i n (ESA CHEOPS Project Scientist for the Guest Observers program) for his guidance and devoted support. We are also grateful to Nikolaus Volgenau (Operations Scientist at LCO) for his kind support during the LCO Director's Discretionary Time preparation. C.d.B. acknowledges support from the Agencia Estatal de Investigación del Ministerio de Ciencia, Innovación y Universidades (MCIU/AEI) under grant WEAVE: EXPLORING THE COSMIC ORIGINAL SYMPHONY, FROM STARS TO GALAXY CLUSTERS and the European Regional Development Fund (ERDF) with reference PID2023-153342NB-I00/10.13039/501100011033, as well as from a Beatriz Galindo Senior Fellowship (BG22/00166) from the MICIU. The University of La Laguna (ULL) and the Department of Economy, Knowledge, and Employment of the Government of the Canary Islands are also gratefully acknowledged for the support provided to C.d.B. (2024/347). 
{A.S.M. acknowledges financial support from the Spanish Ministry of Science and Innovation (MICINN) project PID2020-117493GB-I00 and from the Government of the Canary Islands project ProID2020010129}. 
J.P.M. acknowledges support by the National Science and Technology Council of Taiwan under grant NSTC 112-2112-M-001-032-MY3. 
This research has made use of the SIMBAD database, operated at CDS, Strasbourg, France. 
E.G. gratefully acknowledges support from UK Research and Innovation (UKRI) under the UK government’s Horizon Europe funding guarantee for an ERC Starting Grant [grant number EP/Z000890/1], which funds M.P.B. 
J.S.J. gratefully acknowledges support by FONDECYT grant 1240738 and from the ANID BASAL project FB210003. 
CHEOPS is an ESA mission in partnership with Switzerland with important contributions to the payload and the ground segment from Austria, Belgium, France, Germany, Hungary, Italy, Portugal, Spain, Sweden, and the United Kingdom. The CHEOPS Consortium would like to gratefully acknowledge the support received by all the agencies, offices, universities, and industries involved. Their flexibility and willingness to explore new approaches were essential to the success of this mission. CHEOPS data analyzed in this article will be made available in the CHEOPS mission archive (\href{https://cheops.unige.ch/archive\_browser/}{https://cheops.unige.ch/archive\_browser/}).
This paper includes data collected with the TESS mission, obtained from the MAST data archive at the Space Telescope Science Institute (STScI). Funding for the TESS mission is provided by the NASA Explorer Program. STScI is operated by the Association of Universities for Research in Astronomy, Inc., under NASA contract NAS 5–26555. 
TESS data analysed here can be accessed via \href{https://doi.org/10.17909/t9-nmc8-f686}{doi:10.17909/t9-nmc8-f686} \citep[][]{https://doi.org/10.17909/t9-nmc8-f686}. This research has made use of the Astrophysics Data System, funded by NASA under Cooperative Agreement 80NSSC25M7105.  
This work has made use of data from the European Space Agency (ESA) mission {\it Gaia} (\url{https://www.cosmos.esa.int/gaia}), processed by the {\it Gaia} Data Processing and Analysis Consortium (DPAC, \url{https://www.cosmos.esa.int/web/gaia/dpac/consortium}). Funding for the DPAC has been provided by national institutions, in particular the institutions participating in the {\it Gaia} Multilateral Agreement. 
This work makes use of observations from the Las Cumbres Observatory global telescope network. 
Based on data collected under the NGTS project at the ESO La Silla Paranal Observatory. The NGTS facility is operated by the consortium institutes with support from the UK Science and Technology Facilities Council (STFC) under projects ST/M001962/1, ST/S002642/1 and ST/W003163/1. 
This work uses data obtained with the ASTEP$+$ telescope, at Concordia Station in Antarctica. ASTEP$+$ benefited from the support of the French and Italian polar agencies IPEV and PNRA in the framework of the Concordia station program, from OCA, INSU, Idex UCAJEDI (ANR-15-IDEX-01) and ESA through the Science Faculty of the European Space Research and Technology Centre (ESTEC). We also received funding via Science and Technology Facilities Council (STFC; grants Nos. ST/S00193X/1, ST/W002582/1 and ST/Y001710/1) as well as from European Research Council (ERC) under the European Union's Horizon 2020 research and innovation program (grant agreement No. 803193/BEBOP).
We would like to acknowledge the dedicated staff of the French and Italian polar agencies (IPEV and PNRA) for their dedication and for their work, particularly those that winter-over, which is essential in maintaining the Concordia Station operational throughout the Austral winter, and thanks to whom the ASTEP$+$ telescope could collect data for this publication. 
Based on data obtained from the ESO Science Archive Facility that can be accessed via \href{https://doi.org/10.18727/archive/33}{doi:10.18727/archive/33} \citep[][]{https://doi.org/10.18727/archive/33}, under programmes 0101.A-9012 and 113.26Y7.001 (PI: R. Launhardt), 0103.A-9010, 0107.A-9004, 0108.A-9014 and 0109.A-9014 (PI: O. Zakhozhay), 098.C-0739 and 1101.C-0557 (PI: A. M. Lagrange). 
{M.L. acknowledges support of the Swiss National Science Foundation, grant
PCEFP2\_194576. M.L. and H.P.O.'s contributions have been carried out within the NCCR PlanetS supported by the Swiss National Science Foundation under grants
51NF40\_182901 and 51NF40\_205606.} 
This publication makes use of The Data \& Analysis Center for Exoplanets (DACE), which is a facility based at the University of Geneva (CH) dedicated to extrasolar planets data visualisation, exchange and analysis. DACE is a platform of the Swiss National Centre of Competence in Research (NCCR) PlanetS, federating the Swiss expertise in Exoplanet research. The DACE platform is available at \href{https://dace.unige.ch}{https://dace.unige.ch}. 
This research has made use of the NASA Exoplanet Archive, which is operated by the California Institute of Technology, under contract with the National Aeronautics and Space Administration under the Exoplanet Exploration Program. Data used in this paper can be accessed via \href{https://doi.org/10.26133/NEA2}{10.26133/NEA2} \citep[Composite Planet Data Table;][]{https://doi.org/10.26133/nea2}.

\end{acknowledgments}

\facilities{TESS, CHEOPS, NGTS, ASTEP$+$, LCO, ESO:3.6m, Max Planck:2.2m.}

\software{In addition to those mentioned  explicitly in the text, this work has made use of the following software packages: 
{\sc astropy} \citep{AstroPy13,AstroPy18,AstroPy22}, 
{\sc corner} \citep{foremanmackey2016}, 
{\sc matplotlib} \citep{Matplotlib}, 
{\sc numpy} \citep{Numpy}, 
and {\sc scipy} \citep{SciPy}.}

\begin{contribution}
C.d.B. was responsible for the idea, leading several successful observing proposals, coordinating the project, characterizing the star, computing TTVs, creating figures, contributing to the overall analysis, developing the full interpretation, and writing and submitting the manuscript.  
A.S.M. computed the RV measurements, built and performed the global photometric and RV model, created the related figures, and edited the manuscript. 
A.H. performed the LCO data reduction, contributed to the discussion on the nature of planet c, and edited the manuscript.
J.P.M. implemented the two-component model, fitting the infrared excess, creating a figure, and edited the manuscript.
P.J.W. conducted the NGTS observing plan and data reduction, and contributed to the manuscript edition. E.M.B. and S.G. contributed to the NGTS data reduction. J.F.F. developed the NGTS pipeline. D.R.A., M.P.B., E.G., and S.U.M. planned the NGTS observations. JMcC operated NGTS. M.L. made comments to the manuscript. I.A., D.B., F.H., J.S.J., M.M., L.D.N., A.M.S.S., S.S., S.U., J.I.V., R.G.W., and M.R.B. are members of NGTS.
H.P.O. belongs to CHEOPS and made comments to the manuscript.
T.G. and A.T. allocated the ASTEP observations. 
O.S. is responsible of the ASTEP photometric pipelines. 
M.B. was responsible for the maintenance of ASTEP at the Concordia Station. 
A.A. was in charge of the ASTEP operations.
L.A., P.B., G.D., and D.M. are members of ASTEP+.
I.P. and M.J.H. {conducted the CHEOPS GTO programmes}.
\end{contribution}

\appendix

\section{Stellar Characterization} \label{appendixa}

HD 114082 is currently the bluest spectral type star among those with young transiting exoplanets. We apply the Bayesian inference code of \citet{delburgo2016,delburgo2018} on a grid of PARSEC 1.2S stellar evolution models \citep{bressan2012,chen2014,chen2015,tang2014} 
to derive the fundamental stellar parameters. The arranged grid of models comprises ages from 2 to 13,800 million years with steps of 5\%, 
and metallicites [M/H] from -2.18 to 0.51 with steps of 0.02 dex, adopting the photometric passband calibration of \citet{riello2021} 
and the zero-points of the VEGAMAG system. 
The Bayesian code is fed by the absolute $G$ magnitude $M_{G}$, the color $G_{\rm BP}-G_{\rm RP}$ \citep[][]{gaia2023}, 
and the stellar age \citep{squicciarini2025}. 
Given the relatively small distance of the star, we assume null extinction when deriving $G_{\rm BP}-G_{\rm RP}$ and $M_{G}$ 
from \textit{Gaia} DR3 photometry and astrometry.  
We estimated the stellar rotation period from the rotational broadening (vsini) value of \citet{zakhozhay2022}, assuming that HD 114082 is spinning in the plane of the likely primordial exo-Kuiper Belt \citep[inclined by $\approx$83$^\circ$;][]{engler2023}.
Table \ref{tab:stellar_parameters} shows the fundamental stellar parameters for our target.

\label{sec:star_params}
\small
\begin{table}
    \centering
    \caption{Fundamental Stellar Parameters of HD~114082. 
    \label{tab:stellar_parameters}}
    \begin{tabular}{lcc}
    \hline\hline 
        Parameter & Value & Reference \\
        \hline
        Distance (pc)  & 95.06$\pm$0.20 & 1 \\
        \hline
        M$_G$ (mag) & 3.216 $\pm$ 0.005 & 1 \\       
        BP-RP (mag) & 0.568 $\pm$ 0.005 & 1 \\ 
        Age (Ma)   & 15 $\pm$ 3 & 2\\
        \hline
        Effective temperature (K) & 6610$\pm23$ & 3 \\
        Radius ($R_{\odot}$) & 1.482$\pm0.021$ & 3 \\         
        Mass ($M_{\odot}$) & 1.28$\pm0.07$ & 3 \\               
        Density (g cm$^{-3}$) & 0.55 $\pm$ 0.06 & 3 \\         
        Surface gravity ($\log g$/$cgs$) & 4.20$\pm0.03$ & 3 \\
        Luminosity ($L_{\odot}$) & 3.78$\pm0.07$ & 3 \\
        Bolometric magnitude (mag) & 3.297$\pm0.019$ & 3 \\
        Iron-to-hydrogen abundance  & -0.21$\pm0.25$ & 3 \\ 
        vsini (km s$^{-1}$)         & 39.2$\pm$0.5 & 4 \\
        Rotational period (d)       & 1.85$\pm$0.04 & 3 \\
    \hline
    \end{tabular}
    {\bf References.} (1)~\citet{gaia2023}; (2)~\cite{squicciarini2025}; 
    (3) this work; (4)~\cite{zakhozhay2022}.
\end{table}
\normalsize

\begin{figure*}[t]
	\resizebox{\hsize}{!}{\includegraphics{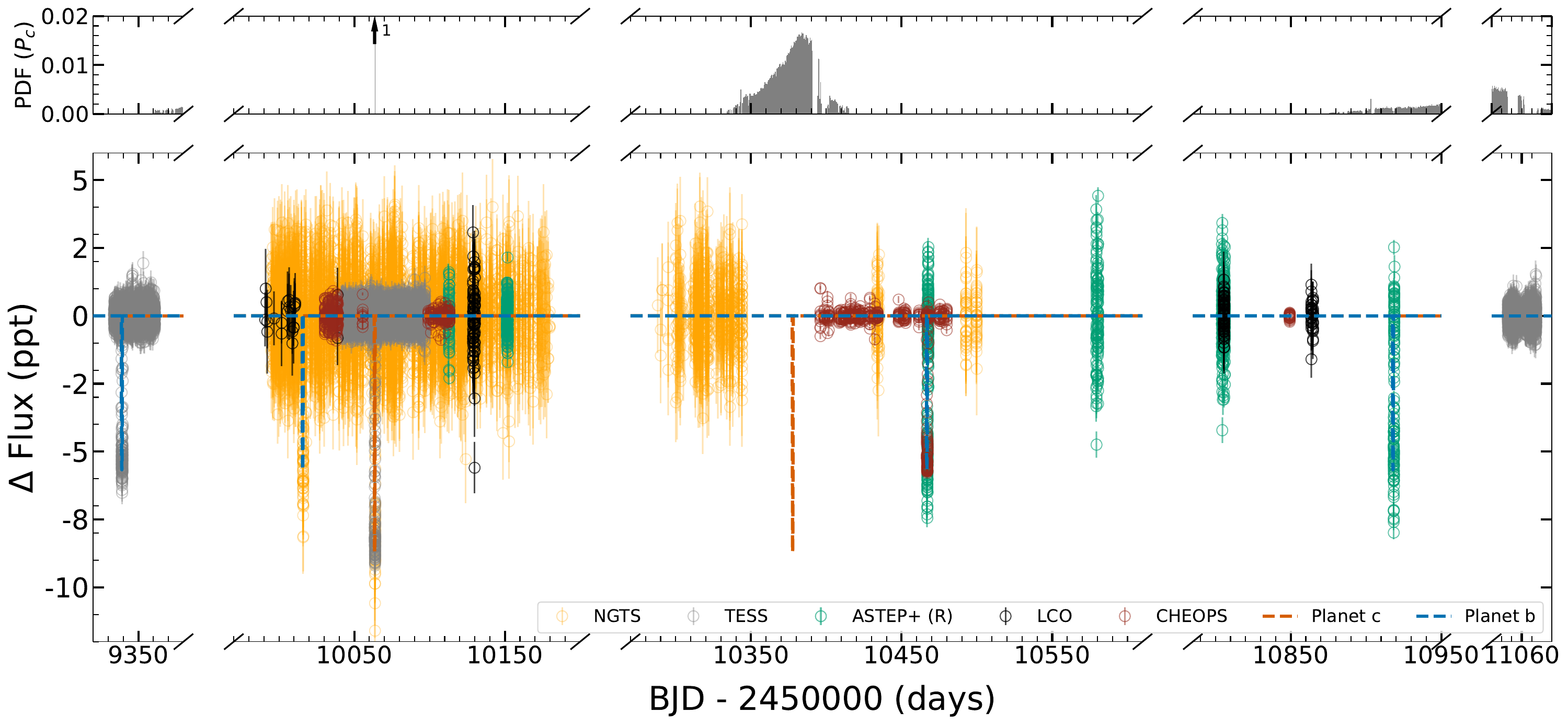}}

    \caption{{\it Top:} PDF of midtransit times for planet c (12 hr bins; $T_{14}$= 13.21 h). The PDF of the observed transit integrates to unity (BJD 246063.5112; $\sigma\approx$ 1.7\,min $\lll$ bin width). {\it Bottom:} Full photometric dataset, derived from different facilities, after detrending from stellar and instrumental effects. The data used to generate this figure are available in machine-readable format (see Figure~\ref{fig:fig1}).
    }
    \label{fig:figA}
\end{figure*}

\subsection{Search for Stellar Companions}
Using the Spectro-Polarimetric High-contrast Exoplanet REsearch (SPHERE) at the Very Large Telescope (VLT), with an angular resolution of $\approx$25 mas, \citet{engler2023} found only background stars within a $\lesssim$6.5$^{\prime\prime}$ cone around the target, which they attributed to its proximity to the Galactic plane (latitude $\approx$+2.49$^{\circ}$). 

\section{Planetary Characterization}
\label{appendixb}

Our global GP model, used for joint fitting, follows the equation below, with the relative flux $\Delta F$ 
in parts-per-thousand (ppt) and the relative radial velocity $\Delta V_{R}$ in m s$^{-1}$:
\begin{equation}\label{eq_full_model}
    \begin{split}
        & \Delta F =  Z_{F,j} + GP_{j} + PS_{F},\\ 
        & \Delta V_{R}   =  Z_{V_R,j} + T_L + T_{B} + GP + PS_{V_R},
    \end{split}
\end{equation}

where $PS_F$ is the planetary signal in the light curve, $PS_{V_R}$ is the planetary signal in the radial velocity (RV) curve, $T_L$ is the linear trend, $T_{B}$ is the trend against bisector span, and $Z_{F,j}$ and $Z_{V_R,j}$ represent the zero-point of each individual source of data (denoted by $j$), with priors $\mathcal{N}[0,5]$ ppt for $\Delta F$, and $\mathcal{N}[0,1000]$ m s$^{-1}$ for $\Delta V_R$. $T_L$ accounts for very long-term variations, with a prior $\mathcal{N}[0,1000]$ m s$^{-1}$ and BJD 2459332.9583 indicating the midpoint of the RV baseline. 

In joint fitting to two-planet systems, $PS_{F}$ constrains---per component---the planet-to-star radius ratio ($R/R_*$), orbital period ($P$), eccentricity ($e$; primarily if secondary transits are observed), inclination ($i$), and scaled semimajor axis $a/R_\star$. The stellar density ($\rho_\star$) is robustly bounded from the ensemble of transit durations, even for moderately eccentric orbits. $PS_{V_R}$, in turn, constrains each planet’s mass (given the stellar mass, $M_\star$), $P$, and $e$, provided the data have sufficient signal-to-noise ratio and are well sampled. We used $\rho_\star$ and $M_\star$, together with effective temperature, 
surface gravity, and iron-to-hydrogen abundance [Fe/H] (which inform limb-darkening coefficients), as priors or to convert relative quantities into absolute planetary values. 

\begin{figure}[h!]
	\resizebox{\hsize}{!}{\includegraphics{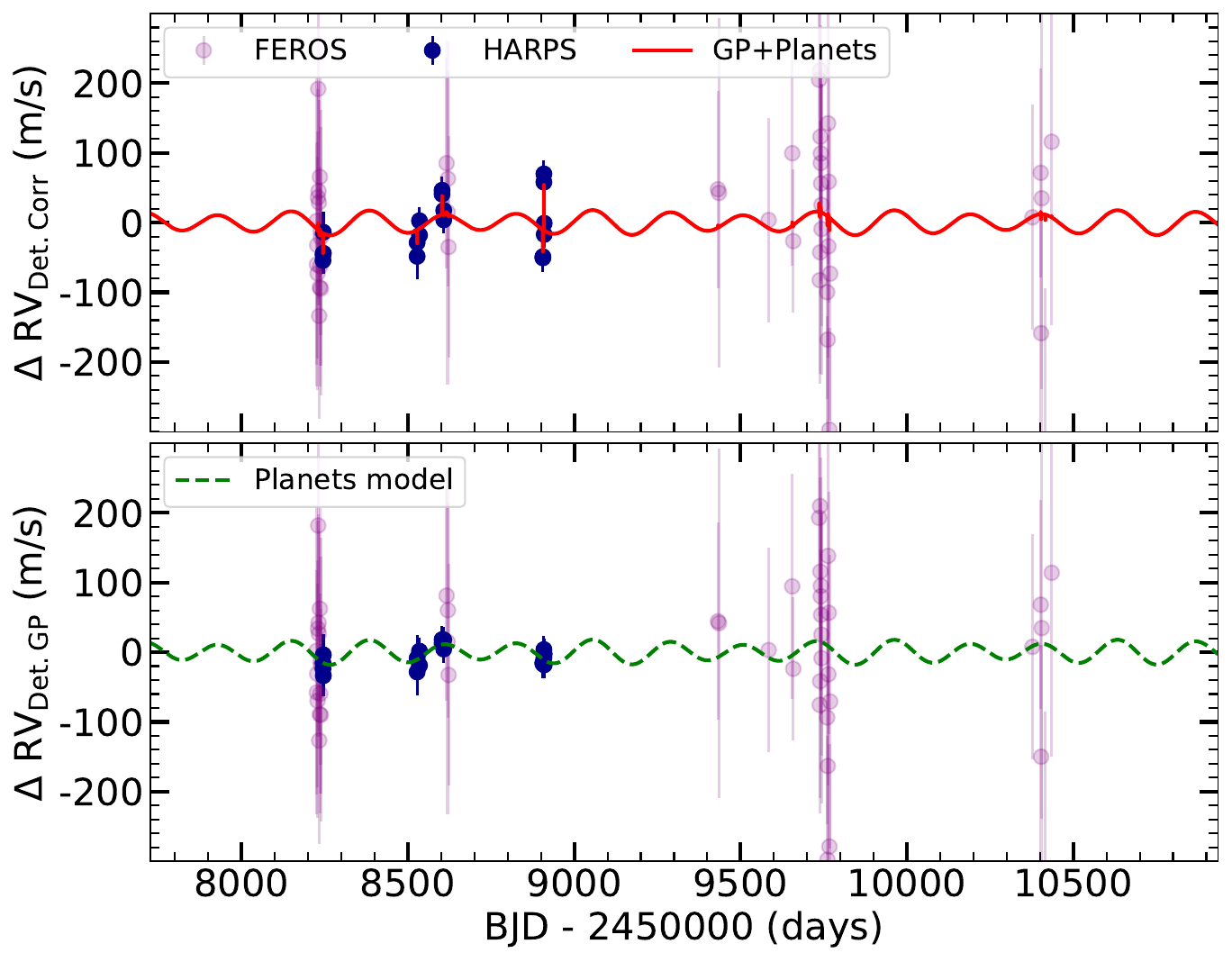}}
    \caption{
    \textit{Top:} RV curve, with HARPS and FEROS data points, after detrending from the long-term trend and the correlation with the bisector span. \textit{Bottom:} RV curve detrended from the GP variations with the planetary model ($PS_{V_R}$). The data used to generate this figure are available in machine-readable format in the online journal (see Figure~\ref{fig:fig2}).
    }
    \label{fig:figB}
\end{figure}

\begin{figure}[h!]
	\resizebox{\hsize}{!}{\includegraphics{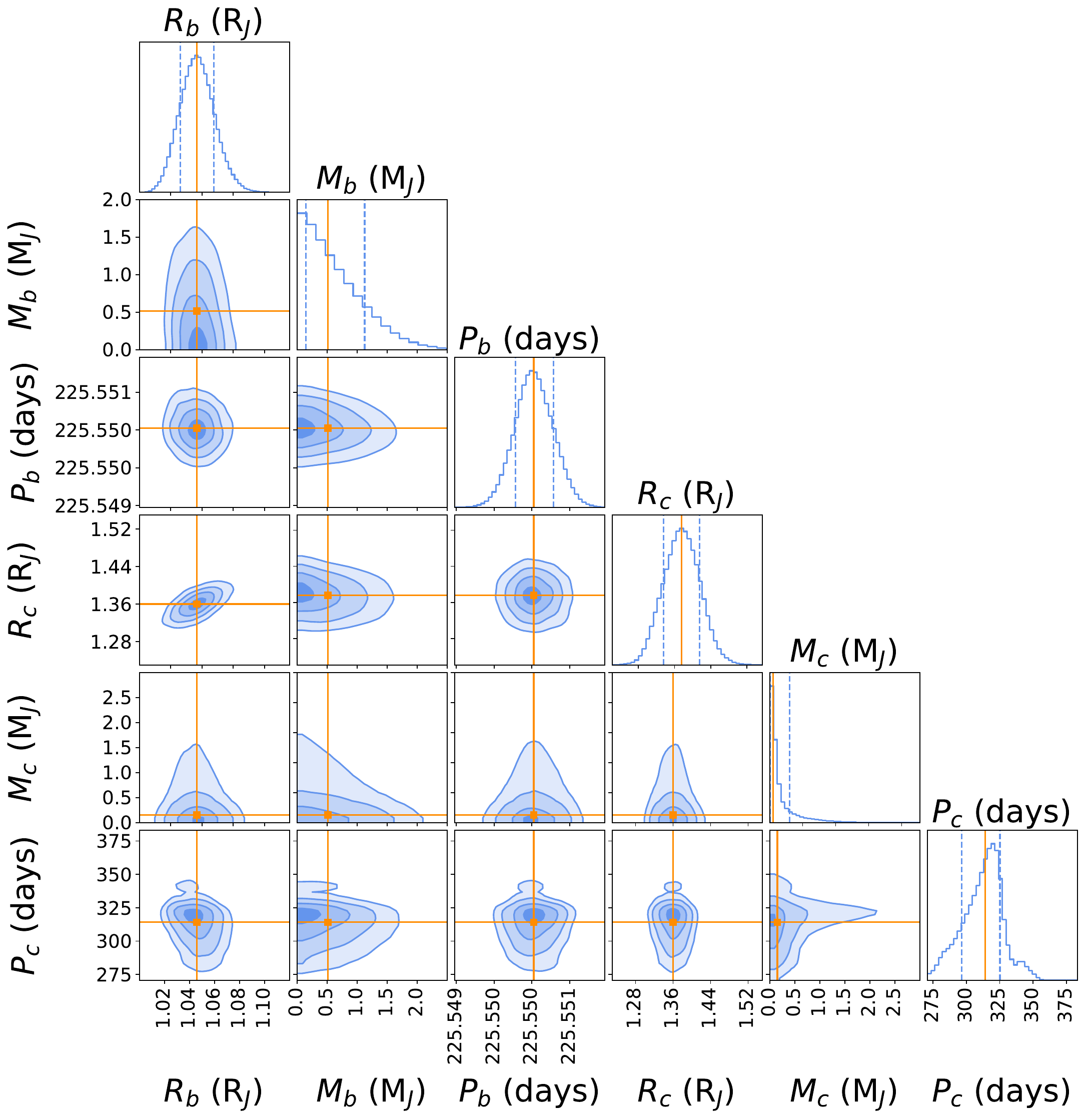}}
    \caption{
    Corner plot of the parameters from the best joint model fit: correlation maps and posterior distributions. 
    }
    \label{fig:figC}
\end{figure}

Figures \ref{fig:figA} and \ref{fig:figB} respectively show the different light curves and the RV curve employed to perform our analysis, while Figure \ref{fig:figC} illustrates the results from our joint modeling. 

\subsection{CCF Bisector Span}

We identified a correlation between the RV measurements and the bisector span of the CCF. In particular, for the more precise HARPS data, we find a Spearman's rank correlation coefficient \citep{spearman1904} of --0.59 (p-value $<$ 0.01). Previous RV surveys for stars above the Kraft break (e.g. \citealt{lagrange2009}), such as HD 114082 (see \S\ref{thehost} and \S\ref{pairformation}), have found correlations between RV and bisector span to be a common result of stellar variability. We account for this effect by including linear terms between RV and bisector span for the HARPS and FEROS data.

\subsection{GP Modeling}

We apply GP regression to model short-term variability of stellar and instrumental origin. This flexible, probabilistic machine learning method defines a probability distribution over functions, employs a kernel to encode relationships between data points, and yields predictions with quantified uncertainty.
For ground-based photometry, most variability arises from systematic effects and scintillation noise, which vary by data source.
Each source is modeled independently with a simple harmonic oscillator (SHO) kernel \citep[][]{Foreman-Mackey2017}. We rely on the same kernel for the RV data. 

Here, the SHO kernel $k_{SHO}(\tau, \alpha_{j}, P_{j}, Q_{j})$ depends on the time interval $\tau$, the standard deviation of the process ($\alpha_{j}$), the oscillator period ($P_{j}$), and the oscillator quality factor ($Q_{j}$), related to the evolution timescale ($L$) as $Q = {\pi {L}\over{P_{\rm s}}}$, where $P_{\rm s}$ is the stellar variability period. For all parameters, log-uniform priors with values $\mathcal{U}[-10,10]$ are used to ensure that they are always positive and have the maximum flexibility.

\subsection{Planetary Models}

The transit signals are modeled using Pytransit \citep[][]{Parviainen2015}, with quadratic limb-darkening parameters derived from the Limb Darkening Toolkit \citep[LDTK: ][]{Parviainen2015b}, while for RV signals sinusoidal functions are employed. We optimize the orbital and physical parameters $T_0$, $P$, $R$, and $b$, and transform them into $i$, $K$, $R/R_*$, and $a/R_\star$, inside the likelihood, using as priors the aforementioned stellar parameters, listed in Section \ref{appendixa}. 
The optimization process implicitly performs an end-to-end uncertainty propagation. For $T_0$, we use the prior $\mathcal{N}[T0,0.1]$ (d), introducing the known transit epochs. 
For $P$, we apply the prior $\mathcal{LU}[20,8000]$ (d). For $R$, we use $\mathcal{LU}[0.007,2.7]$ ($R_{\rm J}$), and for $M$, $\mathcal{LU}[0.007,20]$ ($M_{\rm J}$). 
For $b$, a uniform prior $\mathcal{U}[0,1.2]$ is applied, allowing for grazing transits and even non-transiting solutions. In the case of TESS data, we include a dilution parameter $\mathcal{N}[1.0,0.3]$. 

In order to test circular and eccentric models, we adopt a parametrisation that separates the eccentricity $e$ and the argument of the periapsis $\omega$ into $x= \sqrt{e}\cos{\omega}$ and $y= \sqrt{e}\sin{\omega}$, applying a normal prior $\mathcal{N}[0,0.3]$ on both components of the eccentricity vector. The $(e, \omega)$ parametrisation induces strong covariance at low $e$ due to rotational degeneracy in the likelihood, while the adopted vector maps the parameter space to the unit disk, removes the $(e,\omega)$ covariance ridge at $e\to0$, and allows $\omega$ to be undefined at $e=0$ without sampling artifacts. The independent normal priors $\mathcal{N}(0, 0.3)$ on $x$ and $y$ induce a smooth $\propto \sqrt{e}$ prior on $e$---weakly informative and well-suited for Bayesian model selection.
This parametrisation yields a distribution of eccentricities similar to that proposed by \citet{Kipping2013}, based on the distribution of eccentricities of known planets. 
Thus, it avoids any statistical issue that occurs when trying to fit simultaneously the correlated $e$ and $\omega$ from a single constraint like the transit duration. 
The inferred eccentricities of both planets are consistent with zero ($e_{b}$ < 0.2, $e_{c}$ < 0.3), with Bayesian evidence favoring the circular model (see Table \ref{tab:prior_posterior}). However, we cannot rule out low but non-zero eccentricities.

\subsection{White-noise Terms}

Finally, we include white noise terms for all data sources, to account for possible unmodeled variability. 
We adopt a prior $\mathcal{LU}[0.007,150]$ ppt for all photometric sources and a prior $\mathcal{LU}[5^{-5},22000]$ m s$^{-1}$ for RV data. 

\section{TTVs}
\label{appendixc}

{\sc rebound} \citep[][]{rebound,reboundias15} is an N-body integrator that 
enables us to assess TTVs on planet b induced by its planetary companion. 
Adopting the respective masses of 1.6 and 2.0 $M_{\rm J}$ for planets b and c, i.e. the values of the 95\% confidence upper limits listed in Table \ref{tab:planetary_parameters}, the resulting TTV semiamplitude (determined from all virtual mid-transit times in the time span between the first and the last observed transits) is nearly 2 hr (95\% confidence limit), rejecting the candidate period $P_{1/2}$ of about 112.78 d for planet b from our alias testing observations, as a dip would have been recorded (see Figure \ref{fig:fig1}).  

The transit duration of planet b---the same at two observing epochs, from TESS and CHEOPS---is fixed to derive the midtransit times of partial ($\sim$50\% coverage) NGTS and ASTEP$+$ transits.  
The four measures enable us to estimate a TTV semiamplitude of $\lesssim$ 4 $\pm$ 2 hr for planet b; this far exceeds any conceivable midtransit time jitter and suggests a planet c's mass significantly smaller than that of Jupiter, particularly if the pair is locked in an MMR. Given the small statistics, 
instead of directly including the TTVs in our joint modeling, to further constrain the planetary masses, we choose to refine them subsequently. After feeding {\sc rebound} with approximately 1 million representations that settle the full posterior distributions of the parameters for the preferred circular solution shown in Table \ref{tab:planetary_parameters} and \ref{tab:prior_posterior}, those producing TTVs greater than the conservative value of 10 hr are filtered out. As a result, $M_{95\%}$ for planets b and c decrease from 1.6 and 2.0 $M_{\rm J}$ to 1.5 and 0.24 $M_{\rm J}$, respectively, yielding a TTV semiamplitude on planet b of about 7 hr (95\% confidence limit). 

\section{Nature of Planet C}
\label{appendixd}

Planet c's transit, fully observed by TESS and with the same depth and >50\% coverage by NGTS, cannot be an artifact. Moreover, the RV curve imposes 
upper limits for its planetary mass (as it does for planet b), and the debris disk fences off 
its location, with an inclination between the plane of the exo-Kuiper Belt and those of the planetary orbits. This evidence strongly supports planet c as a dynamically interacting body. In addition to ruling out stellar companions in Appendix \ref{appendixa}, a false-positive analysis is conducted here. 
Using {\sc TRICERATOPS} \citep[][]{giacalone2021}, we find that none of the Gaia DR3 Catalogue stars \citep[][]{gaia2023} within the TESS aperture 
can affect or cause this transit, and rule out that it is due to an unresolved eclipsing binary by comparing its expected shape with that of the observed transit. 
Furthermore, Gaia DR3 yields a renormalized unit weight error (RUWE) of 0.944, signifying that HD 114082 is a single star. 

\section{System Architecture Modeling}
\label{appendixe}

Our model uses the 1D analytical radiative transfer code {\sc artifact} \citep[][]{marshall2023}, exploring parameter space using {\sc emcee} \citep[][]{foremanmackey2013}. The adopted disk architecture consists of a pair of Gaussian belts, each defined by a peak radius and width. The radius and width of the cold belt are fixed at the ALMA-observed values ($r_{\rm cold}$ = 38 au, FWHM $\leq$ 40 au) from \citet{matra2025}. A prior $\mathcal{N}[0.1,10]$ au for the warm belt radius ($r_{\rm warm}$) is applied, with a fixed fractional width of 0.1. Note the dust emission profile is dominated by the stellocentric radius rather than the belt width for symmetric architectures, such that the value of this parameter is somewhat arbitrary. For simplicity, we adopt a single power-law size distribution ($s_{\rm min}$, $q$) up to 3~mm and the same dust composition \citep[astronomical silicate--astrosil;][]{draine2003} for both belts, with each component's dust mass as a free parameter.

We sample the posterior with 60 walkers (10 per free parameter, including a nuisance factor) over 2000 steps each. Walkers are initialized uniformly between prior bounds. Namely, $\mathcal{N}[2,5]$ for $q$, $\mathcal{N}[-2,2]$ in $\log_{10} (\mu m)$ for $s_{\rm min}$, $\mathcal{N}[-1,1]$ in $\log_{10} (\mathrm{au})$ for $r_{\rm warm}$, and $\mathcal{N}[-6,0]$ in $\log_{10} (M_{\oplus})$ for the dust masses. Each realization of the disk model is compared to observed data points using error-weighted least squares fitting. Posteriors are sampled from the final model realizations. Maximum amplitude probability values and associated uncertainties are obtained from the 16th, 50th and 84th percentiles of the marginalized posteriors.

\section{Additional tables} 
\label{appendixf}

Tables \ref{tab:cheops_extended} and \ref{tab:lco_extended} provide relevant information about the CHEOPS and LCO observing campaigns, respectively. Table \ref{tab:prior_posterior}
shows the full set of priors and optimized values for the adopted circular model and the eccentric model.

\begin{longtable}{lcc}
\caption{Information about the Individual CHEOPS Observing Campaigns, Including the Midtime, Duration ($\Delta T$) and File identification. \label{tab:cheops_extended}}\\
\endfirsthead
\multicolumn{3}{c}{\tablename\ \thetable\ -- \textit{Continued from previous page}} \\
\hline
\endhead
\hline \multicolumn{3}{r}{\textit{Continued on next page}} \\
\endfoot
\hline
\endlastfoot
\hline\hline 
 Midtime & $\Delta T$   & File Key \\
 BJD-2450000 &   h       &          \\
\hline
10030.2669869135 & 3.20 & PR100017\_TG005901 \\
10030.5503222360 & 2.93 & PR100017\_TG001301 \\
10031.7926661754 & 1.41 & PR100017\_TG001701 \\
10032.6871276520 & 4.47 & PR100017\_TG005401 \\
10033.0690723600 & 2.27 & PR100017\_TG001901 \\
10033.6774027506 & 1.06 & PR100017\_TG002001 \\
10033.8829572285 & 1.07 & PR100017\_TG002101 \\
10034.4348098873 & 4.09 & PR100017\_TG002201 \\
10035.1503247325 & 5.26 & PR100017\_TG004901 \\
10035.8058705040 & 1.06 & PR100017\_TG002501 \\
10036.9732405210 & 1.06 & PR100017\_TG002701 \\
10037.4662834690 & 1.45 & PR100017\_TG002801 \\
10037.7454614015 & 3.20 & PR100017\_TG004401 \\
10038.0287940680 & 4.95 & PR100017\_TG002901 \\
10038.4537943546 & 5.05 & PR100017\_TG003001 \\
10038.9183782640 & 1.55 & PR100017\_TG003201 \\
10040.3030992053 & 3.20 & PR100017\_TG003901 \\
10040.8176759845 & 2.71 & PR100017\_TG003801 \\
10041.5482439375 & 4.56 & PR100017\_TG003701 \\
10055.3468564620 & 1.55 & PR100017\_TG006201 \\
10069.2429013816 & 1.01 & PR100017\_TG006401 \\
10098.8371326667 & 0.70 & PR100017\_TG007001 \\
10100.1037635445 & 1.55 & PR100017\_TG010401 \\
10100.5532800066 & 2.10 & PR100017\_TG010501 \\
10101.0343075204 & 0.71 & PR100017\_TG007901 \\
10101.5148633397 & 0.92 & PR100017\_TG010701 \\
10102.3641785895 & 1.98 & PR100017\_TG008101 \\
10102.8190290960 & 0.72 & PR100017\_TG008201 \\
10103.9218180776 & 3.41 & PR100017\_TG008401 \\
10105.2912536770 & 2.57 & PR100017\_TG008701 \\
10105.4287548630 & 0.92 & PR100017\_TG011501 \\
10106.7329211803 & 0.92 & PR100017\_TG009001 \\
10107.0148753454 & 4.03 & PR100017\_TG013601 \\
10108.1755002120 & 0.91 & PR100017\_TG009301 \\
10109.3002906220 & 1.71 & PR100017\_TG009501 \\
10109.7548877370 & 3.96 & PR100017\_TG014201 \\
10110.3725049040 & 2.56 & PR100017\_TG014301 \\
10111.0864045927 & 3.21 & PR100017\_TG014401 \\
10112.6440422570 & 3.42 & PR100017\_TG011801 \\
10395.8878197840 & 1.06 & PR140079\_TG022501 \\
10398.5037984470 & 1.54 & PR140079\_TG022502 \\
10400.4774037595 & 1.06 & PR140079\_TG022503 \\
10401.5051822490 & 1.06 & PR140079\_TG022601 \\
10408.3003134946 & 1.41 & PR140079\_TG020501 \\
10409.6642101766 & 1.54 & PR140079\_TG023101 \\
10411.4642164180 & 1.54 & PR140079\_TG020502 \\
10412.6044952828 & 1.54 & PR140079\_TG020503 \\
10415.8989391990 & 3.91 & PR140079\_TG023301 \\
10416.7121263184 & 1.06 & PR140079\_TG023001 \\
10417.8086503630 & 1.06 & PR140079\_TG023102 \\
10418.8072608896 & 1.44 & PR140079\_TG023002 \\
10421.1651356560 & 3.32 & PR140079\_TG023401 \\
10422.2611943750 & 1.05 & PR140079\_TG024101 \\
10424.8642106615 & 0.99 & PR140079\_TG024001 \\
10428.6378258370 & 2.45 & PR140079\_TG024102 \\
10432.0301935254 & 1.54 & PR140079\_TG024002 \\
10433.7239370504 & 3.19 & PR140079\_TG024401 \\
10435.0718401910 & 1.00 & PR140079\_TG024201 \\
10435.7199955634 & 3.19 & PR140079\_TG025401 \\
10436.5519513395 & 1.54 & PR140079\_TG025601 \\
10439.8674448620 & 0.99 & PR140079\_TG025701 \\
10442.5672145397 & 3.19 & PR140079\_TG025801 \\
10445.8561101425 & 1.54 & PR140079\_TG025602 \\
10447.8419948937 & 3.19 & PR140079\_TG025802 \\
10451.3088705390 & 3.22 & PR140079\_TG026001 \\
10452.4276334317 & 1.54 & PR140079\_TG026002 \\
10457.0632674615 & 1.86 & PR140079\_TG026201 \\
10461.2736957060 & 1.82 & PR140079\_TG026202 \\
10463.1604902204 & 4.01 & PR140079\_TG026301 \\
10466.0370702036 & 46.96 & PR240025\_TG000301 \\
10468.4345744494 & 1.82 & PR140079\_TG027501 \\
10472.1338792680 & 1.61 & PR140079\_TG027601 \\
10475.1049446850 & 3.58 & PR140079\_TG027401 \\
10476.8611839170 & 3.28 & PR140079\_TG027402 \\
10477.6461603073 & 1.63 & PR140079\_TG027602 \\
10479.6690668985 & 0.90 & PR140079\_TG027502 \\
10834.8044083250 & 3.55 & PR250012\_TG000201 \\
10836.0448931560 & 5.67 & PR240025\_TG000601 \\
10837.4745472097 & 2.58 & PR250012\_TG000202 \\
10840.5773606063 & 2.05 & PR240025\_TG000501 \\
10842.4855177826 & 3.90 & PR240025\_TG000602 \\
10845.9837828200 & 2.09 & PR240025\_TG000603 \\
10848.7252775384 & 3.02 & PR240025\_TG000604 \\
10848.9342827920 & 4.43 & PR250012\_TG000203 \\
\hline
\end{longtable}

\begin{table*}\centering
\caption{Detailed Information about the Individual LCO Observing Campaigns. \label{tab:lco_extended}}
    \begin{tabular}{lccccccc}
    \hline\hline 
        Date      & Site &     Telescope & Filter       & Exposure time         & Cadence  & Duration & Program Id\\
                               & & & & s  &    s    &    h             \\
    \hline
     25/12/2022 & Siding Spring & 0m4-03 & rp & 4.2 & 35.1 & 0.01 & KEY2020B-007 \\
     26/12/2022 & Siding Spring & 0m4-03 & rp & 4.2 & 17.9 & 0.01 & KEY2020B-007 \\
     15/02/2023 & Siding Spring & 0m4-03 & zs & 15.2 & 29.4 & 0.09 & DDT2022B-007 \\
16/02/2023 & Siding Spring & 0m4-03 & zs & 15.2 & 29.5 & 0.09 & DDT2022B-007 \\
05/03/2023 & Siding Spring & 0m4-05 & V & 15.3 & 29.5 & 0.109 & DDT2022B-007 \\
06/03/2023 & Siding Spring  & 0m4-05 & V & 15.3 & 29.4 & 0.106 & DDT2022B-007 \\
07/03/2023 & Siding Spring & 0m4-05 &V&15.3&29.5&0.11& DDT2022B-007\\
02/04/2023 & Siding Spring &0m4-03 & rp&4.3&18.4&0.01& KEY2020B-007 \\
04/04/2023 & Siding Spring & 0m4-03 & rp&4.3&18.6&0.01& KEY2020B-007 \\
     16/12/2022 & Cerro Tololo & 0m4-09 & rp & 4.3 & 17 & 0.01 & KEY2020B-007 \\
     20/12/2022 & Cerro Tololo & 0m4-09 & rp & 4.3 & 17.1 & 0.01 & KEY2020B-007 \\
15/01/2023 & Cerro Tololo & 0m4-09 & rp&4.3&36&0.01& KEY2020B-007 \\
16/01/2023 & Cerro Tololo & 0m4-09 & rp&4.3&17.2&0.01& KEY2020B-007 \\
27/01/2023 & Cerro Tololo & 0m4-09 & rp&4.3&17.3&0.01& KEY2020B-007 \\
28/01/2023 & Cerro Tololo & 0m4-09 & rp&4.3&17.2&0.01& KEY2020B-007 \\
14/02/2023 & Cerro Tololo & 0m4-09 & zs&15.3&28.8&0.089& DDT2022B-007 \\
15/02/2023 & Cerro Tololo& 0m4-09 & zs&15.3&28.1&0.087& DDT2022B-007 \\
16/02/2023 & Cerro Tololo & 0m4-09 & zs&15.3&28.3&0.086& DDT2022B-007 \\
21/02/2023 & Cerro Tololo & 0m4-09 & zs&15.3&28.3&0.087& DDT2022B-007 \\
24/02/2023 & Cerro Tololo & 0m4-09 & V&20.3&33.1&0.028& DDT2022B-007 \\
26/02/2023 & Cerro Tololo & 0m4-09 & V&20.4&33.2&0.101& DDT2022B-007 \\
02/03/2023 &Cerro Tololo & 0m4-09 & V&15.4&29.3&0.103& DDT2022B-007 \\
03/03/2023 & Cerro Tololo & 0m4-09 & V& 15.3&30.9&0.104& DDT2022B-007\\
04/03/2023 & Cerro Tololo & 0m4-09 & V&15.3&30.9&0.103& DDT2022B-007\\
05/03/2023 & Cerro Tololo & 0m4-09 & V&15.3&30.9&0.103& DDT2022B-007\\
06/03/2023 & Cerro Tololo & 0m4-09 & V&15.3&32.3&0.104& DDT2022B-007\\
07/03/2023 & Cerro Tololo & 0m4-09 & V&15.3&30.7&0.104& DDT2022B-007\\
20/01/2023 &Cerro Tololo & 0m4-12 &rp&4.2&12.7&0.007& KEY2020B-007\\
21/01/2023 &Cerro Tololo & 0m4-12 &rp&4.2&13.2&0.007& KEY2020B-007\\
02/07/2023 &Cerro Tololo & 0m4-12 & V&20.2&28.5&3.733& DDT2022B-007\\
03/07/2023 &Cerro Tololo & 0m4-12 & V&20.2&28.5&3.48& DDT2022B-007\\
09/05/2025 & Cerro Tololo & 0m4-92 & V&2&13.7&6.773& DDT2025A-006\\
06/07/2025 & Cerro Tololo & 0m4-92 & V&2&4.6&1.943& DDT2025A-006\\
07/07/2025 & Cerro Tololo & 0m4-92 & V&2&4.4&1.965& DDT2025A-006\\     
    \hline
    \end{tabular}
\end{table*}

\clearpage

\begin{longtable}{lccc}
\caption{Full Set of Priors and Optimized Values for the Adopted Circular Model and the Eccentric model \label{tab:prior_posterior}}\\
\endfirsthead
\multicolumn{4}{c}{\tablename\ \thetable\ -- \textit{Continued from previous page}} \\
\hline
\endhead
\hline \multicolumn{4}{r}{\textit{Continued on next page}} \\
\endfoot
\hline
\endlastfoot
\hline\hline 
Parameter                & Prior &   Value (circular model) & Value (eccentric model)\\
\hline

\textbf{White noise} & & &  \\
ln $\sigma$ $F_{~\rm TESS}$ [ppt] & $\mathcal{U}$[-5,5] & $<$ --4.5 &  $<$ --4.7  \\
ln $\sigma$ $F_{~\rm NGTS}$ [ppt] & $\mathcal{U}$[-5,5] & $<$ --3.1 &  $<$ --2.9 \\
ln $\sigma$ $F_{~\rm CHEOPS}$ [ppt] & $\mathcal{U}$[-5,5] & --1.37$\pm$0.04 &  --1.38$\pm$0.05 \\
ln $\sigma$ $F_{~\rm ASTEP+}$ [ppt] & $\mathcal{U}$[-5,5] & 0.082$\pm$0.024 & 0.074$\pm$0.025 \\
ln $\sigma$ $F_{~\rm LCO}$ [ppt] & $\mathcal{U}$[-5,5] & --0.04$\pm$0.016 &  --0.055$\pm$0.011 \\
ln $\sigma$ $RV_{~\rm HARPS}$ [m$\cdot$ s$^{-1}$] & $\mathcal{U}$[-10,10] & $<$ 1.9 & $<$ 1.9 \\
ln $\sigma$ $RV_{~\rm FEROS}$ [m$\cdot$ s$^{-1}$] & $\mathcal{U}$[-10,10] & $<$ 2.8 &  $<$ 3.0 \\
\\
\textbf{Zero points} &  \\
V0 $F_{~\rm TESS}$ [ppt] & $\mathcal{N}$[0,5] & 0.03$\pm$0.03 & 0.03$\pm$0.03 \\
V0 $F_{~\rm NGTS}$ [ppt] & $\mathcal{N}$[0,5] & --0.07$\pm$0.017 & --0.03$^{+0.17}_{-0.18}$  \\
V0 $F_{~\rm CHEOPS}$ [ppt] & $\mathcal{N}$[0,5] & 0.081$\pm$0.019& 0.071$\pm$0.019  \\
V0 $F_{~\rm ASTEP+}$ [ppt] & $\mathcal{N}$[0,5] & 0.5$\pm$0.4 & 0.4$\pm$0.4 \\
V0 $F_{~\rm LCO}$ [ppt]       & $\mathcal{N}$[0,5] & --0.15$\pm$0.15 & 0.0$\pm$0.4 \\
V0 $RV_{~\rm HARPS}$ [m$\cdot$ s$^{-1}$] & $\mathcal{N}$[0,1000] &  21$^{+26}_{-24}$ &  25$^{+24}_{-27}$ \\
V0 $RV_{~\rm FEROS}$ [m$\cdot$ s$^{-1}$] & $\mathcal{N}$[0,1000] & 6$^{+24}_{-25}$  & 4$^{+23}_{-24}$ \\
\\
\textbf{Polynomial and correlations} &  \\
Lin $RV$ [m$\cdot$ s$^{-1}$ $\cdot$ d$^{-1}$] & $\mathcal{N}$[0,100] & 0.02$\pm$0.03 & 0.03$\pm$0.03  \\
Lin $RV-BIS_{~\rm HARPS}$ & $\mathcal{N}$[0,100] & --0.11$\pm$0.04 & --0.12$\pm$0.04\\
 Lin $RV-BIS_{~\rm FEROS}$ & $\mathcal{N}$[0,100] & --0.012$\pm$0.014 & --0.010$\pm$0.016  \\
 \\
\textbf{GP hyperparameters}  \\
ln $\sigma$ $F_{~\rm TESS}$ [ppt] & $\mathcal{U}$[-10,10] & --1.06$\pm$0.05 & --1.08$\pm$0.06 \\
ln $P$ $F_{~\rm TESS}$ [d] & $\mathcal{U}$[-10,10] & 0.25$\pm$0.05 & 0.23$\pm$0.06  \\
ln $Ts$ $F_{~\rm TESS}$ [d] & $\mathcal{U}$[-10,10] & --1.66$\pm$0.09 &  --1.69$\pm$0.11   \\
ln $\sigma$ $F_{~\rm NGTS}$ [ppt] & $\mathcal{U}$[-10,10] & 0.48$\pm$0.06 & 0.47$\pm$0.07 \\
ln $P$ $F_{~\rm NGTS}$ [d] & $\mathcal{U}$[-10,10] & 0.14$^{+0.19}_{-0.16}$ & 0.16$^{+0.15}_{-0.19}$ \\
ln $Ts$ $F_{~\rm NGTS}$ [d] & $\mathcal{U}$[-10,10] & --3.2$^{+0.4}_{-0.3}$ &  --3.0$^{+0.3}_{-0.4}$ \\
ln $\sigma$ $F_{~\rm CHEOPS}$ [ppt] & $\mathcal{U}$[-10,10] & --0.99$\pm$0.04  & --1.00$\pm$0.04 \\
ln $P$ $F_{~\rm CHEOPS}$ [d] & $\mathcal{U}$[-10,10] & --3.51$\pm$0.08 &  --3.48$^{+0.08}_{-0.09}$ \\
ln $Ts$ $F_{~\rm CHEOPS}$ [d] & $\mathcal{U}$[-10,10] & --4.76$^{+0.24}_{-0.23}$ &  --4.66$\pm$0.25  \\
ln $\sigma$ $F_{~\rm ASTEP+}$ [ppt] & $\mathcal{U}$[-10,10] & 0.56$^{+0.10}_{-0.11}$  & 0.60$^{+0.15}_{-0.11}$  \\
ln $P$ $F_{~\rm ASTEP+}$ [d] & $\mathcal{U}$[-10,10] & --2.7$^{+0.7}_{-0.5}$ & --3.3$^{+0.9}_{-0.8}$ \\
ln $Ts$ $F_{~\rm ASTEP+}$ [d] & $\mathcal{U}$[-10,10] & --6.5$^{+1.4}_{-1.0}$ & --7.8$^{+1.7}_{-1.5}$ \\
ln $\sigma$ $F_{~\rm LCO}$ [ppt] & $\mathcal{U}$[-10,10] & 1.21$^{+0.12}_{-0.13}$  & 1.04$\pm$0.18 \\
ln $P$ $F_{~\rm LCO}$ [d] & $\mathcal{U}$[-10,10] & {--7.06$^{+0.10}_{-0.22}$} & --1.9$^{+0.4}_{-0.8}$ \\
ln $Ts$ $F_{~\rm LCO}$ [d] & $\mathcal{U}$[-10,10] & --1.1$^{+0.5}_{-0.4}$ & --4.8$^{+3.6}_{-2.0}$ \\
ln $\sigma$ $RV$ [m$\cdot$ s$^{-1}$] & $\mathcal{U}$[-10,10] & 3.6$\pm$0.3 & 3.6$\pm$0.4   \\
ln $P$ $RV$ [d] & $\mathcal{U}$[-10,10] & --0.7$^{+2.0}_{-1.7}$ &  --0.4$^{+2.0}_{-1.9}$ \\
ln $Ts$ $RV$ [d] & $\mathcal{U}$[-10,10] & {--3$\pm$4} & 1$\pm$6 \\
\\
 \textbf{Stellar parameters} \\
 R$_{*}$ [R$_{\odot}$] & $\mathcal{N}$[1.482,0.021] & 1.473 $\pm$ 0.014 & 1.479 $\pm$ 0.018  \\
 M$_{*}$ [M$_{\odot}$] & $\mathcal{N}$[1.28,0.07] & 1.31 $\pm$ 0.04 & 1.31 $\pm$ 0.06 \\ 
 \\
\textbf{Limb darkening} \\
q1 & $\mathcal{N}$[0.240,0.014] & 0.242 $\pm$ 0.011 & 0.239 $\pm$ 0.013 \\
q2 & $\mathcal{N}$[0.265,0.019] & 0.263 $\pm$ 0.014 &  0.262 $\pm$ 0.017\\
\\
\textbf{Dilution parameters}\\
{D}$_{~\rm TESS}$ & $\mathcal{N}$[1.0, 0.3] & 1.08 $\pm$ 0.03 & 1.10$\pm$0.03\\
\\
\textbf{Planet b (circular)} \\
$T0$ [BJD - 2450000] & $\mathcal{N}$[10466.8,0.1] & 10466.7908 $\pm$0.0017 & 10466.7911$^{+0.0016}_{-0.0020}$  \\
ln $P$ [d] & $\mathcal{U}$[3,9] & 5.4185442$\pm$1.8$\times$10$^{-6}$ & 5.4185455$\pm$1.9$\times$10$^{-6}$ \\
ln $R_{\rm p}$ [$R_{\rm J}$] & $\mathcal{U}$[-5,1] & 0.045 $\pm$ 0.013 & 0.047$\pm$0.016  \\
ln $M_{\rm p}$ [$M_{\rm J}$] & $\mathcal{U}$[-5,3] & $<$ 0.5 & $<$ 0.28\\
b & $\mathcal{U}$[0,1.2] & 0.421$^{+0.022}_{-0.024}$ & 0.36$^{+0.08}_{-0.12}$  \\
$\sqrt{e}\cos{\omega}$ & $\mathcal{N}$[0,0.3] & 0 (fixed) & --0.04$^{+0.26}_{-0.25}$\\
$\sqrt{e}\sin{\omega}$ & $\mathcal{N}$[0,0.3] & 0 (fixed) & --0.11$^{+0.12}_{-0.15}$\\
  \\
\textbf{Planet c (circular)} \\
$T0$ [BJD - 2450000] & $\mathcal{N}$[10063.5,0.1] & 10063.5114$\pm$0.0009 & 10063.5111$\pm$0.0011 \\
ln $P$ [d] & $\mathcal{U}$[3,9] & 5.75$^{+0.035}_{-0.058}$ & 5.84$^{+0.48}_{-0.17}$ \\
ln $R_{\rm p}$ [$R_{\rm J}$] & $\mathcal{U}$[-5,1] & 0.307$^{+0.018}_{-0.019}$ & 0.316$\pm$0.022 \\
ln $M_{\rm p}$ [$M_{\rm J}$] & $\mathcal{U}$[-5,3] & $<$ 0.7 & $<$ 0.7 \\
b & $\mathcal{U}$[0,1.2] & 0.752$^{+0.009}_{-0.008}$ & 0.753$\pm$0.011  \\
$\sqrt{e}\cos{\omega}$ & $\mathcal{N}$[0,0.3] & 0 (fixed) & 0.04$^{+0.26}_{-0.29}$\\
$\sqrt{e}\sin{\omega}$ & $\mathcal{N}$[0,0.3] & 0 (fixed) & 0.11$^{+0.27}_{-0.20}$ \\
\\
ln $Z$ & & --23693.1 & --23698.4 \\
\hline

\multicolumn{4}{p{16.25cm}}{{\bf Note.} Uncertainties are 1$\sigma$. Limits show the 95\% confidence interval.} \\
\end{longtable}

\bibliography{references}{}

@misc{https://doi.org/10.18727/archive/33,
  doi = {10.18727/ARCHIVE/33},
  url = {https://doi.eso.org/10.18727/archive/33},
  author = {{European Southern Observatory (ESO)}},
  language = {en},
  title = {HARPS reduced data obtained by standard ESO pipeline processing},
  publisher = {European Southern Observatory (ESO)},
  year = {2014},
  copyright = {Data Access Policy for ESO Data held in the ESO Science Archive Facility}
}

@misc{https://doi.org/10.17909/t9-nmc8-f686,
  doi = {10.17909/T9-NMC8-F686},
  url = {http://archive.stsci.edu/doi/resolve/resolve.html?doi=10.17909/t9-nmc8-f686},
  author = {{TESS Team}},
  title = {TESS Light Curves - All Sectors},
  publisher = {STScI/MAST},
  year = {2021}
}

@misc{https://doi.org/10.26133/nea2,
  doi = {10.26133/NEA2},
  url = {https://catcopy.ipac.caltech.edu/dois/doi.php?id=10.26133/NEA2},
  author = {{NASA Exoplanet Archive}},
  title = {Composite Planet Data Table},
  publisher = {IPAC},
  year = {2019}
}

@ARTICLE{christiansen2025,
       author = {{Christiansen}, Jessie L. and {McElroy}, Douglas L. and {Harbut}, Marcy and {Ciardi}, David R. and {Crane}, Megan and {Good}, John and {Hardegree-Ullman}, Kevin K. and {Kesseli}, Aurora Y. and {Lund}, Michael B. and {Lynn}, Meca and {Muthiar}, Ananda and {Nilsson}, Ricky and {Oluyide}, Toba and {Papin}, Michael and {Rivera}, Amalia and {Swain}, Melanie and {Susemiehl}, Nicholas D. and {Tam}, Raymond and {van Eyken}, Julian and {Beichman}, Charles},
        title = "{The NASA Exoplanet Archive and Exoplanet Follow-up Observing Program: Data, Tools, and Usage}",
      journal = {\psj},
         year = 2025,
        month = aug,
       volume = {6},
       number = {8},
          eid = {186},
        pages = {186},
          doi = {10.3847/PSJ/ade3c2},
archivePrefix = {arXiv},
       eprint = {2506.03299},
 primaryClass = {astro-ph.EP},
       adsurl = {https://ui.adsabs.harvard.edu/abs/2025PSJ.....6..186C}
}

@ARTICLE{karalis2025,
       author = {{Karalis}, Amalia and {Lee}, Eve J. and {Thorngren}, Daniel P.},
        title = "{Separating Super-puffs versus Hot Jupiters among Young Puffy Planets}",
      journal = {\apj},
         year = 2025,
        month = jan,
       volume = {978},
       number = {1},
          eid = {46},
        pages = {46},
          doi = {10.3847/1538-4357/ad946c},
archivePrefix = {arXiv},
       eprint = {2408.16793},
 primaryClass = {astro-ph.EP},
       adsurl = {https://ui.adsabs.harvard.edu/abs/2025ApJ...978...46K}
}

@ARTICLE{tang2025,
       author = {{Tang}, Yao and {Fortney}, Jonathan J. and {Murray-Clay}, Ruth and {Broome}, Madelyn},
        title = "{Understanding the Origins of Super-Puff Planets: A New Mass-Loss Regime Coupled to Planetary Evolution}",
      journal = {arXiv e-prints},
         year = 2025,
        month = oct,
          eid = {arXiv:2510.02201},
        pages = {arXiv:2510.02201},
          doi = {10.48550/arXiv.2510.02201},
archivePrefix = {arXiv},
       eprint = {2510.02201},
 primaryClass = {astro-ph.EP},
       adsurl = {https://ui.adsabs.harvard.edu/abs/2025arXiv251002201T}
}

@ARTICLE{dohnanyi1969,
       author = {{Dohnanyi}, J.~S.},
        title = "{Collisional Model of Asteroids and Their Debris}",
      journal = {\jgr},
         year = 1969,
        month = may,
       volume = {74},
        pages = {2531-2554},
          doi = {10.1029/JB074i010p02531},
       adsurl = {https://ui.adsabs.harvard.edu/abs/1969JGR....74.2531D}
}

@INPROCEEDINGS{mccully2018,
       author = {{McCully}, Curtis and {Volgenau}, Nikolaus H. and {Harbeck}, Daniel-Rolf and {Lister}, Tim A. and {Saunders}, Eric S. and {Turner}, Monica L. and {Siiverd}, Robert J. and {Bowman}, Mark},
        title = "{Real-time processing of the imaging data from the network of Las Cumbres Observatory Telescopes using BANZAI}",
    booktitle = {Software and Cyberinfrastructure for Astronomy V},
         year = 2018,
       editor = {{Guzman}, Juan C. and {Ibsen}, Jorge},
       series = {Society of Photo-Optical Instrumentation Engineers (SPIE) Conference Series},
       volume = {10707},
        month = jul,
          eid = {107070K},
        pages = {107070K},
          doi = {10.1117/12.2314340},
archivePrefix = {arXiv},
       eprint = {1811.04163},
 primaryClass = {astro-ph.IM},
       adsurl = {https://ui.adsabs.harvard.edu/abs/2018SPIE10707E..0KM}
}

@ARTICLE{yoshikawa1989,
       author = {{Yoshikawa}, M.},
        title = "{A survey of the motions of asteroids in the commensurabilities with Jupiter}",
      journal = {\aap},
         year = 1989,
        month = apr,
       volume = {213},
       number = {1-2},
        pages = {436-458},
       adsurl = {https://ui.adsabs.harvard.edu/abs/1989A&A...213..436Y}
}

@INPROCEEDINGS{dransfield2022,
       author = {{Dransfield}, Georgina and {M{\'e}karnia}, Djamel and {Triaud}, Amaury H.~M.~J. and {Guillot}, Tristan and {Abe}, Lyu and {Garc{\'\i}a}, Lionel J. and {Timmermans}, Mathilde and {Crouzet}, Nicolas and {Schmider}, Fran{\c{c}}ois-Xavier and {Agabi}, Abdelkarim and {Suarez}, Olga and {Bendjoya}, Philippe and {Guenther}, Maximilian N. and {Lai}, Olivier and {Merin}, Bruno and {Stee}, Philippe},
        title = "{Observation scheduling and automatic data reduction for the Antarctic Telescope, ASTEP+}",
    booktitle = {Observatory Operations: Strategies, Processes, and Systems IX},
         year = 2022,
       editor = {{Adler}, David S. and {Seaman}, Robert L. and {Benn}, Chris R.},
       series = {Society of Photo-Optical Instrumentation Engineers (SPIE) Conference Series},
       volume = {12186},
        month = aug,
          eid = {121861F},
        pages = {121861F},
          doi = {10.1117/12.2629920},
archivePrefix = {arXiv},
       eprint = {2208.04501},
 primaryClass = {astro-ph.EP},
       adsurl = {https://ui.adsabs.harvard.edu/abs/2022SPIE12186E..1FD}
}

@ARTICLE{marino2026,
       author = {{Marino}, S. and {Matr{\`a}}, L. and {Hughes}, A.~M. and {Ehrhardt}, J. and {Kennedy}, G.~M. and {del Burgo}, C. and {Brennan}, A. and {Han}, Y. and {Jankovic}, M.~R. and {Lovell}, J.~B. and {Mac Manamon}, S. and {Milli}, J. and {Weber}, P. and {Zawadzki}, B. and {Bendahan-West}, R. and {Fehr}, A. and {Mansell}, E. and {Olofsson}, J. and {Pearce}, T.~D. and {Bayo}, A. and {Matthews}, B.~C. and {L{\"o}hne}, T. and {Wyatt}, M.~C. and {{\'A}brah{\'a}m}, P. and {Bonduelle}, M. and {Booth}, M. and {Cataldi}, G. and {Carpenter}, J.~M. and {Chiang}, E. and {Ertel}, S. and {Hales}, A.~S. and {Henning}, Th. and {K{\'o}sp{\'a}l}, {\'A}. and {Krivov}, A.~V. and {Luppe}, P. and {MacGregor}, M.~A. and {Marshall}, J.~P. and {Mo{\'o}r}, A. and {P{\'e}rez}, S. and {Sefilian}, A.~A. and {Sepulveda}, A.~G. and {Wilner}, D.~J.},
        title = "{The ALMA survey to Resolve exoKuiper belt Substructures (ARKS): I. Motivation, sample, data reduction, and results overview}",
      journal = {\aap},
         year = 2026,
        month = jan,
       volume = {705},
          eid = {A195},
        pages = {A195},
          doi = {10.1051/0004-6361/202556489},
archivePrefix = {arXiv},
       eprint = {2601.11708},
 primaryClass = {astro-ph.EP},
       adsurl = {https://ui.adsabs.harvard.edu/abs/2026A&A...705A.195M}
}

@ARTICLE{rebound,
       author = {{Rein}, H. and {Liu}, S. -F.},
        title = "{REBOUND: an open-source multi-purpose N-body code for collisional dynamics}",
      journal = {\aap},
         year = 2012,
        month = jan,
       volume = {537},
          eid = {A128},
        pages = {A128},
          doi = {10.1051/0004-6361/201118085},
archivePrefix = {arXiv},
       eprint = {1110.4876},
 primaryClass = {astro-ph.EP},
       adsurl = {https://ui.adsabs.harvard.edu/abs/2012A&A...537A.128R}
}

@ARTICLE{reboundias15,
       author = {{Rein}, Hanno and {Spiegel}, David S.},
        title = "{IAS15: a fast, adaptive, high-order integrator for gravitational dynamics, accurate to machine precision over a billion orbits}",
      journal = {\mnras},
         year = 2015,
        month = jan,
       volume = {446},
       number = {2},
        pages = {1424-1437},
          doi = {10.1093/mnras/stu2164},
archivePrefix = {arXiv},
       eprint = {1409.4779},
 primaryClass = {astro-ph.EP},
       adsurl = {https://ui.adsabs.harvard.edu/abs/2015MNRAS.446.1424R}
}

@ARTICLE{giacalone2021,
       author = {{Giacalone}, Steven and {Dressing}, Courtney D. and {Jensen}, Eric L.~N. and {Collins}, Karen A. and {Ricker}, George R. and {Vanderspek}, Roland and {Seager}, S. and {Winn}, Joshua N. and {Jenkins}, Jon M. and {Barclay}, Thomas and {Barkaoui}, Khalid and {Cadieux}, Charles and {Charbonneau}, David and {Collins}, Kevin I. and {Conti}, Dennis M. and {Doyon}, Ren{\'e} and {Evans}, Phil and {Ghachoui}, Mourad and {Gillon}, Micha{\"e}l and {Guerrero}, Natalia M. and {Hart}, Rhodes and {Jehin}, Emmanu{\"e}l and {Kielkopf}, John F. and {McLean}, Brian and {Murgas}, Felipe and {Palle}, Enric and {Parviainen}, Hannu and {Pozuelos}, Francisco J. and {Relles}, Howard M. and {Shporer}, Avi and {Socia}, Quentin and {Stockdale}, Chris and {Tan}, Thiam-Guan and {Torres}, Guillermo and {Twicken}, Joseph D. and {Waalkes}, William C. and {Waite}, Ian A.},
        title = "{Vetting of 384 TESS Objects of Interest with TRICERATOPS and Statistical Validation of 12 Planet Candidates}",
      journal = {\aj},
         year = 2021,
        month = jan,
       volume = {161},
       number = {1},
          eid = {24},
        pages = {24},
          doi = {10.3847/1538-3881/abc6af},
archivePrefix = {arXiv},
       eprint = {2002.00691},
primaryClass = {astro-ph.EP},
       adsurl = {https://ui.adsabs.harvard.edu/abs/2021AJ....161...24G}
}

@ARTICLE{broeg2005,
       author = {{Broeg}, Ch. and {Fern{\'a}ndez}, M. and {Neuh{\"a}user}, R.},
        title = "{A new algorithm for differential photometry: computing an optimum artificial comparison star}",
      journal = {Astronomische Nachrichten},
         year = 2005,
        month = feb,
       volume = {326},
       number = {2},
        pages = {134-142},
          doi = {10.1002/asna.200410350},
       adsurl = {https://ui.adsabs.harvard.edu/abs/2005AN....326..134B}
}

@ARTICLE{ishihara2010,
       author = {{Ishihara}, D. and {Onaka}, T. and {Kataza}, H. and {Salama}, A. and {Alfageme}, C. and {Cassatella}, A. and {Cox}, N. and {Garc{\'\i}a-Lario}, P. and {Stephenson}, C. and {Cohen}, M. and {Fujishiro}, N. and {Fujiwara}, H. and {Hasegawa}, S. and {Ita}, Y. and {Kim}, W. and {Matsuhara}, H. and {Murakami}, H. and {M{\"u}ller}, T.~G. and {Nakagawa}, T. and {Ohyama}, Y. and {Oyabu}, S. and {Pyo}, J. and {Sakon}, I. and {Shibai}, H. and {Takita}, S. and {Tanab{\'e}}, T. and {Uemizu}, K. and {Ueno}, M. and {Usui}, F. and {Wada}, T. and {Watarai}, H. and {Yamamura}, I. and {Yamauchi}, C.},
        title = "{The AKARI/IRC mid-infrared all-sky survey}",
      journal = {\aap},
         year = 2010,
        month = may,
       volume = {514},
          eid = {A1},
        pages = {A1},
          doi = {10.1051/0004-6361/200913811},
archivePrefix = {arXiv},
       eprint = {1003.0270},
 primaryClass = {astro-ph.IM},
       adsurl = {https://ui.adsabs.harvard.edu/abs/2010A&A...514A...1I}
}

@ARTICLE{marshall2023,
       author = {{Marshall}, Jonathan P. and {Milli}, J. and {Choquet}, E. and {del Burgo}, C. and {Kennedy}, G.~M. and {Kemper}, F. and {Wyatt}, M.~C. and {Kral}, Q. and {Soummer}, R.},
        title = "{Stirred but not shaken: a multiwavelength view of HD 16743's debris disc}",
      journal = {\mnras},
         year = 2023,
        month = jun,
       volume = {521},
       number = {4},
        pages = {5940-5951},
          doi = {10.1093/mnras/stad913},
archivePrefix = {arXiv},
       eprint = {2303.17128},
 primaryClass = {astro-ph.EP},
       adsurl = {https://ui.adsabs.harvard.edu/abs/2023MNRAS.521.5940M}
}

@ARTICLE{foremanmackey2013,
       author = {{Foreman-Mackey}, Daniel and {Hogg}, David W. and {Lang}, Dustin and {Goodman}, Jonathan},
        title = "{emcee: The MCMC Hammer}",
      journal = {\pasp},
         year = 2013,
        month = mar,
       volume = {125},
       number = {925},
        pages = {306},
          doi = {10.1086/670067},
archivePrefix = {arXiv},
       eprint = {1202.3665},
 primaryClass = {astro-ph.IM},
       adsurl = {https://ui.adsabs.harvard.edu/abs/2013PASP..125..306F}
}

@ARTICLE{draine2003,
       author = {{Draine}, B.~T.},
        title = "{Scattering by Interstellar Dust Grains. I. Optical and Ultraviolet}",
      journal = {\apj},
         year = 2003,
        month = dec,
       volume = {598},
       number = {2},
        pages = {1017-1025},
          doi = {10.1086/379118},
archivePrefix = {arXiv},
       eprint = {astro-ph/0304060},
 primaryClass = {astro-ph},
       adsurl = {https://ui.adsabs.harvard.edu/abs/2003ApJ...598.1017D}
}

@ARTICLE{paunzen2015,
       author = {{Paunzen}, E.},
        title = "{A new catalogue of Str{\"o}mgren-Crawford uvby{\ensuremath{\beta}} photometry}",
      journal = {\aap},
         year = 2015,
        month = aug,
       volume = {580},
          eid = {A23},
        pages = {A23},
          doi = {10.1051/0004-6361/201526413},
archivePrefix = {arXiv},
       eprint = {1506.04568},
 primaryClass = {astro-ph.SR},
       adsurl = {https://ui.adsabs.harvard.edu/abs/2015A&A...580A..23P}
}

@ARTICLE{zeng2019,
       author = {{Zeng}, Li and {Jacobsen}, Stein B. and {Sasselov}, Dimitar D. and {Petaev}, Michail I. and {Vanderburg}, Andrew and {Lopez-Morales}, Mercedes and {Perez-Mercader}, Juan and {Mattsson}, Thomas R. and {Li}, Gongjie and {Heising}, Matthew Z. and {Bonomo}, Aldo S. and {Damasso}, Mario and {Berger}, Travis A. and {Cao}, Hao and {Levi}, Amit and {Wordsworth}, Robin D.},
        title = "{Growth model interpretation of planet size distribution}",
      journal = {Proceedings of the National Academy of Science},
         year = 2019,
        month = may,
       volume = {116},
       number = {20},
        pages = {9723-9728},
          doi = {10.1073/pnas.1812905116},
archivePrefix = {arXiv},
       eprint = {1906.04253},
 primaryClass = {astro-ph.EP},
       adsurl = {https://ui.adsabs.harvard.edu/abs/2019PNAS..116.9723Z}
}

@ARTICLE{kopparapu2014,
       author = {{Kopparapu}, Ravi Kumar and {Ramirez}, Ramses M. and {SchottelKotte}, James and {Kasting}, James F. and {Domagal-Goldman}, Shawn and {Eymet}, Vincent},
        title = "{Habitable Zones around Main-sequence Stars: Dependence on Planetary Mass}",
      journal = {\apjl},
         year = 2014,
        month = jun,
       volume = {787},
       number = {2},
          eid = {L29},
        pages = {L29},
          doi = {10.1088/2041-8205/787/2/L29},
archivePrefix = {arXiv},
       eprint = {1404.5292},
 primaryClass = {astro-ph.EP},
       adsurl = {https://ui.adsabs.harvard.edu/abs/2014ApJ...787L..29K}
}

@ARTICLE{helled2023,
       author = {{Helled}, Ravit},
        title = "{The mass of gas giant planets: Is Saturn a failed gas giant?}",
      journal = {\aap},
         year = 2023,
        month = jul,
       volume = {675},
          eid = {L8},
        pages = {L8},
          doi = {10.1051/0004-6361/202346850},
archivePrefix = {arXiv},
       eprint = {2306.14740},
 primaryClass = {astro-ph.EP},
       adsurl = {https://ui.adsabs.harvard.edu/abs/2023A&A...675L...8H}
}

@ARTICLE{glanz2022,
       author = {{Glanz}, Hila and {Rozner}, Mor and {Perets}, Hagai B. and {Grishin}, Evgeni},
        title = "{Inflated Eccentric Migration of Evolving Gas Giants II - Numerical Methodology and Basic Concepts}",
      journal = {\apj},
         year = 2022,
        month = may,
       volume = {931},
       number = {1},
          eid = {11},
        pages = {11},
          doi = {10.3847/1538-4357/ac6807},
archivePrefix = {arXiv},
       eprint = {2111.12714},
 primaryClass = {astro-ph.EP},
       adsurl = {https://ui.adsabs.harvard.edu/abs/2022ApJ...931...11G}
}

@ARTICLE{benatti2021,
       author = {{Benatti}, S. and {Damasso}, M. and {Borsa}, F. and {Locci}, D. and {Pillitteri}, I. and {Desidera}, S. and {Maggio}, A. and {Micela}, G. and {Wolk}, S. and {Claudi}, R. and {Malavolta}, L. and {Modirrousta-Galian}, D.},
        title = "{Constraints on the mass and on the atmospheric composition and evolution of the low-density young planet DS Tucanae A b}",
      journal = {\aap},
         year = 2021,
        month = jun,
       volume = {650},
          eid = {A66},
        pages = {A66},
          doi = {10.1051/0004-6361/202140416},
archivePrefix = {arXiv},
       eprint = {2103.12922},
 primaryClass = {astro-ph.EP},
       adsurl = {https://ui.adsabs.harvard.edu/abs/2021A&A...650A..66B}
}

@ARTICLE{barragan2024,
       author = {{Barrag{\'a}n}, Oscar and {Yu}, Haochuan and {Freckelton}, Alix Violet and {Meech}, Annabella and {Cretignier}, Michael and {Mortier}, Annelies and {Aigrain}, Suzanne and {Klein}, Baptiste and {O'Sullivan}, Niamh K. and {Gillen}, Edward and {Nielsen}, Louise Dyregaard and {Mallorqu{\'\i}n}, Manuel and {Zicher}, Norbert},
        title = "{TOI-837 b is a young Saturn-sized exoplanet with a massive 70 M$_{{\ensuremath{\oplus}}}$ core}",
      journal = {\mnras},
         year = 2024,
        month = jul,
       volume = {531},
       number = {4},
        pages = {4275-4292},
          doi = {10.1093/mnras/stae1344},
archivePrefix = {arXiv},
       eprint = {2404.13750},
 primaryClass = {astro-ph.EP},
       adsurl = {https://ui.adsabs.harvard.edu/abs/2024MNRAS.531.4275B}
}

@ARTICLE{lebouteiller2011,
       author = {{Lebouteiller}, V. and {Barry}, D.~J. and {Spoon}, H.~W.~W. and {Bernard-Salas}, J. and {Sloan}, G.~C. and {Houck}, J.~R. and {Weedman}, D.~W.},
        title = "{CASSIS: The Cornell Atlas of Spitzer/Infrared Spectrograph Sources}",
      journal = {\apjs},
         year = 2011,
        month = sep,
       volume = {196},
       number = {1},
          eid = {8},
        pages = {8},
          doi = {10.1088/0067-0049/196/1/8},
archivePrefix = {arXiv},
       eprint = {1108.3507},
 primaryClass = {astro-ph.IM},
       adsurl = {https://ui.adsabs.harvard.edu/abs/2011ApJS..196....8L}
}

@ARTICLE{marton2024,
       author = {{Marton}, G. and {Gezer}, I. and {Madar{\'a}sz}, M. and {Dionatos}, O. and {Audard}, M. and {Roquette}, J. and {Hernandez}, D. and {Paladini}, R. and {Altieri}, B.},
        title = "{The new Herschel/PACS Point Source Catalogue}",
      journal = {\aap},
         year = 2024,
        month = aug,
       volume = {688},
          eid = {A203},
        pages = {A203},
          doi = {10.1051/0004-6361/202450032},
archivePrefix = {arXiv},
       eprint = {2406.03116},
 primaryClass = {astro-ph.IM},
       adsurl = {https://ui.adsabs.harvard.edu/abs/2024A&A...688A.203M}
}

@INCOLLECTION{petit2025,
       author = {{Petit}, Antoine C. and {Pichierri}, Gabriele and {Goldberg}, Max and {Morbidelli}, Alessandro},
        title = "{Dynamical Evolution of Planetary Systems}",
    booktitle = {Handbook of Exoplanets},
         year = 2025,
          eid = {145},
        pages = {145},
          doi = {10.1007/978-3-319-55333-7_145},
       adsurl = {https://ui.adsabs.harvard.edu/abs/2025haex.bookE.145P}
}

@ARTICLE{beyer2024,
       author = {{Beyer}, Alexa C. and {White}, Russel J.},
        title = "{The Kraft Break Sharply Divides Low-mass and Intermediate-mass Stars}",
      journal = {\apj},
         year = 2024,
        month = sep,
       volume = {973},
       number = {1},
          eid = {28},
        pages = {28},
          doi = {10.3847/1538-4357/ad6b0d},
archivePrefix = {arXiv},
       eprint = {2408.02638},
 primaryClass = {astro-ph.SR},
       adsurl = {https://ui.adsabs.harvard.edu/abs/2024ApJ...973...28B}
}

@ARTICLE{kraft1967,
       author = {{Kraft}, Robert P.},
        title = "{Studies of Stellar Rotation. V. The Dependence of Rotation on Age among Solar-Type Stars}",
      journal = {\apj},
         year = 1967,
        month = nov,
       volume = {150},
        pages = {551},
          doi = {10.1086/149359},
       adsurl = {https://ui.adsabs.harvard.edu/abs/1967ApJ...150..551K}
}

@ARTICLE{gautam2025,
       author = {{Gautam}, Anuj and {Murphy}, Simon J. and {Bedding}, Timothy R.},
        title = "{Modelling delta Scuti pulsations: A new grid of p, g, and f modes across pre-main-sequence to post-main-sequence evolution}",
      journal = {arXiv e-prints},
         year = 2025,
        month = jul,
          eid = {arXiv:2507.03561},
        pages = {arXiv:2507.03561},
          doi = {10.48550/arXiv.2507.03561},
archivePrefix = {arXiv},
       eprint = {2507.03561},
 primaryClass = {astro-ph.SR},
       adsurl = {https://ui.adsabs.harvard.edu/abs/2025arXiv250703561G}
}

@ARTICLE{hog2000,
       author = {{H{\o}g}, E. and {Fabricius}, C. and {Makarov}, V.~V. and {Urban}, S. and {Corbin}, T. and {Wycoff}, G. and {Bastian}, U. and {Schwekendiek}, P. and {Wicenec}, A.},
        title = "{The Tycho-2 catalogue of the 2.5 million brightest stars}",
      journal = {\aap},
         year = 2000,
        month = mar,
       volume = {355},
        pages = {L27-L30},
       adsurl = {https://ui.adsabs.harvard.edu/abs/2000A&A...355L..27H}
}

@ARTICLE{wright2010,
       author = {{Wright}, Edward L. and {Eisenhardt}, Peter R.~M. and {Mainzer}, Amy K. and {Ressler}, Michael E. and {Cutri}, Roc M. and {Jarrett}, Thomas and {Kirkpatrick}, J. Davy and {Padgett}, Deborah and {McMillan}, Robert S. and {Skrutskie}, Michael and {Stanford}, S.~A. and {Cohen}, Martin and {Walker}, Russell G. and {Mather}, John C. and {Leisawitz}, David and {Gautier}, III, Thomas N. and {McLean}, Ian and {Benford}, Dominic and {Lonsdale}, Carol J. and {Blain}, Andrew and {Mendez}, Bryan and {Irace}, William R. and {Duval}, Valerie and {Liu}, Fengchuan and {Royer}, Don and {Heinrichsen}, Ingolf and {Howard}, Joan and {Shannon}, Mark and {Kendall}, Martha and {Walsh}, Amy L. and {Larsen}, Mark and {Cardon}, Joel G. and {Schick}, Scott and {Schwalm}, Mark and {Abid}, Mohamed and {Fabinsky}, Beth and {Naes}, Larry and {Tsai}, Chao-Wei},
        title = "{The Wide-field Infrared Survey Explorer (WISE): Mission Description and Initial On-orbit Performance}",
      journal = {\aj},
         year = 2010,
        month = dec,
       volume = {140},
       number = {6},
        pages = {1868-1881},
          doi = {10.1088/0004-6256/140/6/1868},
archivePrefix = {arXiv},
       eprint = {1008.0031},
 primaryClass = {astro-ph.IM},
       adsurl = {https://ui.adsabs.harvard.edu/abs/2010AJ....140.1868W}
}

@ARTICLE{skrutskie2006,
       author = {{Skrutskie}, M.~F. and {Cutri}, R.~M. and {Stiening}, R. and {Weinberg}, M.~D. and {Schneider}, S. and {Carpenter}, J.~M. and {Beichman}, C. and {Capps}, R. and {Chester}, T. and {Elias}, J. and {Huchra}, J. and {Liebert}, J. and {Lonsdale}, C. and {Monet}, D.~G. and {Price}, S. and {Seitzer}, P. and {Jarrett}, T. and {Kirkpatrick}, J.~D. and {Gizis}, J.~E. and {Howard}, E. and {Evans}, T. and {Fowler}, J. and {Fullmer}, L. and {Hurt}, R. and {Light}, R. and {Kopan}, E.~L. and {Marsh}, K.~A. and {McCallon}, H.~L. and {Tam}, R. and {Van Dyk}, S. and {Wheelock}, S.},
        title = "{The Two Micron All Sky Survey (2MASS)}",
      journal = {\aj},
         year = 2006,
        month = feb,
       volume = {131},
       number = {2},
        pages = {1163-1183},
          doi = {10.1086/498708},
       adsurl = {https://ui.adsabs.harvard.edu/abs/2006AJ....131.1163S}
}

@ARTICLE{liemansifry2016,
       author = {{Lieman-Sifry}, Jesse and {Hughes}, A. Meredith and {Carpenter}, John M. and {Gorti}, Uma and {Hales}, Antonio and {Flaherty}, Kevin M.},
        title = "{Debris Disks in the Scorpius-Centaurus OB Association Resolved by ALMA}",
      journal = {\apj},
         year = 2016,
        month = sep,
       volume = {828},
       number = {1},
          eid = {25},
        pages = {25},
          doi = {10.3847/0004-637X/828/1/25},
archivePrefix = {arXiv},
       eprint = {1606.07068},
 primaryClass = {astro-ph.EP},
       adsurl = {https://ui.adsabs.harvard.edu/abs/2016ApJ...828...25L}
}

@ARTICLE{matra2025,
       author = {{Matr{\`a}}, L. and {Marino}, S. and {Wilner}, D.~J. and {Kennedy}, G.~M. and {Booth}, M. and {Krivov}, A.~V. and {Williams}, J.~P. and {Hughes}, A.~M. and {del Burgo}, C. and {Carpenter}, J. and {Davies}, C.~L. and {Ertel}, S. and {Kral}, Q. and {Lestrade}, J.-F. and {Marshall}, J.~P. and {Milli}, J. and {{\"O}berg}, K.~I. and {Pawellek}, N. and {Sepulveda}, A.~G. and {Wyatt}, M.~C. and {Matthews}, B.~C. and {MacGregor}, M.},
        title = "{REsolved ALMA and SMA Observations of Nearby Stars (REASONS): A population of 74 resolved planetesimal belts at millimetre wavelengths}",
      journal = {\aap},
         year = 2025,
        month = jan,
       volume = {693},
          eid = {A151},
        pages = {A151},
          doi = {10.1051/0004-6361/202451397},
archivePrefix = {arXiv},
       eprint = {2501.09058},
 primaryClass = {astro-ph.EP},
       adsurl = {https://ui.adsabs.harvard.edu/abs/2025A&A...693A.151M}
}

@ARTICLE{hauschildt1999,
       author = {{Hauschildt}, Peter H. and {Allard}, France and {Baron}, E.},
        title = "{The NextGen Model Atmosphere Grid for 3000<=T$_{eff}$<=10,000 K}",
      journal = {\apj},
         year = 1999,
        month = feb,
       volume = {512},
       number = {1},
        pages = {377-385},
          doi = {10.1086/306745},
archivePrefix = {arXiv},
       eprint = {astro-ph/9807286},
 primaryClass = {astro-ph},
       adsurl = {https://ui.adsabs.harvard.edu/abs/1999ApJ...512..377H}
}

@ARTICLE{horner2020,
       author = {{Horner}, J. and {Kane}, S.~R. and {Marshall}, J.~P. and {Dalba}, P.~A. and {Holt}, T.~R. and {Wood}, J. and {Maynard-Casely}, H.~E. and {Wittenmyer}, R. and {Lykawka}, P.~S. and {Hill}, M. and {Salmeron}, R. and {Bailey}, J. and {L{\"o}hne}, T. and {Agnew}, M. and {Carter}, B.~D. and {Tylor}, C.~C.~E.},
        title = "{Solar System Physics for Exoplanet Research}",
      journal = {\pasp},
         year = 2020,
        month = oct,
       volume = {132},
       number = {1016},
          eid = {102001},
        pages = {102001},
          doi = {10.1088/1538-3873/ab8eb9},
archivePrefix = {arXiv},
       eprint = {2004.13209},
 primaryClass = {astro-ph.EP},
       adsurl = {https://ui.adsabs.harvard.edu/abs/2020PASP..132j2001H}
}

@ARTICLE{caldwell2020,
       author = {{Caldwell}, Douglas A. and {Tenenbaum}, Peter and {Twicken}, Joseph D. and {Jenkins}, Jon M. and {Ting}, Eric and {Smith}, Jeffrey C. and {Hedges}, Christina and {Fausnaugh}, Michael M. and {Rose}, Mark and {Burke}, Christopher},
        title = "{TESS Science Processing Operations Center FFI Target List Products}",
      journal = {Research Notes of the American Astronomical Society},
         year = 2020,
        month = nov,
       volume = {4},
       number = {11},
          eid = {201},
        pages = {201},
          doi = {10.3847/2515-5172/abc9b3},
archivePrefix = {arXiv},
       eprint = {2011.05495},
 primaryClass = {astro-ph.EP},
       adsurl = {https://ui.adsabs.harvard.edu/abs/2020RNAAS...4..201C}
}

@ARTICLE{hoyer2020,
       author = {{Hoyer}, S. and {Guterman}, P. and {Demangeon}, O. and {Sousa}, S.~G. and {Deleuil}, M. and {Meunier}, J.~C. and {Benz}, W.},
        title = "{Expected performances of the Characterising Exoplanet Satellite (CHEOPS). III. Data reduction pipeline: architecture and simulated performances}",
      journal = {\aap},
         year = 2020,
        month = mar,
       volume = {635},
          eid = {A24},
        pages = {A24},
          doi = {10.1051/0004-6361/201936325},
archivePrefix = {arXiv},
       eprint = {1909.08363},
 primaryClass = {astro-ph.IM},
       adsurl = {https://ui.adsabs.harvard.edu/abs/2020A&A...635A..24H}
}

@article{Parviainen2015b,
   author        = {{Parviainen}, H. and {Aigrain}, S.},
   title         = {{LDTK: Limb Darkening Toolkit}},
   journal       = {\mnras},
   year          = 2015,
   month         = nov,
   volume        = {453},
   number        = {4},
   pages         = {3821-3826},
   doi           = {10.1093/mnras/stv1857},
   archiveprefix = {arXiv},
   eprint        = {1508.02634},
   primaryclass  = {astro-ph.EP},
   adsurl        =
{https://ui.adsabs.harvard.edu/abs/2015MNRAS.453.3821P}
}

@inproceedings{Jenkins2016,
  author    = {{Jenkins}, Jon M. and {Twicken}, Joseph D. and {McCauliff}, Sean and {Campbell}, Jennifer and {Sanderfer}, Dwight and {Lung}, David and {Mansouri-Samani}, Masoud and {Girouard}, Forrest and {Tenenbaum}, Peter and {Klaus}, Todd and {Smith}, Jeffrey C. and {Caldwell}, Douglas A. and {Chacon}, A.~D. and {Henze}, Christopher and {Heiges}, Cory and {Latham}, David W. and {Morgan}, Edward and {Swade}, Daryl and {Rinehart}, Stephen and {Vanderspek}, Roland},
  title     = {{The TESS science processing operations center}},
  booktitle = {Software and Cyberinfrastructure for Astronomy IV},
  year      = 2016,
  editor    = {{Chiozzi}, Gianluca and {Guzman}, Juan C.},
  series    = {Society of Photo-Optical Instrumentation Engineers (SPIE) Conference Series},
  volume    = {9913},
  month     = aug,
  eid       = {99133E},
  pages     = {99133E},
  doi       = {10.1117/12.2233418},
  adsurl    = {https://ui.adsabs.harvard.edu/abs/2016SPIE.9913E..3EJ}
}

@article{Smith2012,
  author        = {{Smith}, Jeffrey C. and {Stumpe}, Martin C. and {Van Cleve}, Jeffrey E. and {Jenkins}, Jon M. and {Barclay}, Thomas S. and {Fanelli}, Michael N. and {Girouard}, Forrest R. and {Kolodziejczak}, Jeffery J. and {McCauliff}, Sean D. and {Morris}, Robert L. and {Twicken}, Joseph D.},
  title         = {{Kepler Presearch Data Conditioning II - A Bayesian Approach to Systematic Error Correction}},
  journal       = {\pasp},
  year          = 2012,
  month         = sep,
  volume        = {124},
  number        = {919},
  pages         = {1000},
  doi           = {10.1086/667697},
  archiveprefix = {arXiv},
  eprint        = {1203.1383},
  primaryclass  = {astro-ph.IM},
  adsurl        = {https://ui.adsabs.harvard.edu/abs/2012PASP..124.1000S}
}

@article{Stumpe2012,
  author        = {{Stumpe}, Martin C. and {Smith}, Jeffrey C. and {Van Cleve}, Jeffrey E. and {Twicken}, Joseph D. and {Barclay}, Thomas S. and {Fanelli}, Michael N. and {Girouard}, Forrest R. and {Jenkins}, Jon M. and {Kolodziejczak}, Jeffery J. and {McCauliff}, Sean D. and {Morris}, Robert L.},
  title         = {{Kepler Presearch Data Conditioning I{\textemdash}Architecture and Algorithms for Error Correction in Kepler Light Curves}},
  journal       = {\pasp},
  year          = 2012,
  month         = sep,
  volume        = {124},
  number        = {919},
  pages         = {985},
  doi           = {10.1086/667698},
  archiveprefix = {arXiv},
  eprint        = {1203.1382},
  primaryclass  = {astro-ph.IM},
  adsurl        = {https://ui.adsabs.harvard.edu/abs/2012PASP..124..985S}
}

@article{Stumpe2014,
  author  = {{Stumpe}, M.~C. and {Smith}, J.~C. and {Catanzarite}, J.~H. and 
             {Van Cleve}, J.~E. and {Jenkins}, J.~M. and {Twicken}, J.~D. and 
             {Girouard}, F.~R.},
  title   = {{Multiscale Systematic Error Correction via Wavelet-Based Bandsplitting in Kepler Data}},
  journal = {\pasp},
  year    = 2014,
  month   = jan,
  volume  = 126,
  pages   = {100},
  doi     = {10.1086/674989},
  adsurl  = {http://adsabs.harvard.edu/abs/2014PASP..126..100S}
}

@article{Fellgett1955,
  author  = {{Fellgett}, Peter},
  title   = {{A Proposal for a Radial Velocity Photometer}},
  journal = {Optica Acta},
  year    = 1955,
  month   = apr,
  volume  = {2},
  number  = {1},
  pages   = {9-16},
  doi     = {10.1080/713820996},
  adsurl  = {https://ui.adsabs.harvard.edu/abs/1955AcOpt...2....9F}
}

@article{Baranne1996,
  author   = {{Baranne}, A. and {Queloz}, D. and {Mayor}, M. and {Adrianzyk}, G. and 
              {Knispel}, G. and {Kohler}, D. and {Lacroix}, D. and {Meunier}, J.-P. and 
              {Rimbaud}, G. and {Vin}, A.},
  title    = {{ELODIE: A spectrograph for accurate radial velocity measurements.}},
  journal  = {\aaps},
  year     = 1996,
  month    = oct,
  volume   = 119,
  pages    = {373-390},
  adsurl   = {http://adsabs.harvard.edu/abs/1996A%26AS..119..373B}
}

@INPROCEEDINGS{Pepe2000,
       author = {{Pepe}, Francesco and {Mayor}, Michel and {Delabre}, Bernard and {Kohler}, Dominique and {Lacroix}, Daniel and {Queloz}, Didier and {Udry}, Stephane and {Benz}, Willy and {Bertaux}, Jean-Loup and {Sivan}, Jean-Pierre},
        title = "{HARPS: a new high-resolution spectrograph for the search of extrasolar planets}",
    booktitle = {Optical and IR Telescope Instrumentation and Detectors},
         year = 2000,
       editor = {{Iye}, Masanori and {Moorwood}, Alan F.},
       series = {Society of Photo-Optical Instrumentation Engineers (SPIE) Conference Series},
       volume = {4008},
        month = aug,
        pages = {582-592},
          doi = {10.1117/12.395516},
       adsurl = {https://ui.adsabs.harvard.edu/abs/2000SPIE.4008..582P}
}

@book{Rasmussen2006,
  author    = {{Rasmussen}, C.~E. and {Williams}, C.~K.~I.},
  title     = {{Gaussian Processes for Machine Learning}},
  booktitle = {Gaussian Processes for Machine Learning, by C.E.~Rasmussen and C.K.I.~Williams.~ISBN-13 978-0-262-18253-9},
  year      = 2006,
  adsurl    = {http://adsabs.harvard.edu/abs/2006gpml.book.....R}
}

@article{Delisle2022,
  author        = {{Delisle}, J. -B. and {Unger}, N. and {Hara}, N.~C. and {S{\'e}gransan}, D.},
  title         = {{Efficient modeling of correlated noise. III. Scalable methods for jointly modeling several observables' time series with Gaussian processes}},
  journal       = {\aap},
  year          = 2022,
  month         = mar,
  volume        = {659},
  eid           = {A182},
  pages         = {A182},
  doi           = {10.1051/0004-6361/202141949},
  archiveprefix = {arXiv},
  eprint        = {2201.02440},
  primaryclass  = {astro-ph.EP},
  adsurl        = {https://ui.adsabs.harvard.edu/abs/2022A&A...659A.182D}
}

@article{Parviainen2015,
  author        = {{Parviainen}, Hannu},
  title         = {{PYTRANSIT: fast and easy exoplanet transit modelling in PYTHON}},
  journal       = {\mnras},
  year          = 2015,
  month         = jul,
  volume        = {450},
  number        = {3},
  pages         = {3233-3238},
  doi           = {10.1093/mnras/stv894},
  archiveprefix = {arXiv},
  eprint        = {1504.07433},
  primaryclass  = {astro-ph.EP},
  adsurl        = {https://ui.adsabs.harvard.edu/abs/2015MNRAS.450.3233P}
}

@article{MandelAgol2002,
  author        = {{Mandel}, Kaisey and {Agol}, Eric},
  title         = {{Analytic Light Curves for Planetary Transit Searches}},
  journal       = {\apjl},
  year          = 2002,
  month         = dec,
  volume        = {580},
  number        = {2},
  pages         = {L171-L175},
  doi           = {10.1086/345520},
  archiveprefix = {arXiv},
  eprint        = {astro-ph/0210099},
  primaryclass  = {astro-ph},
  adsurl        = {https://ui.adsabs.harvard.edu/abs/2002ApJ...580L.171M}
}

@inproceedings{Skilling2004,
  author    = {{Skilling}, John},
  title     = {{Nested Sampling}},
  booktitle = {Bayesian Inference and Maximum Entropy Methods in Science and Engineering: 24th International Workshop on Bayesian Inference and Maximum Entropy Methods in Science and Engineering},
  year      = 2004,
  editor    = {{Fischer}, Rainer and {Preuss}, Roland and {Toussaint}, Udo Von},
  series    = {American Institute of Physics Conference Series},
  volume    = {735},
  month     = nov,
  pages     = {395-405},
  doi       = {10.1063/1.1835238},
  adsurl    = {https://ui.adsabs.harvard.edu/abs/2004AIPC..735..395S}
}

@article{Skilling2006,
  author    = {John Skilling},
  title     = {{Nested sampling for general Bayesian computation}},
  volume    = {1},
  journal   = {Bayesian Analysis},
  number    = {4},
  publisher = {International Society for Bayesian Analysis},
  pages     = {833 -- 859},
  year      = {2006},
  doi       = {10.1214/06-BA127},
  url       = {https://doi.org/10.1214/06-BA127}
}

@article{Speagle2020,
  author        = {{Speagle}, Joshua S.},
  title         = {{DYNESTY: a dynamic nested sampling package for estimating Bayesian posteriors and evidences}},
  journal       = {\mnras},
  year          = 2020,
  month         = apr,
  volume        = {493},
  number        = {3},
  pages         = {3132-3158},
  doi           = {10.1093/mnras/staa278},
  archiveprefix = {arXiv},
  eprint        = {1904.02180},
  primaryclass  = {astro-ph.IM},
  adsurl        = {https://ui.adsabs.harvard.edu/abs/2020MNRAS.493.3132S}
}

@software{Koposov2023,
       author = {{Koposov}, Sergey and {Speagle}, Josh and {Barbary}, Kyle and {Ashton}, Gregory and {Bennett}, Ed and {Buchner}, Johannes and {Scheffler}, Carl and {Cook}, Ben and {Talbot}, Colm and {Guillochon}, James and {Cubillos}, Patricio and {Asensio Ramos}, Andr{\'e}s and {Johnson}, Ben and {Lang}, Dustin and {Ilya} and {Dartiailh}, Matthieu and {Nitz}, Alex and {McCluskey}, Andrew and {Archibald}, Anne and {Deil}, Christoph and {Foreman-Mackey}, Dan and {Goldstein}, Danny and {Tollerud}, Erik and {Leja}, Joel and {Kirk}, Matthew and {Pitkin}, Matt and {Sheehan}, Patrick and {Cargile}, Phillip and {Patel}, Ruskin and {Angus}, Ruth},
        title = "{joshspeagle/dynesty: v2.1.1}",
         year = 2023,
        month = apr,
          eid = {10.5281/zenodo.7832419},
          doi = {10.5281/zenodo.7832419},
      version = {v2.1.1},
    publisher = {Zenodo},
       adsurl = {https://ui.adsabs.harvard.edu/abs/2023zndo...7832419K}
}

@article{Foreman-Mackey2017,
  author        = {{Foreman-Mackey}, Daniel and {Agol}, Eric and {Ambikasaran}, Sivaram and
                   {Angus}, Ruth},
  title         = {{Fast and Scalable Gaussian Process Modeling with Applications to Astronomical Time Series}},
  journal       = {\aj},
  year          = {2017},
  month         = {Dec},
  volume        = {154},
  number        = {6},
  eid           = {220},
  pages         = {220},
  doi           = {10.3847/1538-3881/aa9332},
  archiveprefix = {arXiv},
  eprint        = {1703.09710},
  primaryclass  = {astro-ph.IM},
  adsurl        = {https://ui.adsabs.harvard.edu/abs/2017AJ....154..220F}
}

@ARTICLE{kaufer1999,
       author = {{Kaufer}, A. and {Stahl}, O. and {Tubbesing}, S. and {N{\o}rregaard}, P. and {Avila}, G. and {Francois}, P. and {Pasquini}, L. and {Pizzella}, A.},
        title = "{Commissioning FEROS, the new high-resolution spectrograph at La-Silla.}",
      journal = {The Messenger},
         year = 1999,
        month = mar,
       volume = {95},
        pages = {8-12},
       adsurl = {https://ui.adsabs.harvard.edu/abs/1999Msngr..95....8K}
}

@ARTICLE{mayor2003,
       author = {{Mayor}, M. and {Pepe}, F. and {Queloz}, D. and {Bouchy}, F. and {Rupprecht}, G. and {Lo Curto}, G. and {Avila}, G. and {Benz}, W. and {Bertaux}, J. -L. and {Bonfils}, X. and {Dall}, Th. and {Dekker}, H. and {Delabre}, B. and {Eckert}, W. and {Fleury}, M. and {Gilliotte}, A. and {Gojak}, D. and {Guzman}, J.~C. and {Kohler}, D. and {Lizon}, J. -L. and {Longinotti}, A. and {Lovis}, C. and {Megevand}, D. and {Pasquini}, L. and {Reyes}, J. and {Sivan}, J. -P. and {Sosnowska}, D. and {Soto}, R. and {Udry}, S. and {van Kesteren}, A. and {Weber}, L. and {Weilenmann}, U.},
        title = "{Setting New Standards with HARPS}",
      journal = {The Messenger},
         year = 2003,
        month = dec,
       volume = {114},
        pages = {20-24},
       adsurl = {https://ui.adsabs.harvard.edu/abs/2003Msngr.114...20M}
}

@ARTICLE{chenchristine2014,
       author = {{Chen}, Christine H. and {Mittal}, Tushar and {Kuchner}, Marc and {Forrest}, William J. and {Lisse}, Carey M. and {Manoj}, P. and {Sargent}, Benjamin A. and {Watson}, Dan M.},
        title = "{The Spitzer Infrared Spectrograph Debris Disk Catalog. I. Continuum Analysis of Unresolved Targets}",
      journal = {\apjs},
         year = 2014,
        month = apr,
       volume = {211},
       number = {2},
          eid = {25},
        pages = {25},
          doi = {10.1088/0067-0049/211/2/25},
       adsurl = {https://ui.adsabs.harvard.edu/abs/2014ApJS..211...25C}
}

@ARTICLE{mittal2015,
       author = {{Mittal}, Tushar and {Chen}, Christine H. and {Jang-Condell}, Hannah and {Manoj}, P. and {Sargent}, Benjamin A. and {Watson}, Dan M. and {Lisse}, Carey M.},
        title = "{The Spitzer Infrared Spectrograph Debris Disk Catalog. II. Silicate Feature Analysis of Unresolved Targets}",
      journal = {\apj},
         year = 2015,
        month = jan,
       volume = {798},
       number = {2},
          eid = {87},
        pages = {87},
          doi = {10.1088/0004-637X/798/2/87},
       adsurl = {https://ui.adsabs.harvard.edu/abs/2015ApJ...798...87M}
}

@ARTICLE{jangcondell2015,
       author = {{Jang-Condell}, Hannah and {Chen}, Christine H. and {Mittal}, Tushar and {Manoj}, P. and {Watson}, Dan and {Lisse}, Carey M. and {Nesvold}, Erika and {Kuchner}, Marc},
        title = "{Spitzer IRS Spectra of Debris Disks in the Scorpius-Centaurus OB Association}",
      journal = {\apj},
         year = 2015,
        month = aug,
       volume = {808},
       number = {2},
          eid = {167},
        pages = {167},
          doi = {10.1088/0004-637X/808/2/167},
archivePrefix = {arXiv},
       eprint = {1506.05428},
 primaryClass = {astro-ph.EP},
       adsurl = {https://ui.adsabs.harvard.edu/abs/2015ApJ...808..167J}
}

@ARTICLE{hasegawa2022,
       author = {{Hasegawa}, Yasuhiro and {Haworth}, Thomas J. and {Hoadley}, Keri and {Kim}, Jinyoung Serena and {Goto}, Hina and {Juzikenaite}, Aine and {Turner}, Neal J. and {Pascucci}, Ilaria and {Hamden}, Erika T.},
        title = "{Determining Dispersal Mechanisms of Protoplanetary Disks Using Accretion and Wind Mass Loss Rates}",
      journal = {\apjl},
         year = 2022,
        month = feb,
       volume = {926},
       number = {2},
          eid = {L23},
        pages = {L23},
          doi = {10.3847/2041-8213/ac50aa},
archivePrefix = {arXiv},
       eprint = {2112.02831},
 primaryClass = {astro-ph.EP},
       adsurl = {https://ui.adsabs.harvard.edu/abs/2022ApJ...926L..23H}
}

@ARTICLE{bergez-casalou2024,
       author = {{Bergez-Casalou}, C. and {Kral}, Q.},
        title = "{Observing planetary gaps in the gas of debris disks}",
      journal = {\aap},
         year = 2024,
        month = dec,
       volume = {692},
          eid = {A156},
        pages = {A156},
          doi = {10.1051/0004-6361/202452097},
archivePrefix = {arXiv},
       eprint = {2411.14241},
 primaryClass = {astro-ph.EP},
       adsurl = {https://ui.adsabs.harvard.edu/abs/2024A&A...692A.156B}
}

@ARTICLE{ikoma2025,
       author = {{Ikoma}, Masahiro and {Kobayashi}, Hiroshi},
        title = "{Formation of Giant Planets}",
      journal = {\araa},
         year = 2025,
        month = aug,
       volume = {63},
       number = {1},
        pages = {217-258},
          doi = {10.1146/annurev-astro-052722-094843},
archivePrefix = {arXiv},
       eprint = {2504.04090},
 primaryClass = {astro-ph.EP},
       adsurl = {https://ui.adsabs.harvard.edu/abs/2025ARA&A..63..217I}
}

@ARTICLE{armitage2024,
       author = {{Armitage}, Philip J.},
        title = "{Planet formation theory: an overview}",
      journal = {arXiv e-prints},
         year = 2024,
        month = dec,
          eid = {arXiv:2412.11064},
        pages = {arXiv:2412.11064},
          doi = {10.48550/arXiv.2412.11064},
archivePrefix = {arXiv},
       eprint = {2412.11064},
 primaryClass = {astro-ph.EP},
       adsurl = {https://ui.adsabs.harvard.edu/abs/2024arXiv241211064A}
}

@ARTICLE{fulton2021,
       author = {{Fulton}, Benjamin J. and {Rosenthal}, Lee J. and {Hirsch}, Lea A. and {Isaacson}, Howard and {Howard}, Andrew W. and {Dedrick}, Cayla M. and {Sherstyuk}, Ilya A. and {Blunt}, Sarah C. and {Petigura}, Erik A. and {Knutson}, Heather A. and {Behmard}, Aida and {Chontos}, Ashley and {Crepp}, Justin R. and {Crossfield}, Ian J.~M. and {Dalba}, Paul A. and {Fischer}, Debra A. and {Henry}, Gregory W. and {Kane}, Stephen R. and {Kosiarek}, Molly and {Marcy}, Geoffrey W. and {Rubenzahl}, Ryan A. and {Weiss}, Lauren M. and {Wright}, Jason T.},
        title = "{California Legacy Survey. II. Occurrence of Giant Planets beyond the Ice Line}",
      journal = {\apjs},
         year = 2021,
        month = jul,
       volume = {255},
       number = {1},
          eid = {14},
        pages = {14},
          doi = {10.3847/1538-4365/abfcc1},
archivePrefix = {arXiv},
       eprint = {2105.11584},
 primaryClass = {astro-ph.EP},
       adsurl = {https://ui.adsabs.harvard.edu/abs/2021ApJS..255...14F}
}

@INPROCEEDINGS{drazkowska2023,
       author = {{Dr{\k{a}}{\.z}kowska}, J. and {Bitsch}, B. and {Lambrechts}, M. and {Mulders}, G.~D. and {Harsono}, D. and {Vazan}, A. and {Liu}, B. and {Ormel}, C.~W. and {Kretke}, K. and {Morbidelli}, A.},
        title = "{Planet Formation Theory in the Era of ALMA and Kepler: from Pebbles to Exoplanets}",
    booktitle = {Protostars and Planets VII},
         year = 2023,
       editor = {{Inutsuka}, S. and {Aikawa}, Y. and {Muto}, T. and {Tomida}, K. and {Tamura}, M.},
       series = {Astronomical Society of the Pacific Conference Series},
       volume = {534},
        month = jul,
        pages = {717},
          doi = {10.48550/arXiv.2203.09759},
archivePrefix = {arXiv},
       eprint = {2203.09759},
 primaryClass = {astro-ph.EP},
       adsurl = {https://ui.adsabs.harvard.edu/abs/2023ASPC..534..717D}
}

@ARTICLE{youdin2025,
       author = {{Youdin}, Andrew N. and {Zhu}, Zhaohuan},
        title = "{Formation of Giant Planets}",
      journal = {arXiv e-prints},
         year = 2025,
        month = jan,
          eid = {arXiv:2501.13214},
        pages = {arXiv:2501.13214},
          doi = {10.48550/arXiv.2501.13214},
archivePrefix = {arXiv},
       eprint = {2501.13214},
 primaryClass = {astro-ph.EP},
       adsurl = {https://ui.adsabs.harvard.edu/abs/2025arXiv250113214Y}
}

@ARTICLE{bergezcasalou2023,
       author = {{Bergez-Casalou}, C. and {Bitsch}, B. and {Raymond}, S.~N.},
        title = "{Simultaneous gas accretion onto a pair of giant planets: Impact on their final mass and on the protoplanetary disk structure}",
      journal = {\aap},
         year = 2023,
        month = jan,
       volume = {669},
          eid = {A129},
        pages = {A129},
          doi = {10.1051/0004-6361/202244988},
archivePrefix = {arXiv},
       eprint = {2211.16239},
 primaryClass = {astro-ph.EP},
       adsurl = {https://ui.adsabs.harvard.edu/abs/2023A&A...669A.129B}
}

@ARTICLE{carrera2019,
       author = {{Carrera}, Daniel and {Raymond}, Sean N. and {Davies}, Melvyn B.},
        title = "{Planet-planet scattering as the source of the highest eccentricity exoplanets}",
      journal = {\aap},
         year = 2019,
        month = sep,
       volume = {629},
          eid = {L7},
        pages = {L7},
          doi = {10.1051/0004-6361/201935744},
archivePrefix = {arXiv},
       eprint = {1903.02564},
 primaryClass = {astro-ph.EP},
       adsurl = {https://ui.adsabs.harvard.edu/abs/2019A&A...629L...7C}
}

@ARTICLE{sandhaus2025,
       author = {{Sandhaus}, Phoebe and {Dawson}, Rebekah I. and {MacDonald}, Mariah and {Shakespeare}, Cody J. and {Morrison}, Sarah},
        title = "{Effects of Outer Giant Planets on In Situ Formation of Inner Super-Earths}",
      journal = {\apj},
         year = 2025,
        month = sep,
       volume = {990},
       number = {1},
          eid = {61},
        pages = {61},
          doi = {10.3847/1538-4357/adf1a6},
       adsurl = {https://ui.adsabs.harvard.edu/abs/2025ApJ...990...61S}
}

@ARTICLE{kong2024,
       author = {{Kong}, Zhihui and {Johansen}, Anders and {Lambrechts}, Michiel and {Jiang}, Jonathan H. and {Zhu}, Zong-Hong},
        title = "{How the presence of a giant planet affects the outcome of terrestrial planet formation simulations}",
      journal = {\aap},
         year = 2024,
        month = jul,
       volume = {687},
          eid = {A121},
        pages = {A121},
          doi = {10.1051/0004-6361/202349043},
archivePrefix = {arXiv},
       eprint = {2405.04228},
 primaryClass = {astro-ph.EP},
       adsurl = {https://ui.adsabs.harvard.edu/abs/2024A&A...687A.121K}
}

@ARTICLE{Maxted2022,
       author = {{Maxted}, P.~F.~L. and {Ehrenreich}, D. and {Wilson}, T.~G. and {Alibert}, Y. and {Cameron}, A. Collier and {Hoyer}, S. and {Sousa}, S.~G. and {Olofsson}, G. and {Bekkelien}, A. and {Deline}, A. and {Delrez}, L. and {Bonfanti}, A. and {Borsato}, L. and {Alonso}, R. and {Anglada Escud{\'e}}, G. and {Barrado}, D. and {Barros}, S.~C.~C. and {Baumjohann}, W. and {Beck}, M. and {Beck}, T. and {Benz}, W. and {Billot}, N. and {Biondi}, F. and {Bonfils}, X. and {Brandeker}, A. and {Broeg}, C. and {B{\'a}rczy}, T. and {Cabrera}, J. and {Charnoz}, S. and {Corral Van Damme}, C. and {Csizmadia}, Sz and {Davies}, M.~B. and {Deleuil}, M. and {Demangeon}, O.~D.~S. and {Demory}, B. -O. and {Erikson}, A. and {Flor{\'e}n}, H.~G. and {Fortier}, A. and {Fossati}, L. and {Fridlund}, M. and {Futyan}, D. and {Gandolfi}, D. and {Gillon}, M. and {Guedel}, M. and {Guterman}, P. and {Heng}, K. and {Isaak}, K.~G. and {Kiss}, L. and {Laskar}, J. and {Lecavelier des Etangs}, A. and {Lendl}, M. and {Lovis}, C. and {Magrin}, D. and {Nascimbeni}, V. and {Ottensamer}, R. and {Pagano}, I. and {Pall{\'e}}, E. and {Peter}, G. and {Piotto}, G. and {Pollacco}, D. and {Pozuelos}, F.~J. and {Queloz}, D. and {Ragazzoni}, R. and {Rando}, N. and {Rauer}, H. and {Reimers}, C. and {Ribas}, I. and {Salmon}, S. and {Santos}, N.~C. and {Scandariato}, G. and {Simon}, A.~E. and {Smith}, A.~M.~S. and {Steller}, M. and {Swayne}, M.~I. and {Szab{\'o}}, Gy M. and {S{\'e}gransan}, D. and {Thomas}, N. and {Udry}, S. and {Van Grootel}, V. and {Walton}, N.~A.},
        title = "{Analysis of Early Science observations with the CHaracterising ExOPlanets Satellite (CHEOPS) using PYCHEOPS}",
      journal = {\mnras},
         year = 2022,
        month = jul,
       volume = {514},
       number = {1},
        pages = {77-104},
          doi = {10.1093/mnras/stab3371},
archivePrefix = {arXiv},
       eprint = {2111.08828},
 primaryClass = {astro-ph.EP},
       adsurl = {https://ui.adsabs.harvard.edu/abs/2022MNRAS.514...77M}
}

@ARTICLE{Brown2013,
       author = {{Brown}, T.~M. and {Baliber}, N. and {Bianco}, F.~B. and {Bowman}, M. and {Burleson}, B. and {Conway}, P. and {Crellin}, M. and {Depagne}, {\'E}. and {De Vera}, J. and {Dilday}, B. and {Dragomir}, D. and {Dubberley}, M. and {Eastman}, J.~D. and {Elphick}, M. and {Falarski}, M. and {Foale}, S. and {Ford}, M. and {Fulton}, B.~J. and {Garza}, J. and {Gomez}, E.~L. and {Graham}, M. and {Greene}, R. and {Haldeman}, B. and {Hawkins}, E. and {Haworth}, B. and {Haynes}, R. and {Hidas}, M. and {Hjelstrom}, A.~E. and {Howell}, D.~A. and {Hygelund}, J. and {Lister}, T.~A. and {Lobdill}, R. and {Martinez}, J. and {Mullins}, D.~S. and {Norbury}, M. and {Parrent}, J. and {Paulson}, R. and {Petry}, D.~L. and {Pickles}, A. and {Posner}, V. and {Rosing}, W.~E. and {Ross}, R. and {Sand}, D.~J. and {Saunders}, E.~S. and {Shobbrook}, J. and {Shporer}, A. and {Street}, R.~A. and {Thomas}, D. and {Tsapras}, Y. and {Tufts}, J.~R. and {Valenti}, S. and {Vander Horst}, K. and {Walker}, Z. and {White}, G. and {Willis}, M.},
        title = "{Las Cumbres Observatory Global Telescope Network}",
      journal = {\pasp},
         year = 2013,
        month = sep,
       volume = {125},
       number = {931},
        pages = {1031},
          doi = {10.1086/673168},
archivePrefix = {arXiv},
       eprint = {1305.2437},
 primaryClass = {astro-ph.IM},
       adsurl = {https://ui.adsabs.harvard.edu/abs/2013PASP..125.1031B}
}

@ARTICLE{squicciarini2025,
       author = {{Squicciarini}, V. and {Mazoyer}, J. and {Lagrange}, A. -M. and {Chomez}, A. and {Delorme}, P. and {Flasseur}, O. and {Kiefer}, F. and {Bergeon}, S. and {Albert}, D. and {Meunier}, N.},
        title = "{The COBREX archival survey: Improved constraints on the occurrence rate of wide-orbit substellar companions: I. A uniform re-analysis of 400 stars from the GPIES survey}",
      journal = {\aap},
         year = 2025,
        month = jan,
       volume = {693},
          eid = {A54},
        pages = {A54},
          doi = {10.1051/0004-6361/202452310},
archivePrefix = {arXiv},
       eprint = {2411.06157},
 primaryClass = {astro-ph.EP},
       adsurl = {https://ui.adsabs.harvard.edu/abs/2025A&A...693A..54S}
}

@ARTICLE{tang2014,
       author = {{Tang}, Jing and {Bressan}, Alessandro and {Rosenfield}, Philip and {Slemer}, Alessandra and {Marigo}, Paola and {Girardi}, L{\'e}o and {Bianchi}, Luciana},
        title = "{New PARSEC evolutionary tracks of massive stars at low metallicity: testing canonical stellar evolution in nearby star-forming dwarf galaxies}",
      journal = {\mnras},
         year = 2014,
        month = dec,
       volume = {445},
       number = {4},
        pages = {4287-4305},
          doi = {10.1093/mnras/stu2029},
archivePrefix = {arXiv},
       eprint = {1410.1745},
 primaryClass = {astro-ph.SR},
       adsurl = {https://ui.adsabs.harvard.edu/abs/2014MNRAS.445.4287T}
}

@ARTICLE{chen2015,
       author = {{Chen}, Yang and {Bressan}, Alessandro and {Girardi}, L{\'e}o and {Marigo}, Paola and {Kong}, Xu and {Lanza}, Antonio},
        title = "{PARSEC evolutionary tracks of massive stars up to 350 M$_{{\ensuremath{\odot}}}$ at metallicities 0.0001 {\ensuremath{\leq}} Z {\ensuremath{\leq}} 0.04}",
      journal = {\mnras},
         year = 2015,
        month = sep,
       volume = {452},
       number = {1},
        pages = {1068-1080},
          doi = {10.1093/mnras/stv1281},
archivePrefix = {arXiv},
       eprint = {1506.01681},
 primaryClass = {astro-ph.SR},
       adsurl = {https://ui.adsabs.harvard.edu/abs/2015MNRAS.452.1068C}
}

@ARTICLE{chen2014,
       author = {{Chen}, Yang and {Girardi}, L{\'e}o and {Bressan}, Alessandro and {Marigo}, Paola and {Barbieri}, Mauro and {Kong}, Xu},
        title = "{Improving PARSEC models for very low mass stars}",
      journal = {\mnras},
         year = 2014,
        month = nov,
       volume = {444},
       number = {3},
        pages = {2525-2543},
          doi = {10.1093/mnras/stu1605},
archivePrefix = {arXiv},
       eprint = {1409.0322},
 primaryClass = {astro-ph.SR},
       adsurl = {https://ui.adsabs.harvard.edu/abs/2014MNRAS.444.2525C}
}

@ARTICLE{bressan2012,
       author = {{Bressan}, Alessandro and {Marigo}, Paola and {Girardi}, L{\'e}o. and {Salasnich}, Bernardo and {Dal Cero}, Claudia and {Rubele}, Stefano and {Nanni}, Ambra},
        title = "{PARSEC: stellar tracks and isochrones with the PAdova and TRieste Stellar Evolution Code}",
      journal = {\mnras},
         year = 2012,
        month = nov,
       volume = {427},
       number = {1},
        pages = {127-145},
          doi = {10.1111/j.1365-2966.2012.21948.x},
archivePrefix = {arXiv},
       eprint = {1208.4498},
 primaryClass = {astro-ph.SR},
       adsurl = {https://ui.adsabs.harvard.edu/abs/2012MNRAS.427..127B}
}

@ARTICLE{riello2021,
       author = {{Riello}, M. and {De Angeli}, F. and {Evans}, D.~W. and {Montegriffo}, P. and {Carrasco}, J.~M. and {Busso}, G. and {Palaversa}, L. and {Burgess}, P.~W. and {Diener}, C. and {Davidson}, M. and {Rowell}, N. and {Fabricius}, C. and {Jordi}, C. and {Bellazzini}, M. and {Pancino}, E. and {Harrison}, D.~L. and {Cacciari}, C. and {van Leeuwen}, F. and {Hambly}, N.~C. and {Hodgkin}, S.~T. and {Osborne}, P.~J. and {Altavilla}, G. and {Barstow}, M.~A. and {Brown}, A.~G.~A. and {Castellani}, M. and {Cowell}, S. and {De Luise}, F. and {Gilmore}, G. and {Giuffrida}, G. and {Hidalgo}, S. and {Holland}, G. and {Marinoni}, S. and {Pagani}, C. and {Piersimoni}, A.~M. and {Pulone}, L. and {Ragaini}, S. and {Rainer}, M. and {Richards}, P.~J. and {Sanna}, N. and {Walton}, N.~A. and {Weiler}, M. and {Yoldas}, A.},
        title = "{Gaia Early Data Release 3. Photometric content and validation}",
      journal = {\aap},
         year = 2021,
        month = may,
       volume = {649},
          eid = {A3},
        pages = {A3},
          doi = {10.1051/0004-6361/202039587},
archivePrefix = {arXiv},
       eprint = {2012.01916},
 primaryClass = {astro-ph.IM},
       adsurl = {https://ui.adsabs.harvard.edu/abs/2021A&A...649A...3R}
}

@ARTICLE{delBurgo2016,
       author = {{del Burgo}, C. and {Allende Prieto}, C.},
        title = "{Accurate parameters for HD 209458 and its planet from HST spectrophotometry}",
      journal = {\mnras},
         year = 2016,
        month = dec,
       volume = {463},
       number = {2},
        pages = {1400-1408},
          doi = {10.1093/mnras/stw2005},
archivePrefix = {arXiv},
       eprint = {1703.01449},
 primaryClass = {astro-ph.SR},
       adsurl = {https://ui.adsabs.harvard.edu/abs/2016MNRAS.463.1400D}
}

@ARTICLE{delBurgo2018,
       author = {{del Burgo}, C. and {Allende Prieto}, C.},
        title = "{Testing models of stellar structure and evolution - I. Comparison with detached eclipsing binaries}",
      journal = {\mnras},
         year = 2018,
        month = sep,
       volume = {479},
       number = {2},
        pages = {1953-1973},
          doi = {10.1093/mnras/sty1371},
       adsurl = {https://ui.adsabs.harvard.edu/abs/2018MNRAS.479.1953D}
}

@ARTICLE{avallone2022,
       author = {{Avallone}, Ellis A. and {Tayar}, Jamie N. and {van Saders}, Jennifer L. and {Berger}, Travis A. and {Claytor}, Zachary R. and {Beaton}, Rachael L. and {Teske}, Johanna and {Godoy-Rivera}, Diego and {Pan}, Kaike},
        title = "{Rotation Distributions around the Kraft Break with TESS and Kepler: The Influences of Age, Metallicity, and Binarity}",
      journal = {\apj},
         year = 2022,
        month = may,
       volume = {930},
       number = {1},
          eid = {7},
        pages = {7},
          doi = {10.3847/1538-4357/ac60a1},
archivePrefix = {arXiv},
       eprint = {2203.15116},
 primaryClass = {astro-ph.SR},
       adsurl = {https://ui.adsabs.harvard.edu/abs/2022ApJ...930....7A}
}

@ARTICLE{benz2021,
       author = {{Benz}, W. and {Broeg}, C. and {Fortier}, A. and {Rando}, N. and {Beck}, T. and {Beck}, M. and {Queloz}, D. and {Ehrenreich}, D. and {Maxted}, P.~F.~L. and {Isaak}, K.~G. and {Billot}, N. and {Alibert}, Y. and {Alonso}, R. and {Ant{\'o}nio}, C. and {Asquier}, J. and {Bandy}, T. and {B{\'a}rczy}, T. and {Barrado}, D. and {Barros}, S.~C.~C. and {Baumjohann}, W. and {Bekkelien}, A. and {Bergomi}, M. and {Biondi}, F. and {Bonfils}, X. and {Borsato}, L. and {Brandeker}, A. and {Busch}, M. -D. and {Cabrera}, J. and {Cessa}, V. and {Charnoz}, S. and {Chazelas}, B. and {Collier Cameron}, A. and {Corral Van Damme}, C. and {Cortes}, D. and {Davies}, M.~B. and {Deleuil}, M. and {Deline}, A. and {Delrez}, L. and {Demangeon}, O. and {Demory}, B.~O. and {Erikson}, A. and {Farinato}, J. and {Fossati}, L. and {Fridlund}, M. and {Futyan}, D. and {Gandolfi}, D. and {Garcia Munoz}, A. and {Gillon}, M. and {Guterman}, P. and {Gutierrez}, A. and {Hasiba}, J. and {Heng}, K. and {Hernandez}, E. and {Hoyer}, S. and {Kiss}, L.~L. and {Kovacs}, Z. and {Kuntzer}, T. and {Laskar}, J. and {Lecavelier des Etangs}, A. and {Lendl}, M. and {L{\'o}pez}, A. and {Lora}, I. and {Lovis}, C. and {L{\"u}ftinger}, T. and {Magrin}, D. and {Malvasio}, L. and {Marafatto}, L. and {Michaelis}, H. and {de Miguel}, D. and {Modrego}, D. and {Munari}, M. and {Nascimbeni}, V. and {Olofsson}, G. and {Ottacher}, H. and {Ottensamer}, R. and {Pagano}, I. and {Palacios}, R. and {Pall{\'e}}, E. and {Peter}, G. and {Piazza}, D. and {Piotto}, G. and {Pizarro}, A. and {Pollaco}, D. and {Ragazzoni}, R. and {Ratti}, F. and {Rauer}, H. and {Ribas}, I. and {Rieder}, M. and {Rohlfs}, R. and {Safa}, F. and {Salatti}, M. and {Santos}, N.~C. and {Scandariato}, G. and {S{\'e}gransan}, D. and {Simon}, A.~E. and {Smith}, A.~M.~S. and {Sordet}, M. and {Sousa}, S.~G. and {Steller}, M. and {Szab{\'o}}, G.~M. and {Szoke}, J. and {Thomas}, N. and {Tschentscher}, M. and {Udry}, S. and {Van Grootel}, V. and {Viotto}, V. and {Walter}, I. and {Walton}, N.~A. and {Wildi}, F. and {Wolter}, D.},
        title = "{The CHEOPS mission}",
      journal = {Experimental Astronomy},
         year = 2021,
        month = feb,
       volume = {51},
       number = {1},
        pages = {109-151},
          doi = {10.1007/s10686-020-09679-4},
archivePrefix = {arXiv},
       eprint = {2009.11633},
 primaryClass = {astro-ph.IM},
       adsurl = {https://ui.adsabs.harvard.edu/abs/2021ExA....51..109B}
}

@ARTICLE{best2024,
       author = {{Best}, Marcy and {Sefilian}, Antranik A. and {Petrovich}, Cristobal},
        title = "{The Influence of Cold Jupiters in the Formation of Close-in Planets. I. Planetesimal Transport}",
      journal = {\apj},
         year = 2024,
        month = jan,
       volume = {960},
       number = {1},
          eid = {89},
        pages = {89},
          doi = {10.3847/1538-4357/ad0965},
archivePrefix = {arXiv},
       eprint = {2304.02045},
 primaryClass = {astro-ph.EP},
       adsurl = {https://ui.adsabs.harvard.edu/abs/2024ApJ...960...89B}
}

@INPROCEEDINGS{crouzet2020,
       author = {{Crouzet}, Nicolas and {Agabi}, Abdelkrim and {Guillot}, Tristan and {Abe}, Lyu and {Schmider}, Fran{\c{c}}ois-Xavier and {M{\'e}karnia}, Djamel and {Triaud}, Amaury H.~M.~J. and {Bresson}, Yves and {Mauclert}, Nicolas and {Bailet}, Christophe and {Breeveld}, Dennis and {Blommaert}, Sander and {Shortt}, Brian and {Daban}, Jean-Baptiste and {Lagrange}, Anne-Marie and {Touz{\'e}}, Romain and {Dufour}, Justin and {Stee}, Valentin and {Caruana}, Jocelyn},
        title = "{Towards ASTEP+, a two-color photometric telescope at Dome C, Antarctica}",
    booktitle = {Ground-based and Airborne Instrumentation for Astronomy VIII},
         year = 2020,
       editor = {{Evans}, Christopher J. and {Bryant}, Julia J. and {Motohara}, Kentaro},
       series = {Society of Photo-Optical Instrumentation Engineers (SPIE) Conference Series},
       volume = {11447},
        month = dec,
          eid = {114470O},
        pages = {114470O},
          doi = {10.1117/12.2562550},
       adsurl = {https://ui.adsabs.harvard.edu/abs/2020SPIE11447E..0OC}
}

@ARTICLE{engler2023,
       author = {{Engler}, N. and {Milli}, J. and {Gratton}, R. and {Ulmer-Moll}, S. and {Vigan}, A. and {Lagrange}, A. -M. and {Kiefer}, F. and {Rubini}, P. and {Grandjean}, A. and {Schmid}, H.~M. and {Messina}, S. and {Squicciarini}, V. and {Olofsson}, J. and {Th{\'e}bault}, P. and {van Holstein}, R.~G. and {Janson}, M. and {M{\'e}nard}, F. and {Marshall}, J.~P. and {Chauvin}, G. and {Lendl}, M. and {Bhowmik}, T. and {Boccaletti}, A. and {Bonnefoy}, M. and {del Burgo}, C. and {Choquet}, E. and {Desidera}, S. and {Feldt}, M. and {Fusco}, T. and {Girard}, J. and {Gisler}, D. and {Hagelberg}, J. and {Langlois}, M. and {Maire}, A. -L. and {Mesa}, D. and {Meyer}, M.~R. and {Rabou}, P. and {Rodet}, L. and {Schmidt}, T. and {Zurlo}, A.},
        title = "{The high-albedo, low polarization disk around HD 114082 that harbors a Jupiter-sized transiting planet. Constraints from VLT/SPHERE completed with TESS, Gaia, and radial velocities}",
      journal = {\aap},
         year = 2023,
        month = apr,
       volume = {672},
          eid = {A1},
        pages = {A1},
          doi = {10.1051/0004-6361/202244380},
archivePrefix = {arXiv},
       eprint = {2211.11767},
 primaryClass = {astro-ph.EP},
       adsurl = {https://ui.adsabs.harvard.edu/abs/2023A&A...672A...1E}
}

@ARTICLE{gaia2023,
       author = {{Gaia Collaboration} and {Vallenari}, A. and {Brown}, A.~G.~A. and {Prusti}, T. and {de Bruijne}, J.~H.~J. and {Arenou}, F. and {Babusiaux}, C. and {Biermann}, M. and {Creevey}, O.~L. and {Ducourant}, C. and {Evans}, D.~W. and {Eyer}, L. and {Guerra}, R. and {Hutton}, A. and {Jordi}, C. and {Klioner}, S.~A. and {Lammers}, U.~L. and {Lindegren}, L. and {Luri}, X. and {Mignard}, F. and {Panem}, C. and {Pourbaix}, D. and {Randich}, S. and {Sartoretti}, P. and {Soubiran}, C. and {Tanga}, P. and {Walton}, N.~A. and {Bailer-Jones}, C.~A.~L. and {Bastian}, U. and {Drimmel}, R. and {Jansen}, F. and {Katz}, D. and {Lattanzi}, M.~G. and {van Leeuwen}, F. and {Bakker}, J. and {Cacciari}, C. and {Casta{\~n}eda}, J. and {De Angeli}, F. and {Fabricius}, C. and {Fouesneau}, M. and {Fr{\'e}mat}, Y. and {Galluccio}, L. and {Guerrier}, A. and {Heiter}, U. and {Masana}, E. and {Messineo}, R. and {Mowlavi}, N. and {Nicolas}, C. and {Nienartowicz}, K. and {Pailler}, F. and {Panuzzo}, P. and {Riclet}, F. and {Roux}, W. and {Seabroke}, G.~M. and {Sordo}, R. and {Th{\'e}venin}, F. and {Gracia-Abril}, G. and {Portell}, J. and {Teyssier}, D. and {Altmann}, M. and {Andrae}, R. and {Audard}, M. and {Bellas-Velidis}, I. and {Benson}, K. and {Berthier}, J. and {Blomme}, R. and {Burgess}, P.~W. and {Busonero}, D. and {Busso}, G. and {C{\'a}novas}, H. and {Carry}, B. and {Cellino}, A. and {Cheek}, N. and {Clementini}, G. and {Damerdji}, Y. and {Davidson}, M. and {de Teodoro}, P. and {Nu{\~n}ez Campos}, M. and {Delchambre}, L. and {Dell'Oro}, A. and {Esquej}, P. and {Fern{\'a}ndez-Hern{\'a}ndez}, J. and {Fraile}, E. and {Garabato}, D. and {Garc{\'\i}a-Lario}, P. and {Gosset}, E. and {Haigron}, R. and {Halbwachs}, J. -L. and {Hambly}, N.~C. and {Harrison}, D.~L. and {Hern{\'a}ndez}, J. and {Hestroffer}, D. and {Hodgkin}, S.~T. and {Holl}, B. and {Jan{\ss}en}, K. and {Jevardat de Fombelle}, G. and {Jordan}, S. and {Krone-Martins}, A. and {Lanzafame}, A.~C. and {L{\"o}ffler}, W. and {Marchal}, O. and {Marrese}, P.~M. and {Moitinho}, A. and {Muinonen}, K. and {Osborne}, P. and {Pancino}, E. and {Pauwels}, T. and {Recio-Blanco}, A. and {Reyl{\'e}}, C. and {Riello}, M. and {Rimoldini}, L. and {Roegiers}, T. and {Rybizki}, J. and {Sarro}, L.~M. and {Siopis}, C. and {Smith}, M. and {Sozzetti}, A. and {Utrilla}, E. and {van Leeuwen}, M. and {Abbas}, U. and {{\'A}brah{\'a}m}, P. and {Abreu Aramburu}, A. and {Aerts}, C. and {Aguado}, J.~J. and {Ajaj}, M. and {Aldea-Montero}, F. and {Altavilla}, G. and {{\'A}lvarez}, M.~A. and {Alves}, J. and {Anders}, F. and {Anderson}, R.~I. and {Anglada Varela}, E. and {Antoja}, T. and {Baines}, D. and {Baker}, S.~G. and {Balaguer-N{\'u}{\~n}ez}, L. and {Balbinot}, E. and {Balog}, Z. and {Barache}, C. and {Barbato}, D. and {Barros}, M. and {Barstow}, M.~A. and {Bartolom{\'e}}, S. and {Bassilana}, J. -L. and {Bauchet}, N. and {Becciani}, U. and {Bellazzini}, M. and {Berihuete}, A. and {Bernet}, M. and {Bertone}, S. and {Bianchi}, L. and {Binnenfeld}, A. and {Blanco-Cuaresma}, S. and {Blazere}, A. and {Boch}, T. and {Bombrun}, A. and {Bossini}, D. and {Bouquillon}, S. and {Bragaglia}, A. and {Bramante}, L. and {Breedt}, E. and {Bressan}, A. and {Brouillet}, N. and {Brugaletta}, E. and {Bucciarelli}, B. and {Burlacu}, A. and {Butkevich}, A.~G. and {Buzzi}, R. and {Caffau}, E. and {Cancelliere}, R. and {Cantat-Gaudin}, T. and {Carballo}, R. and {Carlucci}, T. and {Carnerero}, M.~I. and {Carrasco}, J.~M. and {Casamiquela}, L. and {Castellani}, M. and {Castro-Ginard}, A. and {Chaoul}, L. and {Charlot}, P. and {Chemin}, L. and {Chiaramida}, V. and {Chiavassa}, A. and {Chornay}, N. and {Comoretto}, G. and {Contursi}, G. and {Cooper}, W.~J. and {Cornez}, T. and {Cowell}, S. and {Crifo}, F. and {Cropper}, M. and {Crosta}, M. and {Crowley}, C. and {Dafonte}, C. and {Dapergolas}, A. and {David}, M. and {David}, P. and {de Laverny}, P. and {De Luise}, F. and {De March}, R. and {De Ridder}, J. and {de Souza}, R. and {de Torres}, A. and {del Peloso}, E.~F. and {del Pozo}, E. and {Delbo}, M. and {Delgado}, A. and {Delisle}, J. -B. and {Demouchy}, C. and {Dharmawardena}, T.~E. and {Di Matteo}, P. and {Diakite}, S. and {Diener}, C. and {Distefano}, E. and {Dolding}, C. and {Edvardsson}, B. and {Enke}, H. and {Fabre}, C. and {Fabrizio}, M. and {Faigler}, S. and {Fedorets}, G. and {Fernique}, P. and {Fienga}, A. and {Figueras}, F. and {Fournier}, Y. and {Fouron}, C. and {Fragkoudi}, F. and {Gai}, M. and {Garcia-Gutierrez}, A. and {Garcia-Reinaldos}, M. and {Garc{\'\i}a-Torres}, M. and {Garofalo}, A. and {Gavel}, A. and {Gavras}, P. and {Gerlach}, E. and {Geyer}, R. and {Giacobbe}, P. and {Gilmore}, G. and {Girona}, S. and {Giuffrida}, G. and {Gomel}, R. and {Gomez}, A. and {Gonz{\'a}lez-N{\'u}{\~n}ez}, J. and {Gonz{\'a}lez-Santamar{\'\i}a}, I. and {Gonz{\'a}lez-Vidal}, J.~J. and {Granvik}, M. and {Guillout}, P. and {Guiraud}, J. and {Guti{\'e}rrez-S{\'a}nchez}, R. and {Guy}, L.~P. and {Hatzidimitriou}, D. and {Hauser}, M. and {Haywood}, M. and {Helmer}, A. and {Helmi}, A. and {Sarmiento}, M.~H. and {Hidalgo}, S.~L. and {Hilger}, T. and {H{\l}adczuk}, N. and {Hobbs}, D. and {Holland}, G. and {Huckle}, H.~E. and {Jardine}, K. and {Jasniewicz}, G. and {Jean-Antoine Piccolo}, A. and {Jim{\'e}nez-Arranz}, {\'O}. and {Jorissen}, A. and {Juaristi Campillo}, J. and {Julbe}, F. and {Karbevska}, L. and {Kervella}, P. and {Khanna}, S. and {Kontizas}, M. and {Kordopatis}, G. and {Korn}, A.~J. and {K{\'o}sp{\'a}l}, {\'A}. and {Kostrzewa-Rutkowska}, Z. and {Kruszy{\'n}ska}, K. and {Kun}, M. and {Laizeau}, P. and {Lambert}, S. and {Lanza}, A.~F. and {Lasne}, Y. and {Le Campion}, J. -F. and {Lebreton}, Y. and {Lebzelter}, T. and {Leccia}, S. and {Leclerc}, N. and {Lecoeur-Taibi}, I. and {Liao}, S. and {Licata}, E.~L. and {Lindstr{\o}m}, H.~E.~P. and {Lister}, T.~A. and {Livanou}, E. and {Lobel}, A. and {Lorca}, A. and {Loup}, C. and {Madrero Pardo}, P. and {Magdaleno Romeo}, A. and {Managau}, S. and {Mann}, R.~G. and {Manteiga}, M. and {Marchant}, J.~M. and {Marconi}, M. and {Marcos}, J. and {Marcos Santos}, M.~M.~S. and {Mar{\'\i}n Pina}, D. and {Marinoni}, S. and {Marocco}, F. and {Marshall}, D.~J. and {Martin Polo}, L. and {Mart{\'\i}n-Fleitas}, J.~M. and {Marton}, G. and {Mary}, N. and {Masip}, A. and {Massari}, D. and {Mastrobuono-Battisti}, A. and {Mazeh}, T. and {McMillan}, P.~J. and {Messina}, S. and {Michalik}, D. and {Millar}, N.~R. and {Mints}, A. and {Molina}, D. and {Molinaro}, R. and {Moln{\'a}r}, L. and {Monari}, G. and {Mongui{\'o}}, M. and {Montegriffo}, P. and {Montero}, A. and {Mor}, R. and {Mora}, A. and {Morbidelli}, R. and {Morel}, T. and {Morris}, D. and {Muraveva}, T. and {Murphy}, C.~P. and {Musella}, I. and {Nagy}, Z. and {Noval}, L. and {Oca{\~n}a}, F. and {Ogden}, A. and {Ordenovic}, C. and {Osinde}, J.~O. and {Pagani}, C. and {Pagano}, I. and {Palaversa}, L. and {Palicio}, P.~A. and {Pallas-Quintela}, L. and {Panahi}, A. and {Payne-Wardenaar}, S. and {Pe{\~n}alosa Esteller}, X. and {Penttil{\"a}}, A. and {Pichon}, B. and {Piersimoni}, A.~M. and {Pineau}, F. -X. and {Plachy}, E. and {Plum}, G. and {Poggio}, E. and {Pr{\v{s}}a}, A. and {Pulone}, L. and {Racero}, E. and {Ragaini}, S. and {Rainer}, M. and {Raiteri}, C.~M. and {Rambaux}, N. and {Ramos}, P. and {Ramos-Lerate}, M. and {Re Fiorentin}, P. and {Regibo}, S. and {Richards}, P.~J. and {Rios Diaz}, C. and {Ripepi}, V. and {Riva}, A. and {Rix}, H. -W. and {Rixon}, G. and {Robichon}, N. and {Robin}, A.~C. and {Robin}, C. and {Roelens}, M. and {Rogues}, H.~R.~O. and {Rohrbasser}, L. and {Romero-G{\'o}mez}, M. and {Rowell}, N. and {Royer}, F. and {Ruz Mieres}, D. and {Rybicki}, K.~A. and {Sadowski}, G. and {S{\'a}ez N{\'u}{\~n}ez}, A. and {Sagrist{\`a} Sell{\'e}s}, A. and {Sahlmann}, J. and {Salguero}, E. and {Samaras}, N. and {Sanchez Gimenez}, V. and {Sanna}, N. and {Santove{\~n}a}, R. and {Sarasso}, M. and {Schultheis}, M. and {Sciacca}, E. and {Segol}, M. and {Segovia}, J.~C. and {S{\'e}gransan}, D. and {Semeux}, D. and {Shahaf}, S. and {Siddiqui}, H.~I. and {Siebert}, A. and {Siltala}, L. and {Silvelo}, A. and {Slezak}, E. and {Slezak}, I. and {Smart}, R.~L. and {Snaith}, O.~N. and {Solano}, E. and {Solitro}, F. and {Souami}, D. and {Souchay}, J. and {Spagna}, A. and {Spina}, L. and {Spoto}, F. and {Steele}, I.~A. and {Steidelm{\"u}ller}, H. and {Stephenson}, C.~A. and {S{\"u}veges}, M. and {Surdej}, J. and {Szabados}, L. and {Szegedi-Elek}, E. and {Taris}, F. and {Taylor}, M.~B. and {Teixeira}, R. and {Tolomei}, L. and {Tonello}, N. and {Torra}, F. and {Torra}, J. and {Torralba Elipe}, G. and {Trabucchi}, M. and {Tsounis}, A.~T. and {Turon}, C. and {Ulla}, A. and {Unger}, N. and {Vaillant}, M.~V. and {van Dillen}, E. and {van Reeven}, W. and {Vanel}, O. and {Vecchiato}, A. and {Viala}, Y. and {Vicente}, D. and {Voutsinas}, S. and {Weiler}, M. and {Wevers}, T. and {Wyrzykowski}, {\L}. and {Yoldas}, A. and {Yvard}, P. and {Zhao}, H. and {Zorec}, J. and {Zucker}, S. and {Zwitter}, T.},
        title = "{Gaia Data Release 3. Summary of the content and survey properties}",
      journal = {\aap},
         year = 2023,
        month = jun,
       volume = {674},
          eid = {A1},
        pages = {A1},
          doi = {10.1051/0004-6361/202243940},
archivePrefix = {arXiv},
       eprint = {2208.00211},
 primaryClass = {astro-ph.GA},
       adsurl = {https://ui.adsabs.harvard.edu/abs/2023A&A...674A...1G}
}

@ARTICLE{gaia2016,
       author = {{Gaia Collaboration} and {Prusti}, T. and {de Bruijne}, J.~H.~J. and {Brown}, A.~G.~A. and {Vallenari}, A. and {Babusiaux}, C. and {Bailer-Jones}, C.~A.~L. and {Bastian}, U. and {Biermann}, M. and {Evans}, D.~W. and {Eyer}, L. and {Jansen}, F. and {Jordi}, C. and {Klioner}, S.~A. and {Lammers}, U. and {Lindegren}, L. and {Luri}, X. and {Mignard}, F. and {Milligan}, D.~J. and {Panem}, C. and {Poinsignon}, V. and {Pourbaix}, D. and {Randich}, S. and {Sarri}, G. and {Sartoretti}, P. and {Siddiqui}, H.~I. and {Soubiran}, C. and {Valette}, V. and {van Leeuwen}, F. and {Walton}, N.~A. and {Aerts}, C. and {Arenou}, F. and {Cropper}, M. and {Drimmel}, R. and {H{\o}g}, E. and {Katz}, D. and {Lattanzi}, M.~G. and {O'Mullane}, W. and {Grebel}, E.~K. and {Holland}, A.~D. and {Huc}, C. and {Passot}, X. and {Bramante}, L. and {Cacciari}, C. and {Casta{\~n}eda}, J. and {Chaoul}, L. and {Cheek}, N. and {De Angeli}, F. and {Fabricius}, C. and {Guerra}, R. and {Hern{\'a}ndez}, J. and {Jean-Antoine-Piccolo}, A. and {Masana}, E. and {Messineo}, R. and {Mowlavi}, N. and {Nienartowicz}, K. and {Ord{\'o}{\~n}ez-Blanco}, D. and {Panuzzo}, P. and {Portell}, J. and {Richards}, P.~J. and {Riello}, M. and {Seabroke}, G.~M. and {Tanga}, P. and {Th{\'e}venin}, F. and {Torra}, J. and {Els}, S.~G. and {Gracia-Abril}, G. and {Comoretto}, G. and {Garcia-Reinaldos}, M. and {Lock}, T. and {Mercier}, E. and {Altmann}, M. and {Andrae}, R. and {Astraatmadja}, T.~L. and {Bellas-Velidis}, I. and {Benson}, K. and {Berthier}, J. and {Blomme}, R. and {Busso}, G. and {Carry}, B. and {Cellino}, A. and {Clementini}, G. and {Cowell}, S. and {Creevey}, O. and {Cuypers}, J. and {Davidson}, M. and {De Ridder}, J. and {de Torres}, A. and {Delchambre}, L. and {Dell'Oro}, A. and {Ducourant}, C. and {Fr{\'e}mat}, Y. and {Garc{\'\i}a-Torres}, M. and {Gosset}, E. and {Halbwachs}, J. -L. and {Hambly}, N.~C. and {Harrison}, D.~L. and {Hauser}, M. and {Hestroffer}, D. and {Hodgkin}, S.~T. and {Huckle}, H.~E. and {Hutton}, A. and {Jasniewicz}, G. and {Jordan}, S. and {Kontizas}, M. and {Korn}, A.~J. and {Lanzafame}, A.~C. and {Manteiga}, M. and {Moitinho}, A. and {Muinonen}, K. and {Osinde}, J. and {Pancino}, E. and {Pauwels}, T. and {Petit}, J. -M. and {Recio-Blanco}, A. and {Robin}, A.~C. and {Sarro}, L.~M. and {Siopis}, C. and {Smith}, M. and {Smith}, K.~W. and {Sozzetti}, A. and {Thuillot}, W. and {van Reeven}, W. and {Viala}, Y. and {Abbas}, U. and {Abreu Aramburu}, A. and {Accart}, S. and {Aguado}, J.~J. and {Allan}, P.~M. and {Allasia}, W. and {Altavilla}, G. and {{\'A}lvarez}, M.~A. and {Alves}, J. and {Anderson}, R.~I. and {Andrei}, A.~H. and {Anglada Varela}, E. and {Antiche}, E. and {Antoja}, T. and {Ant{\'o}n}, S. and {Arcay}, B. and {Atzei}, A. and {Ayache}, L. and {Bach}, N. and {Baker}, S.~G. and {Balaguer-N{\'u}{\~n}ez}, L. and {Barache}, C. and {Barata}, C. and {Barbier}, A. and {Barblan}, F. and {Baroni}, M. and {Barrado y Navascu{\'e}s}, D. and {Barros}, M. and {Barstow}, M.~A. and {Becciani}, U. and {Bellazzini}, M. and {Bellei}, G. and {Bello Garc{\'\i}a}, A. and {Belokurov}, V. and {Bendjoya}, P. and {Berihuete}, A. and {Bianchi}, L. and {Bienaym{\'e}}, O. and {Billebaud}, F. and {Blagorodnova}, N. and {Blanco-Cuaresma}, S. and {Boch}, T. and {Bombrun}, A. and {Borrachero}, R. and {Bouquillon}, S. and {Bourda}, G. and {Bouy}, H. and {Bragaglia}, A. and {Breddels}, M.~A. and {Brouillet}, N. and {Br{\"u}semeister}, T. and {Bucciarelli}, B. and {Budnik}, F. and {Burgess}, P. and {Burgon}, R. and {Burlacu}, A. and {Busonero}, D. and {Buzzi}, R. and {Caffau}, E. and {Cambras}, J. and {Campbell}, H. and {Cancelliere}, R. and {Cantat-Gaudin}, T. and {Carlucci}, T. and {Carrasco}, J.~M. and {Castellani}, M. and {Charlot}, P. and {Charnas}, J. and {Charvet}, P. and {Chassat}, F. and {Chiavassa}, A. and {Clotet}, M. and {Cocozza}, G. and {Collins}, R.~S. and {Collins}, P. and {Costigan}, G. and {Crifo}, F. and {Cross}, N.~J.~G. and {Crosta}, M. and {Crowley}, C. and {Dafonte}, C. and {Damerdji}, Y. and {Dapergolas}, A. and {David}, P. and {David}, M. and {De Cat}, P. and {de Felice}, F. and {de Laverny}, P. and {De Luise}, F. and {De March}, R. and {de Martino}, D. and {de Souza}, R. and {Debosscher}, J. and {del Pozo}, E. and {Delbo}, M. and {Delgado}, A. and {Delgado}, H.~E. and {di Marco}, F. and {Di Matteo}, P. and {Diakite}, S. and {Distefano}, E. and {Dolding}, C. and {Dos Anjos}, S. and {Drazinos}, P. and {Dur{\'a}n}, J. and {Dzigan}, Y. and {Ecale}, E. and {Edvardsson}, B. and {Enke}, H. and {Erdmann}, M. and {Escolar}, D. and {Espina}, M. and {Evans}, N.~W. and {Eynard Bontemps}, G. and {Fabre}, C. and {Fabrizio}, M. and {Faigler}, S. and {Falc{\~a}o}, A.~J. and {Farr{\`a}s Casas}, M. and {Faye}, F. and {Federici}, L. and {Fedorets}, G. and {Fern{\'a}ndez-Hern{\'a}ndez}, J. and {Fernique}, P. and {Fienga}, A. and {Figueras}, F. and {Filippi}, F. and {Findeisen}, K. and {Fonti}, A. and {Fouesneau}, M. and {Fraile}, E. and {Fraser}, M. and {Fuchs}, J. and {Furnell}, R. and {Gai}, M. and {Galleti}, S. and {Galluccio}, L. and {Garabato}, D. and {Garc{\'\i}a-Sedano}, F. and {Gar{\'e}}, P. and {Garofalo}, A. and {Garralda}, N. and {Gavras}, P. and {Gerssen}, J. and {Geyer}, R. and {Gilmore}, G. and {Girona}, S. and {Giuffrida}, G. and {Gomes}, M. and {Gonz{\'a}lez-Marcos}, A. and {Gonz{\'a}lez-N{\'u}{\~n}ez}, J. and {Gonz{\'a}lez-Vidal}, J.~J. and {Granvik}, M. and {Guerrier}, A. and {Guillout}, P. and {Guiraud}, J. and {G{\'u}rpide}, A. and {Guti{\'e}rrez-S{\'a}nchez}, R. and {Guy}, L.~P. and {Haigron}, R. and {Hatzidimitriou}, D. and {Haywood}, M. and {Heiter}, U. and {Helmi}, A. and {Hobbs}, D. and {Hofmann}, W. and {Holl}, B. and {Holland}, G. and {Hunt}, J.~A.~S. and {Hypki}, A. and {Icardi}, V. and {Irwin}, M. and {Jevardat de Fombelle}, G. and {Jofr{\'e}}, P. and {Jonker}, P.~G. and {Jorissen}, A. and {Julbe}, F. and {Karampelas}, A. and {Kochoska}, A. and {Kohley}, R. and {Kolenberg}, K. and {Kontizas}, E. and {Koposov}, S.~E. and {Kordopatis}, G. and {Koubsky}, P. and {Kowalczyk}, A. and {Krone-Martins}, A. and {Kudryashova}, M. and {Kull}, I. and {Bachchan}, R.~K. and {Lacoste-Seris}, F. and {Lanza}, A.~F. and {Lavigne}, J. -B. and {Le Poncin-Lafitte}, C. and {Lebreton}, Y. and {Lebzelter}, T. and {Leccia}, S. and {Leclerc}, N. and {Lecoeur-Taibi}, I. and {Lemaitre}, V. and {Lenhardt}, H. and {Leroux}, F. and {Liao}, S. and {Licata}, E. and {Lindstr{\o}m}, H.~E.~P. and {Lister}, T.~A. and {Livanou}, E. and {Lobel}, A. and {L{\"o}ffler}, W. and {L{\'o}pez}, M. and {Lopez-Lozano}, A. and {Lorenz}, D. and {Loureiro}, T. and {MacDonald}, I. and {Magalh{\~a}es Fernandes}, T. and {Managau}, S. and {Mann}, R.~G. and {Mantelet}, G. and {Marchal}, O. and {Marchant}, J.~M. and {Marconi}, M. and {Marie}, J. and {Marinoni}, S. and {Marrese}, P.~M. and {Marschalk{\'o}}, G. and {Marshall}, D.~J. and {Mart{\'\i}n-Fleitas}, J.~M. and {Martino}, M. and {Mary}, N. and {Matijevi{\v{c}}}, G. and {Mazeh}, T. and {McMillan}, P.~J. and {Messina}, S. and {Mestre}, A. and {Michalik}, D. and {Millar}, N.~R. and {Miranda}, B.~M.~H. and {Molina}, D. and {Molinaro}, R. and {Molinaro}, M. and {Moln{\'a}r}, L. and {Moniez}, M. and {Montegriffo}, P. and {Monteiro}, D. and {Mor}, R. and {Mora}, A. and {Morbidelli}, R. and {Morel}, T. and {Morgenthaler}, S. and {Morley}, T. and {Morris}, D. and {Mulone}, A.~F. and {Muraveva}, T. and {Musella}, I. and {Narbonne}, J. and {Nelemans}, G. and {Nicastro}, L. and {Noval}, L. and {Ord{\'e}novic}, C. and {Ordieres-Mer{\'e}}, J. and {Osborne}, P. and {Pagani}, C. and {Pagano}, I. and {Pailler}, F. and {Palacin}, H. and {Palaversa}, L. and {Parsons}, P. and {Paulsen}, T. and {Pecoraro}, M. and {Pedrosa}, R. and {Pentik{\"a}inen}, H. and {Pereira}, J. and {Pichon}, B. and {Piersimoni}, A.~M. and {Pineau}, F. -X. and {Plachy}, E. and {Plum}, G. and {Poujoulet}, E. and {Pr{\v{s}}a}, A. and {Pulone}, L. and {Ragaini}, S. and {Rago}, S. and {Rambaux}, N. and {Ramos-Lerate}, M. and {Ranalli}, P. and {Rauw}, G. and {Read}, A. and {Regibo}, S. and {Renk}, F. and {Reyl{\'e}}, C. and {Ribeiro}, R.~A. and {Rimoldini}, L. and {Ripepi}, V. and {Riva}, A. and {Rixon}, G. and {Roelens}, M. and {Romero-G{\'o}mez}, M. and {Rowell}, N. and {Royer}, F. and {Rudolph}, A. and {Ruiz-Dern}, L. and {Sadowski}, G. and {Sagrist{\`a} Sell{\'e}s}, T. and {Sahlmann}, J. and {Salgado}, J. and {Salguero}, E. and {Sarasso}, M. and {Savietto}, H. and {Schnorhk}, A. and {Schultheis}, M. and {Sciacca}, E. and {Segol}, M. and {Segovia}, J.~C. and {Segransan}, D. and {Serpell}, E. and {Shih}, I. -C. and {Smareglia}, R. and {Smart}, R.~L. and {Smith}, C. and {Solano}, E. and {Solitro}, F. and {Sordo}, R. and {Soria Nieto}, S. and {Souchay}, J. and {Spagna}, A. and {Spoto}, F. and {Stampa}, U. and {Steele}, I.~A. and {Steidelm{\"u}ller}, H. and {Stephenson}, C.~A. and {Stoev}, H. and {Suess}, F.~F. and {S{\"u}veges}, M. and {Surdej}, J. and {Szabados}, L. and {Szegedi-Elek}, E. and {Tapiador}, D. and {Taris}, F. and {Tauran}, G. and {Taylor}, M.~B. and {Teixeira}, R. and {Terrett}, D. and {Tingley}, B. and {Trager}, S.~C. and {Turon}, C. and {Ulla}, A. and {Utrilla}, E. and {Valentini}, G. and {van Elteren}, A. and {Van Hemelryck}, E. and {van Leeuwen}, M. and {Varadi}, M. and {Vecchiato}, A. and {Veljanoski}, J. and {Via}, T. and {Vicente}, D. and {Vogt}, S. and {Voss}, H. and {Votruba}, V. and {Voutsinas}, S. and {Walmsley}, G. and {Weiler}, M. and {Weingrill}, K. and {Werner}, D. and {Wevers}, T. and {Whitehead}, G. and {Wyrzykowski}, {\L}. and {Yoldas}, A. and {{\v{Z}}erjal}, M. and {Zucker}, S. and {Zurbach}, C. and {Zwitter}, T. and {Alecu}, A. and {Allen}, M. and {Allende Prieto}, C. and {Amorim}, A. and {Anglada-Escud{\'e}}, G. and {Arsenijevic}, V. and {Azaz}, S. and {Balm}, P. and {Beck}, M. and {Bernstein}, H. -H. and {Bigot}, L. and {Bijaoui}, A. and {Blasco}, C. and {Bonfigli}, M. and {Bono}, G. and {Boudreault}, S. and {Bressan}, A. and {Brown}, S. and {Brunet}, P. -M. and {Bunclark}, P. and {Buonanno}, R. and {Butkevich}, A.~G. and {Carret}, C. and {Carrion}, C. and {Chemin}, L. and {Ch{\'e}reau}, F. and {Corcione}, L. and {Darmigny}, E. and {de Boer}, K.~S. and {de Teodoro}, P. and {de Zeeuw}, P.~T. and {Delle Luche}, C. and {Domingues}, C.~D. and {Dubath}, P. and {Fodor}, F. and {Fr{\'e}zouls}, B. and {Fries}, A. and {Fustes}, D. and {Fyfe}, D. and {Gallardo}, E. and {Gallegos}, J. and {Gardiol}, D. and {Gebran}, M. and {Gomboc}, A. and {G{\'o}mez}, A. and {Grux}, E. and {Gueguen}, A. and {Heyrovsky}, A. and {Hoar}, J. and {Iannicola}, G. and {Isasi Parache}, Y. and {Janotto}, A. -M. and {Joliet}, E. and {Jonckheere}, A. and {Keil}, R. and {Kim}, D. -W. and {Klagyivik}, P. and {Klar}, J. and {Knude}, J. and {Kochukhov}, O. and {Kolka}, I. and {Kos}, J. and {Kutka}, A. and {Lainey}, V. and {LeBouquin}, D. and {Liu}, C. and {Loreggia}, D. and {Makarov}, V.~V. and {Marseille}, M.~G. and {Martayan}, C. and {Martinez-Rubi}, O. and {Massart}, B. and {Meynadier}, F. and {Mignot}, S. and {Munari}, U. and {Nguyen}, A. -T. and {Nordlander}, T. and {Ocvirk}, P. and {O'Flaherty}, K.~S. and {Olias Sanz}, A. and {Ortiz}, P. and {Osorio}, J. and {Oszkiewicz}, D. and {Ouzounis}, A. and {Palmer}, M. and {Park}, P. and {Pasquato}, E. and {Peltzer}, C. and {Peralta}, J. and {P{\'e}turaud}, F. and {Pieniluoma}, T. and {Pigozzi}, E. and {Poels}, J. and {Prat}, G. and {Prod'homme}, T. and {Raison}, F. and {Rebordao}, J.~M. and {Risquez}, D. and {Rocca-Volmerange}, B. and {Rosen}, S. and {Ruiz-Fuertes}, M.~I. and {Russo}, F. and {Sembay}, S. and {Serraller Vizcaino}, I. and {Short}, A. and {Siebert}, A. and {Silva}, H. and {Sinachopoulos}, D. and {Slezak}, E. and {Soffel}, M. and {Sosnowska}, D. and {Strai{\v{z}}ys}, V. and {ter Linden}, M. and {Terrell}, D. and {Theil}, S. and {Tiede}, C. and {Troisi}, L. and {Tsalmantza}, P. and {Tur}, D. and {Vaccari}, M. and {Vachier}, F. and {Valles}, P. and {Van Hamme}, W. and {Veltz}, L. and {Virtanen}, J. and {Wallut}, J. -M. and {Wichmann}, R. and {Wilkinson}, M.~I. and {Ziaeepour}, H. and {Zschocke}, S.},
        title = "{The Gaia mission}",
      journal = {\aap},
         year = 2016,
        month = nov,
       volume = {595},
          eid = {A1},
        pages = {A1},
          doi = {10.1051/0004-6361/201629272},
archivePrefix = {arXiv},
       eprint = {1609.04153},
 primaryClass = {astro-ph.IM},
       adsurl = {https://ui.adsabs.harvard.edu/abs/2016A&A...595A...1G}
}

@ARTICLE{gagne2018,
       author = {{Gagn{\'e}}, Jonathan and {Mamajek}, Eric E. and {Malo}, Lison and {Riedel}, Adric and {Rodriguez}, David and {Lafreni{\`e}re}, David and {Faherty}, Jacqueline K. and {Roy-Loubier}, Olivier and {Pueyo}, Laurent and {Robin}, Annie C. and {Doyon}, Ren{\'e}},
        title = "{BANYAN. XI. The BANYAN {\ensuremath{\Sigma}} Multivariate Bayesian Algorithm to Identify Members of Young Associations with 150 pc}",
      journal = {\apj},
         year = 2018,
        month = mar,
       volume = {856},
       number = {1},
          eid = {23},
        pages = {23},
          doi = {10.3847/1538-4357/aaae09},
archivePrefix = {arXiv},
       eprint = {1801.09051},
 primaryClass = {astro-ph.SR},
       adsurl = {https://ui.adsabs.harvard.edu/abs/2018ApJ...856...23G}
}

@ARTICLE{garcia2023,
       author = {{Garc{\'\i}a}, R.~A. and {Gourv{\`e}s}, C. and {Santos}, A.~R.~G. and {Strugarek}, A. and {Godoy-Rivera}, D. and {Mathur}, S. and {Delsanti}, V. and {Breton}, S.~N. and {Beck}, P.~G. and {Brun}, A.~S. and {Mathis}, S.},
        title = "{Stellar spectral-type (mass) dependence of the dearth of close-in planets around fast-rotating stars. Architecture of Kepler confirmed single-exoplanet systems compared to star-planet evolution models}",
      journal = {\aap},
         year = 2023,
        month = nov,
       volume = {679},
          eid = {L12},
        pages = {L12},
          doi = {10.1051/0004-6361/202346933},
archivePrefix = {arXiv},
       eprint = {2311.00108},
 primaryClass = {astro-ph.EP},
       adsurl = {https://ui.adsabs.harvard.edu/abs/2023A&A...679L..12G}
}

@ARTICLE{kral2020,
       author = {{Kral}, Quentin and {Matr{\`a}}, Luca and {Kennedy}, Grant M. and {Marino}, Sebastian and {Wyatt}, Mark C.},
        title = "{Survey of planetesimal belts with ALMA: gas detected around the Sun-like star HD 129590}",
      journal = {\mnras},
         year = 2020,
        month = sep,
       volume = {497},
       number = {3},
        pages = {2811-2830},
          doi = {10.1093/mnras/staa2038},
archivePrefix = {arXiv},
       eprint = {2005.05841},
 primaryClass = {astro-ph.EP},
       adsurl = {https://ui.adsabs.harvard.edu/abs/2020MNRAS.497.2811K}
}

@INPROCEEDINGS{rando2020,
       author = {{Rando}, N. and {Asquier}, J. and {Corral Van Damme}, C. and {Isaak}, K. and {Ratti}, F. and {Safa}, F. and {Southworth}, R. and {Broeg}, C. and {Beck}, T. and {Benz}, W. and {Fortier}, A. and {Beck}, M. and {Borges}, A. and {Palacios}, R. and {Martin}, J. and {De Miguel}, D. and {Pizarro}, A.},
        title = "{CHEOPS, the ESA mission for exo-planets characterization: early operations and commissioning results}",
    booktitle = {Space Telescopes and Instrumentation 2020: Optical, Infrared, and Millimeter Wave},
         year = 2020,
       editor = {{Lystrup}, Makenzie and {Perrin}, Marshall D.},
       series = {Society of Photo-Optical Instrumentation Engineers (SPIE) Conference Series},
       volume = {11443},
        month = dec,
          eid = {1144314},
        pages = {1144314},
          doi = {10.1117/12.2567296},
       adsurl = {https://ui.adsabs.harvard.edu/abs/2020SPIE11443E..14R}
}
\bibliographystyle{aasjournalv7}

\end{document}